\documentclass[a4paper,11pt]{article}
\pdfoutput=1 
\usepackage{jheppub} 
\usepackage{bbm,bm,graphicx,mathtools,color,hyperref,slashed}
\usepackage{amssymb,amsmath}

\usepackage[caption=false]{subfig}

\usepackage[all]{xy}

\graphicspath{{./Figures/}{./figures_TM/}}

\newcommand{\diff}{\mathrm{d}}
\newcommand{\p}{\partial}
\newcommand{\ve}{\varepsilon}
\newcommand{\Diff}{{\mathcal{D}}}

\newcommand{\be}{\begin{equation}}      
\newcommand{\ee}{\end{equation}}      
\newcommand{\bea}{\begin{eqnarray}}      
\newcommand{\eea}{\end{eqnarray}}

\newcommand{\tr}{\mathrm{tr}}

\newcommand{\im}{\mathrm{i}}

\newcommand{\calA}{\mathcal{A}}

\newcommand{\calU}{\mathcal{U}}

\newcommand{\rmc}{\mathrm{c}}
\newcommand{\rme}{\mathrm{e}}

\newcommand{\rmA}{\mathrm{A}}
\newcommand{\rmL}{\mathrm{L}}
\newcommand{\rmR}{\mathrm{R}}
\newcommand{\rmV}{\mathrm{V}}

\newcommand{\R}{\mathbb{R}}
\newcommand{\Z}{\mathbb{Z}}

\title{Fractional $\theta$ angle, 't~Hooft anomaly, and quantum instantons in charge-$q$ multi-flavor Schwinger model}

\author[1,2,3]{Tatsuhiro Misumi,}
\author[4]{Yuya Tanizaki,}
\author[4]{and Mithat \"{U}nsal}

\affiliation[1]{Department of Mathematical Science, Akita University, Akita 010-8502, Japan}
\affiliation[2]{iTHEMS Program, RIKEN, Wako 351-0198, Japan}
\affiliation[3]{Research and Education Center for Natural Sciences, Keio University, Kanagawa 223-8521, Japan}
\affiliation[4]{Department of Physics, North Carolina State University, Raleigh, NC 27607, USA}

\emailAdd{misumi@phys.akita-u.ac.jp}
\emailAdd{ytaniza@ncsu.edu}
\emailAdd{unsal.mithat@gmail.com}

\abstract{
This work examines non-perturbative dynamics of a $2$-dimensional QFT by using discrete 't~Hooft anomaly, semi-classics with circle compactification and bosonization. We focus on  charge-$q$ $N$-flavor Schwinger model,  and also Wess-Zumino-Witten model.
We first apply the recent developments of discrete 't~Hooft anomaly matching to theories on   $\mathbb{R}^2$ and its compactification to $\mathbb{R} \times S^1_L$. 
We  then compare the 't~Hooft anomaly with dynamics of the models by explicitly constructing eigenstates and calculating physical quantities on the cylinder spacetime with periodic and flavor-twisted boundary conditions.  We find different boundary conditions realize different anomalies. 
Especially under the twisted boundary conditions, there are $Nq$ vacua associated with discrete chiral symmetry breaking. Chiral condensates for this case have fractional $\theta$ dependence $\mathrm{e}^{\mathrm{i} \theta/Nq}$, which provides the $Nq$-branch structure with soft fermion mass.  
We show that these behaviors at a small circumference cannot be explained by usual instantons but should be understood by ``quantum'' instantons, which saturate the BPS bound between classical action and quantum-induced effective potential.
The effects of the quantum-instantons  match  the exact results obtained via bosonization within the region of applicability of semi-classics. We also argue  that large-$N$ limit of the Schwinger model with twisted boundary conditions satisfy volume independence. 
}

\begin{document}
\maketitle

\section{Introduction and Summary}\label{sec:intro}

Low-energy behaviors of quantum gauge theories are still one of the biggest and the most interesting problems in contemporary theoretical physics. Despite the fact that we are getting descriptions of confinement, chiral symmetry breaking, dynamical mass   generation in compactified gauge theories on $\R^3 \times S^1_L$ within semi-classics~\cite{Dunne:2016nmc},  
 it is still a very hard task to make those ideas into  useful tools on $\R^4$
and obtain  reliable computations of physical quantities. 
Two-dimensional quantum field theories  host a number of exactly solvable cases, and may provide useful perspective to deepen such ideas.   In that regards, it  provides 
a useful play-ground to understand the non-perturbative dynamics and behavior of the theory upon compactification. 
With these goals in mind, we examine certain two-dimensional QFTs by using discrete 't~Hooft anomaly, semi-classics (including Hamiltonian formalism)  and bosonization.

Schwinger model is one of such an example~\cite{Schwinger:1962tp}. It is a $2$-dimensional QED with one massless Dirac fermion, and the photon excitation becomes massive despite the gauge invariance~\cite{Schwinger:1962tp, Schwinger:1962tn}. 
Similarities with $4$-dimensional QCD are not limited to this phenomenon, and this $2$d QED model also shows charge screening/confinement, presence of instantons and $\theta$ vacua, and so on~\cite{Lowenstein:1971fc,Casher:1974vf, Coleman:1975pw}. 
Furthermore, massless Schwinger model is exactly solvable on various spacetime, such as cylinder~\cite{Manton:1985jm, Hetrick:1988yg}, two-sphere~\cite{Jayewardena:1988td}, and two-torus~\cite{Sachs:1991en}. 
Because of this exact solvability, variants of Schwinger models have been used as a benchmark to test methods against the fermion sign problem in numerical Monte Carlo simulations~\cite{Banuls:2015sta, Banuls:2016lkq, Buyens:2016ecr, Buyens:2016hhu, Tanizaki:2016xcu, Alexandru:2018ngw}. 

We would like to note that charge-$1$ $1$-flavor Schwinger model is analogous to $4$d QCD with a $1$-flavor fermion. Chiral symmetry does not appear even if we turn off the fermion mass because of Adler-Bell-Jackiw (ABJ) anomaly~\cite{Adler:1969gk, Bell:1969ts}, and the non-vanishing chiral condensate does not break global symmetry. 
The situation becomes completely different if we consider $N\ge 2$ flavors of fermions~\cite{Coleman:1976uz, Affleck:1985wa}, essentially because the theory has $SU(N)_\rmL\times SU(N)_\rmR$ chiral symmetry. 
Because of Coleman-Mermin-Wagner theorem~\cite{Coleman:1973ci, mermin1966absence}, the chiral condensate must vanish, $\langle \overline{\psi}\psi\rangle=0$, but still the system shows the algebraic-long-range order, or conformal behavior, as in Kosterlitz-Thouless phase~\cite{Kosterlitz:1973xp}. 
Fermion mass breaks this chiral symmetry explicitly, and thus it becomes an interesting question to ask how the fermion mass changes the vacuum structure.

In this paper, we take one step further and consider charge-$q$ $N$-flavor Schwinger models.  
This extension recently gets some attention since $q=2$ case is found to appear on the high-temperature domain wall of $\mathcal{N}=1$ $SU(2)$ super Yang-Mills theory in Refs.~\cite{Anber:2018jdf, Anber:2018xek}. Also, in Ref.~\cite{Armoni:2018bga}, the authors propose the string construction of the model with $q=2$ and $N=8$ to find the potential between $D$-brane and orientifold plane in a non-supersymmetric setup. 
In both papers, the recent development of 't~Hooft anomaly matching plays an important role in their analysis, but full structure of anomaly is not yet studied. 

In this work, we will first figure out the full structure of 't~Hooft anomaly by gauging the whole internal symmetry. 
Then, we shall discuss its physical consequences with the help of semiclassics after circle compactifications, where we explicitly construct eigenstates under periodic and flavor-twisted boundary conditions. This leads to explicit calculations of chiral condensate and Polyakov loop.
For the twisted boundary condition, we find $Nq$ vacua associated with discrete chiral symmetry breaking and chiral condensate with fractional $\theta$ dependence $\rme^{\im \theta/Nq}$, leading to the $Nq$-branch structure with soft fermion mass. 
We will emphasize that the (fractional) quantum instantons, which saturate the BPS bound between classical action and quantum-induced effective potential, have a direct consequence on the physical quantities.
In addition to these outcomes, we will derive the expression of chiral condensate valid for all the range of the circumference, gain new insights into the volume independence,  and investigate the twist-compactified WZW models as dual theories of the Schwinger models.

In the following, let us summarize the main results of each section. 

In Sec.~\ref{sec:general_Schwinger}, we discuss symmetry and anomaly of charge-$q$ $N$-flavor massless Schwinger model. 
Symmetry group of this theory consists of $1$-form symmetry $G^{[1]}=\mathbb{Z}_q^{[1]}$ and $0$-form chiral symmetry $G^{[0]}$, 
\be
G=G^{[1]}\times G^{[0]}=\mathbb{Z}_q^{[1]}\times {SU(N)_\rmL\times SU(N)_\rmR\times (\mathbb{Z}_{q N})_\rmR\over (\mathbb{Z}_N)_\rmV\times (\mathbb{Z}_N)_\rmR}. 
\ee
Unlike the case of charge-$1$ Schwinger model, ABJ anomaly does not spoil $U(1)_\rmR$ chiral symmetry completely, and there is a discrete remnant $(\mathbb{Z}_{Nq})_\rmR$, whose $\mathbb{Z}_N$ subgroup is the same with the center of $SU(N)_\rmR$. 
We will find the 't~Hooft anomaly of $G$ by identifying the $3$d topological action that cancels the anomaly by anomaly-inflow mechanism. 
Through this computation, we find that there is an interesting subgroup, 
\be
G_{\rm sub}=\mathbb{Z}_q^{[1]}\times {SU(N)_\rmV \over (\mathbb{Z}_{N})_\rmV}\times (\mathbb{Z}_{qN})_\rmR \subset G, 
\ee
which has $\mathbb{Z}_{Nq}$ discrete 't~Hooft anomaly including $\mathbb{Z}_{Nq}$ two-form gauge fields, and this anomaly is important to discuss the IR realization of chiral symmetry. 
This $\mathbb{Z}_{Nq}$ 't~Hooft anomaly is the refinement of $\mathbb{Z}_q$ anomaly discussed in previous studies~\cite{Anber:2018jdf, Anber:2018xek, Armoni:2018bga}. 

Using the help of non-Abelian bosonization, we identify how these anomalies are matched in two-dimensions. The result can be summarized in the following table:
 \begin{center}
\begin{tabular}{ | l | l | l | } 
 \hline
 $(q,N)$ & Mass gap  & Symmetry breaking\\ 
 \hline 
 $(1,1)$  & $e^2/\pi^2$ & No symmetry \\  
 \hline
 $(q,1)$  & $q^2e^2/\pi^2$ & $ (\mathbb{Z}_{q})_\rmR\to 1$ with condensate $\langle \overline{\psi}\psi\rangle$ \\ 
 \hline 
 $(1, N)$ & $SU(N)_1$ WZW CFT & Symmetry is unbroken \\
 \hline
 $(q,N)$ & $SU(N)_1$ WZW CFT & $(\mathbb{Z}_{N q})_\rmR\to (\mathbb{Z}_{N})_\rmR$ with condensate $\langle \mathrm{det}(\overline{\psi}^f\psi^{f'})\rangle$ \\
\hline
\end{tabular}
\end{center}
If we consider the charge-$q$ model, the discrete chiral symmetry is broken so that we have $q$ disconnected components of vacua. 
When there are multi-flavor fermions, low-energy properties on each vacuum is given by the level-$1$ $SU(N)$ Wess-Zumino-Witten conformal field theory. 

In this paper, we put charge-$q$ $N$-flavor Schwinger model on the cylinder $\mathbb{R}\times S^1$, with the circumference $L$. 
For $N\ge 2$, the 't~Hooft anomaly on the cylinder depends on the fermion boundary condition, and the result can be summarized as follows:
 \begin{center}
\begin{tabular}{ | l | l | l | l | } 
 \hline
 Fermion b.c. & Anomaly  & Prediction on chiral SSB & Remnant of $2$d CFT\\ 
 \hline 
 Thermal  & $\mathbb{Z}_q$ & $ (\mathbb{Z}_{Nq})_\rmR\to (\mathbb{Z}_N)_\rmR$ & None   \\  
 \hline
 Flavor-twisted  & $\mathbb{Z}_{Nq}$& $ (\mathbb{Z}_{Nq})_\rmR\to 1$ & Extra $(\mathbb{Z}_N)_{\rmR}$ SSB\\ 
 \hline
\end{tabular}
\end{center}
When we take the thermal, or periodic, boundary condition on fermionic fields, only the anomaly involving $\mathbb{Z}_q^{[1]}$ one-form symmetry survives under $S^1$-compactification~\cite{Gaiotto:2017yup}. 
Although this is already interesting since we can predict the spontaneous symmetry breaking of discrete chiral symmetry as $ (\mathbb{Z}_{Nq})_\rmR\to (\mathbb{Z}_N)_\rmR$, we are loosing complete information about continuous chiral symmetry $SU(N)_\rmL\times SU(N)_\rmR$. 
In other words, we find no remnant of $2$-dimensional conformal behavior with thermal compactification. 
Taking the flavor-twisted boundary condition, the story becomes more interesting as we can keep the $\mathbb{Z}_{Nq}$ anomaly of $G_{\mathrm{sub}}$~\cite{Tanizaki:2017qhf}. 
Anomaly predicts the discrete chiral symmetry breaking, $(\mathbb{Z}_{Nq})_\rmR\to 1$, and this extra $\mathbb{Z}_N$ symmetry breaking is expected to be a remnant of algebraic long-range order on $\mathbb{R}^2$. 
This connection to conformal behavior will be explicitly shown by studying $SU(N)$ Wess-Zumino-Witten model with twisted boundary condition, but it is postponed to Sec.~\ref{sec:twisted_wzw} after detailed studies on Schwinger models on $\mathbb{R}\times S^1$.

After some preparation in Sec.~\ref{sec:holonomy} by computing holonomy effective potentials, we construct the ground states of charge-$q$ $N$-flavor Schwinger model on $\mathbb{R}\times S^1$ with both boundary conditions in Sec.~\ref{sec:ChSSB}. 
Interestingly, the number of classical minima of the classical holonomy potential, which is $Nq$ in the charge $q \geq 1, N \geq 1$ is equal to the number of ground states in the compactified quantum theory. This phenomena is similar to extended 
$\mathcal{N} =2$ supersymmetric quantum mechanics~\cite{Witten:1982df}, where the number of classical and quantum vacua are the same.  In our case, this fact arises due to subtle new effects involving the zero mode structure of quantum instantons.
Then, we  construct the $\theta$ vacua $|\theta, k\rangle$ with discrete label $k$. 
To contrast the difference between thermal and flavor-twisted compactifications, let us here quote the results only for $N\ge 2$.  

In the thermal boundary condition, the fermion bilinear does not condense, $\langle \overline{\psi}^f_\rmL\psi^f_\rmR\rangle=0$, for any flavors $f=0,1,\ldots, N-1$,  
and the leading condensate is the determinant condensate:
\be
{\langle {\theta,k}| {\rm det}\overline{\psi}^{f}_{\rm L} \psi^{g}_{\rm R} | {\theta,k} \rangle \over{ {\langle {\theta,k}| {\theta,k} \rangle}}}
= \rme^{\im {\theta+2\pi k\over q}} {N! \over{ L^{N}}} \exp\left(-{N\pi\over{Lm_{\gamma}}}\right),\quad (k=0,1,\ldots, q-1), 
\label{eq:intro_cond_pbc}
\ee
with the photon mass $m_\gamma^2=N q^2 e^2/\pi$. 
As anomaly predicted, the discrete chiral symmetry is broken as $(\mathbb{Z}_{Nq})_\rmR\to (\mathbb{Z}_N)_\rmR$, and we have $q$ vacua $|\theta, k\rangle$ with $k=0,1,\ldots, q-1$ with the fractional $\theta$ dependence $\rme^{\im \theta/q}$. 

Taking the flavor-twisted boundary condition, instead, the fermion bilinear condensation appears, 
\be
{\langle \theta,k | \overline{\psi}_{\rm L}^{f} \psi_{\rm R}^{f} | \theta,k \rangle
\over{\langle{\theta,k} | {\theta,k} \rangle}}
={1\over{N L}} \rme^{\im{\theta+2\pi k\over Nq}}\exp\left(-{\pi\over{N Lm_{\gamma}}}\right), \quad (k=0,1,\ldots, Nq-1). 
\label{eq:intro_cond_tbc}
\ee
Discrete chiral symmetry is spontaneously broken as $(\mathbb{Z}_{Nq})_\rmR\to 1$, and the fractional $\theta$ dependence becomes $\rme^{\im \theta/Nq}$. 

Even though the theta vacua $|\theta, k\rangle$ satisfy the cluster decomposition properties about $2$d local correlators, such as those of chiral condensates $\overline{\psi}_\rmL\psi_\rmR$, this is not true for correlators of Polyakov loop $P=\exp(\im\int_{S^1} a)$. 
Indeed, in both boundary conditions, we find that 
\be
\lim_{\tau\to \infty}{\langle \theta,k| P(\tau)^\dagger P(0) | \theta,k\rangle\over \langle \theta,k|\theta,k\rangle}=\exp\left(-{\pi m_\gamma L\over 2q^2 N}\right), 
\ee
while $\langle \theta,k|P|\theta,k\rangle=0$. 
Correspondingly, $\mathbb{Z}_q^{[1]}$ one-form symmetry is spontaneously broken in $2$d decompactification limit. 
We further clarify that the impossibility to achieve the cluster decomposition for both $\overline{\psi}_\rmL\psi_\rmR$ and $P$ is exactly the way anomaly matching is satisfied in this theory on $\mathbb{R}\times S^1$. 

In Sec.~\ref{sec:fractional_quantum_instanton}, we revisit the computation of chiral symmetry breaking on $\mathbb{R}\times S^1$ with semiclassical approximation of path integral, and we rediscover importance of fractional ``quantum'' instanton, or fracton by Smilga~\cite{Smilga:1993sn} and by Shifman and Smilga~\cite{Shifman:1994ce}. 
The exponents of chiral condensates in (\ref{eq:intro_cond_pbc}) and (\ref{eq:intro_cond_tbc}) depend on the gauge coupling $e$ as 
\be
S_\mathcal{F}={\#\over eL}, 
\ee
with a numerical constant $\#$ that depends on $q,N$ and boundary conditions, and this is unusual as field-theoretic instanton action, which is typically $\sim 1/e^2$. 
We show that this has to occur as a BPS bound of Maxwell kinetic term, $\sim 1/e^2$, and quantum-induced $1$-loop potential, $\sim 1$. 
It is notable that this semiclassical object has a direct consequence on physical observables.

So far, we have limited ourselves to the massless Schwinger models. In Sec.~\ref{sec:massive_fermion}, we discuss the effect of flavor-degenerate soft fermion mass $m_\psi$. 
We first show that the charge conjugation $\mathsf{C}$ at $\theta=\pi$ has a mixed anomaly with other symmetries if $qN\ge 2$ is even, and that global inconsistency exists with $\mathsf{C}$ and other symmetries between $\theta=0$ and $\pi$ if $qN\ge 3$ is odd. 
This explains the spontaneous breakdown of $\mathsf{C}$ at $\theta=\pi$. Indeed, taking the flavor-twisted boundary condition, we have $Nq$ branch structure with the (meta-stable) ground-state energies,
\be
E_k(\theta)=-2m_\psi \exp\left(-{\pi\over Nm_\gamma L}\right) \cos \left({\theta+2\pi k\over Nq}\right). 
\ee
For $-\pi<\theta<\pi$, the ground state is uniquely determined as $|\theta,0\rangle$, but $|\theta,0\rangle$ and $|\theta,-1\rangle$ are degenerate at $\theta=\pi$. 
We obtain this result both by mass perturbation and by dilute gas approximation of fractional quantum instantons. 

This multi-branch ground state energies can be observed, since $E_n(\theta)-E_0(\theta)$ is nothing but the string tension of charge-$n$ test particle. 
Especially, string tension for $n=\pm 1$ vanishes for $\theta=\pi$, while others do not vanish, and this is consistent with anomaly or global inconsistency. 

In Sec.~\ref{sec:vol_indep}, we discuss the large-$N$ volume independence of multi-flavor Schwinger model. 
We argue that the large-$N$ volume independence fails for the thermal boundary condition, while it is intact with the flavor-twisted boundary condition. 

This paper is organized as follows. 
In Sec.~\ref{sec:general_Schwinger}, we discuss symmetry and anomaly of  charge-$q$ $N$-flavor massless Schwinger model. 
In Sec.~\ref{sec:holonomy}, we compute the holonomy effective potential on $\mathbb{R}\times S^1$ with thermal and flavor-twisted boundary conditions. 
In Sec.~\ref{sec:ChSSB}, we perform the quantum-mechanical treatment of this setup, and discuss properties of the ground states, especially about chiral condensate and Polyakov loop.  
In Sec.~\ref{sec:fractional_quantum_instanton}, we provide their semiclassical interpretation as quantum instanton. 
In Sec.~\ref{sec:massive_fermion}, we discuss the effect of soft fermion mass. 
In Sec.~\ref{sec:vol_indep}, we discuss the large-$N$ volume independence of charge-$q$ $N$-flavor Schwinger model. 
In Sec.~\ref{sec:twisted_wzw}, we study the $SU(N)$ Wess-Zumino-Witten model in twisted boundary condition, and see the connection between conformal behavior in $2$d and ground-state degeneracy on $\mathbb{R}\times S^1$. 
We conclude in Sec.~\ref{sec:conclusion}. 
We fix the convention of $2$d Dirac spinor in Appendix~\ref{sec:convention_spinor}. 
In Appendix~\ref{sec:holonomy_bosonization}, we derive the holonomy effective potential in the language of Abelian bosonization.


\section{Anomaly of charge-$q$ multi-flavor Schwinger model}\label{sec:general_Schwinger} 

The Schwinger model is a $(1+1)$ dimensional quantum electrodynamics (QED) of one massless Dirac fermion with minimal electric charge~\cite{Schwinger:1962tp}. 
This model has acquired a lot of attention because it can be exactly solved, while the theory contains many nonperturbative phenomena similar to those of QCD: mass gap of photons, nonvanishing chiral condensate, and so on. 
Furthermore, various correlation functions can be computed not only on $\mathbb{R}^2$, but also on other two-dimensional manifolds, like cylinder $\mathbb{R}\times S^1$~\cite{Manton:1985jm, Hetrick:1988yg} and torus $T^2$~\cite{Sachs:1991en}. 
Despite its interesting features, the low-energy properties of the usual Schwinger model are rather trivial. 
This is mainly because the Schwinger model does not have global symmetries except for Poincare symmetry, and thus interesting phenomena like spontaneous symmetry breaking do not occur at all. 

We therefore consider generalization of Schwinger model to have an interesting low-energy physics while keeping its solvability. 
In this section, we discuss general properties of charge-$q$ $N$-flavor massless Schwinger model, especially by paying attention to symmetry and its 't~Hooft anomaly. 
This generalization of Schwinger model has been recently discussed in Refs.~\cite{Anber:2018jdf, Anber:2018xek, Armoni:2018bga}.

\subsection{Symmetry of charge-$q$ $N$-flavor Schwinger model}\label{sec:symmetry}

The Euclidean action $S$ of charge-$q$ $N$-flavor massless Schwinger model is given by 
\be
S={1\over 2e^2}\int_{M_2} |\diff a|^2+{\im \theta\over 2\pi}\int_{M_2} \diff a+\sum_{f=1}^{N}\int_{M_2} \diff^2 x\,  \overline{\psi}^f \gamma^{\mu}(\p_{\mu}+ q \im a_{\mu})\psi^f. 
\label{eq:Lagrangian_Schwinger}
\ee
Here, $a$ is the $U(1)$ gauge field, which is canonically normalized as $\int \diff a \in 2\pi\mathbb{Z}$ for any closed two-manifolds, $e$ is the gauge coupling with the mass dimension $1$, and $\psi^f$ and $\overline{\psi}^f$ are two-dimensional Dirac fermions with the flavor label $f=1,\ldots, N$.  When the flavor structure is evident, the flavor indices are suppressed below. 
For convention of two-dimensional spinors, see Appendix~\ref{sec:convention_spinor}. 
These fermions have charge $q\in \mathbb{Z}$ under the $U(1)$ gauge group. 
The first term is the Maxwell kinetic term of the photon fields, and the second one is the topological theta term with $\theta\sim \theta+2\pi$. 
$U(1)$ gauge transformation of this theory is given by 
\be
a\mapsto a+\diff \lambda,\; \psi\mapsto \rme^{-q\im \lambda}\psi,\; \overline{\psi}\mapsto \overline{\psi} \rme^{q\im \lambda},
\ee
where the gauge parameter $\lambda$ is $2\pi$-periodic compact scalar fields. 

Let us identify the internal global symmetry of this theory, including higher-form symmetry. 
We will show that the theory has the $0$-form symmetry
\be
G^{[0]}={SU(N)_\rmL\times SU(N)_\rmR\times (\mathbb{Z}_{q N})_\rmR\over (\mathbb{Z}_N)_\rmV\times (\mathbb{Z}_N)_\rmR}, 
\label{eq:0form_symmetry}
\ee
and the $1$-form symmetry $G^{[1]}=\mathbb{Z}_q$. Let us denote these symmetries at the same time as 
\be
G=G^{[1]}\times G^{[0]}=\mathbb{Z}_q^{[1]}\times {SU(N)_\rmL\times SU(N)_\rmR\times (\mathbb{Z}_{q N})_\rmR\over (\mathbb{Z}_N)_\rmV\times (\mathbb{Z}_N)_\rmR}. 
\label{eq:symmetry}
\ee

First, we discuss the $0$-form symmetry. Since the Dirac fermions $\psi$ are massless, the Lagrangian is invariant under independent unitary transformations on the right-handed fermions $\psi_\rmR$ and the left-handed fermions $\psi_\rmL$. Therefore, the Lagrangian is invariant under 
\be
U(N)_\rmL\times U(N)_\rmR={SU(N)_\rmL\times U(1)_\rmL\over (\mathbb{Z}_N)_\rmL}\times {SU(N)_\rmR\times U(1)_\rmR \over (\mathbb{Z}_N)_\rmR}. 
\ee
Since the vector-like $U(1)$ symmetry is gauged, the global symmetry of the classical action is given as\footnote{Here, we rewrite the Abelian symmetry as $U(1)_\rmL\times U(1)_\rmR=U(1)_\rmV\times U(1)_\rmR$ using the vector-like symmetry $U(1)_\rmV$, but we do not introduce the axial symmetry $U(1)_\rmA$, $\psi\mapsto \rme^{\im \beta \gamma}\psi$. This is because the $\pi$ rotations of $U(1)_\rmV$ and $U(1)_\rmA$ both give the $\mathbb{Z}_2$ fermion parity, and thus the group structure becomes slightly complicated as $U(1)_\rmL\times U(1)_\rmR=[U(1)_\rmV\times U(1)_\rmA]/\mathbb{Z}_2$. Similarly, we use $(\mathbb{Z}_N)_\rmL\times (\mathbb{Z}_N)_\rmR=(\mathbb{Z}_N)_\rmV\times (\mathbb{Z}_N)_\rmR$ to find (\ref{eq:classical_symmetry}). Similar identification of symmetry turns out to be useful also for $4$-dimensional QCD~\cite{Tanizaki:2018wtg}. } 
\be
G^{[0]}_{\mathrm{classical}}=
{SU(N)_\rmL\times SU(N)_\rmR\times U(1)_\rmR\over (\mathbb{Z}_N)_\rmV\times (\mathbb{Z}_N)_\rmR}. 
\label{eq:classical_symmetry}
\ee
This is, however, not the symmetry of quantum theory, since the path-integral measure $\Diff\overline{\psi}\Diff \psi$ is not invariant~\cite{Fujikawa:1979ay, Fujikawa:1980eg} due to ABJ anomaly~\cite{Adler:1969gk, Bell:1969ts}. 
In this case, $U(1)_\rmR$ transformation $\psi_\rmR\mapsto \rme^{\im \alpha}\psi_\rmR$, $\overline{\psi}_{\rmR}\mapsto \rme^{-\im \alpha}\overline{\psi}_\rmR$ changes the path-integral measure as 
\be
\Diff \overline{\psi}\Diff \psi\mapsto \Diff \overline{\psi}\Diff \psi \exp\left(\im \alpha{N q\over 2\pi}\int \diff a\right). 
\ee
Because $\int \diff a\in 2\pi \mathbb{Z}$, this transformation is the symmetry only if $\alpha$ is quantized to integer multiples of $2\pi/(qN)$. 
Therefore, $G^{[0]}_{\mathrm{classical}}$ is explicitly broken down to (\ref{eq:0form_symmetry}). 
The ABJ anomaly indicates that the theta angle can be shifted as $\theta\mapsto \theta+qN\alpha$ by performing chiral transformation once we fix the UV regularization, and thus we can set $\theta=0$ without loss of generality if massless fermions exist.

Next, we discuss the $1$-form symmetry~\cite{Gaiotto:2014kfa}. 
Before discussing its mathematical construction, let us explain the physical meaning of $\mathbb{Z}_q$ one-form symmetry. 
We consider the Wilson loop of charge $k$, 
\be
W_k(C)=\exp\left(\im k \int_C a\right), 
\ee
and we are interested in its behavior as $C$ gets larger. Taking $C$ as rectangle with area $T\times R$, its expectation value measures the potential between test particles with charge $k$ and $-k$ at separated points;
\be
-{1\over T}\ln \langle W_k(C)\rangle=V_k(R). 
\ee
Since there are dynamical particles with charge $q$ that repeat pair creation/annihilation from vacuum, the electric charge of test particles makes sense only as $k\bmod q$ after quantization.
It is therefore natural to expect that we have a symmetry operation to measure the test charge modulo $q$. 
The one-form symmetry $\mathbb{Z}_q^{[1]}$ justifies this physical intuition. 

In order to construct the $\mathbb{Z}_q^{[1]}$ one-form transformation, we introduce sufficiently fine local patches $\{\calU_i\}_i$ of our two-dimensional spacetime $M_2$. 
Gauge field $a$ is a set of $\mathbb{R}$-valued one-form fields $a_i$ on $\calU_i$ with the connection formula, 
\be
a_j=a_i-\im g_{ij}^{-1}\diff g_{ij},
\ee
where $g_{ij}$ is the $U(1)$-valued transition function on the double overlap $\calU_{ij}=\calU_i\cap \calU_j$. 
The Dirac fields $\psi_f$ are also spinor-valued fields $(\psi_f)_i$ on each local path $\calU_i$ with the connection formula, 
\be
(\psi_f)_j=(g_{ij})^{-q} (\psi_f)_i. 
\ee 
Since we take sufficiently fine cover of $M_2$, the double overlaps $\calU_{ij}$ can be regarded as the codimension-$1$ submanifolds (i.e. walls) of $M_2$, and we assign the $U(1)$ transformation $g_{ij}$ on each wall. 
We require the cocycle condition on the triple overlap $\calU_{ijk}=\calU_i\cap \calU_j\cap \calU_k$ as 
\be
g_{ij}g_{jk}g_{ki}=1, 
\ee
which gives the canonical normalization condition, $\int \diff a\in2\pi \mathbb{Z}$. 
This condition says that the $U(1)$ transformations on the walls must satisfy the group multiplication law at the junction $\calU_{ijk}$ of three walls, $\calU_{ij}$, $\calU_{jk}$, and $\calU_{ki}$. 
We can readily find that we can construct the codimension-$2$ defect by driving a hole at the junction and inserting Aharonov-Bohm flux quantized to $2\pi/q$. 
That is, we perform the transformation, 
\be
g_{ij}\mapsto g'_{ij}=g_{ij}\rme^{2\pi\im n_{ij}/q}, 
\ee
so that 
\be
g'_{ij}g'_{jk}g'_{ki}=1
\ee
outside the defect but 
\be
g'_{ij}g'_{jk}g'_{ki}=\rme^{2\pi\im/q}. 
\ee
at the defect. Since the connection formulas of gauge fields and dynamical fermions are unaffected by this transformation, insertion of this defect is a topological operation, i.e. a symmetry transformation. 
Since the test particle can feel the Aharonov-Bohm flux around this defect, the  above transformation acts on the Wilson loop as
\be
W_k(C)\mapsto W_k(C) \rme^{2\pi\im k/q}, 
\ee
when $C$ links to the defect. This operation, $\mathbb{Z}_q^{[1]}$, thus measures the charge of test particle modulo $q$, as we have expected from physical arguments. 
We therefore identify the internal symmetry group as (\ref{eq:symmetry}). 

\subsection{'t~Hooft anomaly of symmetry}
\label{sec:anomaly}

Let us briefly review 't~Hooft anomaly matching~\cite{tHooft:1979rat, Frishman:1980dq} in a modern terminology~\cite{Wen:2013oza,  Kapustin:2014zva, Wang:2014pma}. 
We introduce the background gauge field $\calA$ for symmetry $G$, and denote its partition function as $Z_{M_2}[\calA]$. 
In general, this partition function cannot become gauge-invariant under the gauge-transformation of background fields, $\calA\mapsto \calA+\delta_{\theta}\calA$, and it contains the phase ambiguity, 
\be
Z_{M_2}[\calA+\delta_{\theta}\calA]=Z_{M_2}[\calA]\exp\left(\im \int_{M_2} f(\theta,\calA)\right). 
\ee
As indicated in the above expression, if the phase ambiguity depends only on the background fields and their gauge-transformation parameter, we call it an 't~Hooft anomaly (of Dijkgraaf-Witten type). 
An important observation is that the anomaly can be canceled by the boundary contribution of $3$-dimensional topological $G$-gauge theory $S_3[\calA]$, 
\be
\delta_{\theta}S_3[\calA]=\int_{M_2}f(\theta,\calA), 
\ee
so that $Z_{M_2}[\calA]\exp(-\im S_3[A])$ is gauge invariant. 
This shows the 't~Hooft anomaly matching condition  by anomaly-inflow mechanism~\cite{Callan:1984sa}. 
See, for example, Refs.~\cite{Witten:2016cio, Tachikawa:2016cha, Gaiotto:2017yup, Tanizaki:2017bam, Komargodski:2017dmc, Komargodski:2017smk, Shimizu:2017asf, Wang:2017loc,Gaiotto:2017tne, Tanizaki:2017qhf, Tanizaki:2017mtm, Yamazaki:2017dra, Guo:2017xex,  Sulejmanpasic:2018upi, Tanizaki:2018xto, Cordova:2018acb, Anber:2018tcj, Anber:2018jdf, Tanizaki:2018wtg,  Anber:2018xek, Armoni:2018bga, Hongo:2018rpy, Yonekura:2019vyz} for recent applications in various contexts. 

In Refs.~\cite{Anber:2018jdf, Anber:2018xek, Armoni:2018bga}, 't~Hooft anomaly of charge-$q$ $N$-flavor Schwinger model has been partly discussed, but the description there is not complete. 
In this subsection, we are going to give the complete description regarding the 't~Hooft anomaly of internal symmetry $G$. 

\subsubsection{Background gauge fields of internal symmetry $G$}
\label{sec:background_gauge_field}

To find the 't~Hooft anomaly of charge-$q$ $N$-flavor Schwinger model, we first have to construct the background gauge fields $\calA$ of $G$. 
It consists of 
\begin{itemize}
\item $A_\rmR$: $SU(N)_\rmR$ one-form gauge field,
\item $A_\rmL$: $SU(N)_\rmL$ one-form gauge field,
\item $A_\chi=(A_\chi^{(1)},A^{(0)}_\chi)$: $(\mathbb{Z}_{qN})_\rmR$ one-form gauge field, 
\item $B_\rmV=(B^{(2)}_\rmV, B^{(1)}_\rmV)$: $(\mathbb{Z}_{qN})$ two-form gauge field, 
\item $B_\rmR=(B^{(2)}_\rmR, B^{(1)}_\rmR)$: $(\mathbb{Z}_N)_\rmR$ two-form gauge field. 
\end{itemize}
Here, we regard that the $\mathbb{Z}_n$ $p$-form gauge field $C$ as a pair of $U(1)$ $p$-form and $(p-1)$-form gauge fields $(C^{(p)},C^{(p-1)})$ that satisfy the constraint, $n C^{(p)}=\diff C^{(p-1)}$~\cite{Banks:2010zn}. 
Following \cite{Kapustin:2014gua}, we embed $SU(N)_{\rmR/\rmL}$ gauge fields into $U(N)_{\rmR/\rmL}$ gauge fields, which locally looks as 
\be
\widetilde{A}_\rmR=A_\rmR+{1\over N} B^{(1)}_\rmV+{1\over N}B^{(1)}_\rmR,\; 
\widetilde{A}_\rmL=A_\rmL+{1\over N} B^{(1)}_\rmV. 
\label{eq:U(N)_local}
\ee
With these background fields, the fermion kinetic term is replaced as 
\be
\overline{\psi}_\rmR \gamma^{\mu}(\p_\mu+\im [q a_{\mu}+\widetilde{A}_{\rmR,\mu}+A^{(1)}_{\chi,\mu}])\psi_\rmR+\overline{\psi}_\rmL \gamma^{\mu}(\p_\mu+\im [q a_{\mu}+\widetilde{A}_{\rmL,\mu}])\psi_\rmL
\label{eq:fermion_gauged_kinetic}
\ee

We now postulate the invariance under one-form gauge transformations to find the correct topological structure~\cite{Kapustin:2014gua}. On two-form gauge fields, they are defined as 
\bea
&&B^{(2)}_\rmV\mapsto B^{(2)}_\rmV+\diff \lambda_\rmV,\; B^{(1)}_\rmV\mapsto B^{(1)}_\rmV+q N \lambda_\rmV, \nonumber \\
&&B^{(2)}_\rmR \mapsto B^{(2)}_\rmR+\diff \lambda_\rmR,\; B^{(1)}_\rmV\mapsto B^{(1)}_\rmR+ N \lambda_\rmR,  
\label{eq:1form_trans_01}
\eea
where the gauge parameters $\lambda_{\rmV/\rmR}$ are $U(1)$ one-form gauge fields. 
To make consistency with the local expression (\ref{eq:U(N)_local}) of $U(N)$ gauge fields, we find that 
\be
\widetilde{A}_\rmR\mapsto \widetilde{A}_\rmR+ q \lambda_\rmV+\lambda_\rmR,\; 
\widetilde{A}_\rmL\mapsto \widetilde{A}_\rmL + q \lambda_\rmV. 
\label{eq:1form_trans_02}
\ee
In order to make the gauged fermion kinetic term (\ref{eq:fermion_gauged_kinetic}) be invariant under $1$-form transformations, we have to require that 
\be
a\mapsto a- \lambda_\rmV,\; A^{(1)}_\chi\mapsto A^{(1)}_\chi-\lambda_\rmR. 
\label{eq:1form_trans_03}
\ee
Since the transformation (\ref{eq:1form_trans_03}) is not consistent with $qN A^{(1)}_\chi=\diff A^{(0)}_\chi$, we should replace this constraint equation as $q(N A^{(1)}_\chi+B^{(1)}_\rmR)=\diff A^{(0)}_\chi$. 

Now, the field strength $\diff a$ is no longer gauge invariant, and it should be replaced as $\diff a+B^{(2)}_\rmV$ so that the Maxwell term becomes 
\be
{1\over 2e^2}\int_{M_2}|\diff a+B^{(2)}_\rmV|^2+{\im \theta\over 2\pi}\int_{M_2}(\diff a+B^{(2)}_\rmV).
\ee
Combined with the fermion kinetic term (\ref{eq:fermion_gauged_kinetic}), we obtain the gauged action $S_{\mathrm{gauged}}$.  
Fixing a UV regularization scheme, we can compute the partition function as 
\be
Z_{M_2}[\calA]=\int \Diff a \Diff \overline{\psi}\Diff \psi \exp\left(-S_{\rm gauged}\right).  
\ee
Since the background gauge fields $\calA$ are chiral, the fermion path integral potentially suffers from non-Abelian chiral anomaly. 

\subsubsection{Computation of anomaly by Stora-Zumino procedure}

To find the chiral anomaly, the easiest way is to use the descent equation of Stora-Zumino chain~\cite{Stora:1983ct, Zumino:1983ew}. 
It starts from computing $4$-dimensional Abelian anomaly density, 
\bea
\Omega_4&=&{1\over 4\pi}\tr[(q \diff a+\widetilde{F}_\rmR+F_\chi)^2]-{1\over 4\pi}\tr[(q\diff a+\widetilde{F}_\rmL)^2]\nonumber\\
&=&{1\over 4\pi}\tr[(\widetilde{F}_\rmR+F_\chi)^2-\widetilde{F}_\rmL^2]+{1\over 2\pi}q\diff a\wedge (\tr[\widetilde{F}_\rmR+F_\chi]-\tr[\widetilde{F}_\rmL])\nonumber\\
&=&{1\over 4\pi}\tr[(\widetilde{F}_\rmR+F_\chi)^2-\widetilde{F}_\rmL^2]+{q\over 2\pi}\diff a\wedge \diff (B^{(1)}_\rmR+ N A^{(1)}_\chi). 
\eea
Here, $\widetilde{F}_{\rmR/\rmL}=\diff \widetilde{A}_{\rmR/\rmL}+\im \widetilde{A}_{\rmR/\rmL}^2$, and $F_\chi=\diff A^{(1)}_\chi$. 
Since $q(N A^{(1)}_\chi+B^{(1)}_\rmR)=\diff A^{(0)}_\chi$, the second term of the last line vanishes identically. Therefore, we obtain 
\be
\Omega_4={1\over 4\pi}\tr[(\widetilde{F}_\rmR+F_\chi)^2-\widetilde{F}_\rmL^2]. 
\ee

Let us apply the descent procedure to $\Omega_4$. We should find $\Omega_3^0$, which satisfies $\diff \Omega_3^0=\Omega_4$. 
$\Omega_3^0$ is not uniquely determined, and different ones correspond to different regularization scheme of symmetry generators, so we just have to pick up a favorite one. 
It is easy to check that either of the following ones satisfies the equation; 
\bea
\Omega^{0}_3&=&{1\over 4\pi}\tr\left[(\widetilde{A}_\rmR+A^{(1)}_\chi)\diff\left(\diff (\widetilde{A}_\rmR+A^{(1)}_\chi)+{2\im \over 3}\widetilde{A}_\rmR^2\right)-\widetilde{A}_\rmL\left(\diff \widetilde{A}_\rmL+{2\im \over 3} \widetilde{A}_\rmL^2\right)\right],
\label{eq:tHooft_anomaly_LR}
\eea
or
\bea
\Omega_3^0&=&{1\over 4\pi}\tr\left[\widetilde{A}_\rmR\left(\diff \widetilde{A}_\rmR+{2\im \over 3} \widetilde{A}_\rmR^2\right)-\widetilde{A}_\rmL\left(\diff \widetilde{A}_\rmL+{2\im \over 3} \widetilde{A}_\rmL^2\right)\right]\nonumber\\
&&+{1\over 2\pi}A^{(1)}_\chi\wedge (qN B^{(2)}_\rmV+N B^{(2)}_\rmR)+{N\over 4\pi}A^{(1)}_\chi\wedge \diff A^{(1)}_\chi. 
\label{eq:tHooft_anomaly}
\eea
In the first expression (\ref{eq:tHooft_anomaly_LR}), we take the $\rmL$-$\rmR$ scheme for the whole expression. In the second one (\ref{eq:tHooft_anomaly}), we take the $\rmL$-$\rmR$ scheme for the non-Abelian part, and take the $\rmV$-$\rmA$ scheme for the linear term in terms of $A^{(1)}_\chi$. 
The difference between them is expressed by the total derivative. 

The final step of the descent procedure shows that the three-dimensional topological action,
\be
S_3[\calA]=\int_{M_3} \Omega^0_3, 
\ee
satisfies the anomaly-inflow mechanism so that 
\be
Z_{M_2}[\calA]\exp(-\im S_3[\calA])
\ee
is gauge invariant for $\p M_3=M_2$, when we take the consistent regularizations. 
Therefore, Eq.~(\ref{eq:tHooft_anomaly_LR}), or (\ref{eq:tHooft_anomaly}), characterizes the 't~Hooft anomaly of the charge-$q$ $N$-flavor Schwinger model. 

\subsubsection{Discrete 't~Hooft anomaly and four-fermion interaction}

In this part, let us pay attention to a subgroup $G_{\rm sub}$ of the symmetry $G$: 
\be
G_{\rm sub}=\mathbb{Z}_q^{[1]}\times {SU(N)_\rmV \over (\mathbb{Z}_{N})_\rmV}\times (\mathbb{Z}_{qN})_\rmR \subset G. 
\ee
That is, continuous chiral symmetry $SU(N)_\rmL\times SU(N)_\rmR$ is restricted to its diagonal subgroup $SU(N)_\rmV$, while we keep the discrete axial symmetry $(\mathbb{Z}_{qN})_\rmR$. 
Any fermion bilinear operators $\overline{\psi}_\rmR \psi_\rmL$, $\overline{\psi}_\rmL\psi_\rmR$ break the discrete chiral symmetry completely, but we can consider four-fermion operators which are invariant under $G_{\rm sub}$:
\be
(\overline{\psi}_\rmR\psi_\rmL) (\overline{\psi}_\rmL\psi_\rmR),\quad 
\sum_{a=1}^{N^2-1} (\overline{\psi}_\rmR T^a \psi_\rmL)(\overline{\psi}_\rmL T^a \psi_\rmR), 
\label{eq:4Fermi_int}
\ee
where $T^a$ are generators of $SU(N)$. 
By adding this four-fermion interaction to the Lagrangian, we can explicitly break $G$ down to $G_{\rm sub}$. 
In the context of $4$-dimensional QCD with fundamental fermions, this operator was important to discuss the exotic scenario of chiral symmetry breaking, called Stern phase~\cite{Tanizaki:2018wtg, Stern:1997ri, Stern:1998dy, Kogan:1998zc, Kanazawa:2015kca}. 
Also, this restriction of symmetry is important to discuss the application of $2$-flavor Schwinger model ($\simeq$ $SU(2)$ level-$1$ Wess-Zumino-Witten (WZW) model) to $(1+1)$-dimensional anti-ferromagnetic quantum spin chain in the context of Haldane conjecture~\cite{Haldane:1983ru, Haldane:1982rj, Affleck:1986pq, Affleck:1987ch}. 
Its generalization to $N$-flavor case is important when we consider the generalization of Haldane conjecture to $SU(N)$ anti-ferromagnetic spin chain~\cite{Tanizaki:2018xto, Bykov:2011ai, Lajko:2017wif, Yao:2018kel, Ohmori:2018qza}. 

We can readily find the anomaly of $G_{\rm sub}$. We have to set $A_\rmR=A_\rmL\equiv A_\rmV$ for $SU(N)$ gauge fields, and $B_\rmR=0$. 
This sets $\widetilde{A}_\rmR=\widetilde{A}_\rmL\equiv \widetilde{A}_\rmV$ with the constraint $\tr[\widetilde{A}_\rmV]=B^{(1)}_\rmV$, and $qN A^{(1)}_\chi=\diff A^{(0)}_\chi$, and we substitute it into (\ref{eq:tHooft_anomaly}). 
As a consequence, the anomaly is characterized by 
\be
S_3[\calA_{\rm sub}]=\int_{M_3}{qN \over 2\pi}A^{(1)}_\chi\wedge B^{(2)}_\rmV.
\label{eq:discrete_anomaly}
\ee
This action is quantized to $\mathbb{Z}_{qN}$ phase, and thus we find the discrete anomaly of $G_{\rm sub}$. 

Let us make a remark on a related anomaly, which is found in previous studies~\cite{Anber:2018jdf, Anber:2018xek, Armoni:2018bga}. In those papers, authors only perform gauging of $\mathbb{Z}_q^{[1]}$ and not of $SU(N)_\rmV/\mathbb{Z}_N$. As a consequence, anomalous breaking of discrete chiral symmetry occurs as $\mathbb{Z}_{Nq}\to \mathbb{Z}_N$ in Refs.~\cite{Anber:2018jdf, Anber:2018xek, Armoni:2018bga}. 
In our case, we find a stronger discrete anomaly, since (\ref{eq:discrete_anomaly}) says that the discrete chiral symmetry $\mathbb{Z}_{Nq}$ is completely anomalously broken by background gauge fields. 

\subsection{Bosonization and anomaly matching}
\label{sec:bosonization}

In two spacetime dimension, the statistics does not make much sense, and we can interchange descriptions of one field theory with bosonic fundamental field and with fermionic fundamental field. This is called Bose-Fermi duality in two dimension. 

In this section, we provide the bosonic description of charge-$q$ $N$-flavor Schwinger model, and check its anomaly matching explicitly. 
First, we discuss $N=1$ case and $q=1$ case separately, and go into the general case armed with that knowledge. 

\subsubsection{$N=1$: Charge-$q$ Schwinger model}

One-flavor Dirac fermion $\overline{\psi}\slashed{\p}\psi$ can be mapped to the free boson ${1\over 8\pi}|\diff \phi|^2$ with $2\pi$-periodic compact scalar field by Abelian bosonization~\cite{Coleman:1974bu}. 
The correspondence of operators are the following: The $U(1)_{\rmV/\rmA}$ conserved currents become 
\be
\overline{\psi}\gamma^{\mu}\psi\leftrightarrow {1\over 2\pi}\ve^{\mu\nu}\p_\nu\phi,\quad  
\overline{\psi}\gamma \gamma^{\mu}\psi \leftrightarrow {1\over 2\pi}\p^\mu \phi. 
\ee
The scalar fermion bilinear operator is related as 
\be
\overline{\psi}_\rmR \psi_\rmL \leftrightarrow c(\mu)\rme^{\im \phi}
\ee
with some renormalization constant $c(\mu)\propto \mu$, where $\mu$ is the renormalization scale, up to some normal ordering. 

Applying this Abelian bosonization to the Schwinger model (\ref{eq:Lagrangian_Schwinger}) with $N=1$ flavor, we obtain the bosonized action
\be
S=\int_{M_2}\left({1\over 2e^2}|\diff a|^2+{1\over 8\pi}|\diff \phi|^2+ {\im\over 2\pi}(q \phi+\theta)\wedge \diff a\right).
\label{eq:1flavor_bosonized}
\ee
The discrete axial symmetry $(\mathbb{Z}_q)_\rmR$ becomes the shift symmetry on the scalar field $\phi$:
\be
\phi\mapsto \phi+{2\pi\over q}. 
\ee
Indeed, the change of the action is $\Delta S=\im \int \diff a\in 2\pi \im \mathbb{Z}$, as we expect it from ABJ anomaly, and thus it does not affect path integral weight. 
We can also explicitly check the discrete anomaly (\ref{eq:discrete_anomaly}) in the following way: 
Let us rewrite the bosonized action as 
\be
S=\int_{M_2}\left({1\over 2e^2}|\diff a|^2+{1\over 8\pi}|\diff \phi|^2+ {\im\theta\over 2\pi}\diff a\right)+{\im q\over 2\pi}\int_{M_3}\diff \phi\wedge \diff a, 
\ee
where $M_3$ is an arbitrary $3$-manifold with $\p M_3=M_2$. We gauge $\mathbb{Z}_q^{[1]}$ and $(\mathbb{Z}_q)_\rmR$ by the minimal coupling procedure, $\diff a\to (\diff a+B^{(2)}_\rmV)$ and $\diff \phi\to (\diff \phi+A^{(1)}_\chi)$, and then the gauged action becomes 
\bea
S_{\rm gauged}&=&\int_{M_2}\left({1\over 2e^2}|\diff a+B^{(2)}_\rmV|^2+{1\over 8\pi}|\diff \phi+A^{(1)}_\chi|^2+ {\im\theta\over 2\pi}(\diff a+B^{(2)}_\rmV)\right)\nonumber\\
&&+{\im q\over 2\pi}\int_{M_3}(\diff \phi+A^{(1)}_\chi)\wedge (\diff a+B^{(2)}_\rmV)\nonumber\\
&=& \int_{M_2}\left({1\over 2e^2}|\diff a+B^{(2)}_\rmV|^2+{1\over 8\pi}|\diff \phi+A^{(1)}_\chi|^2+ {\im\theta\over 2\pi}(\diff a+B^{(2)}_\rmV)\right)\nonumber\\
&&+{\im\over 2\pi}\int_{M_3} (q \diff \phi\wedge \diff a+\diff A^{(0)}_\chi\wedge \diff a+\diff \phi\wedge \diff B^{(1)}_\rmV)\nonumber\\
&&+\im {q\over 2\pi}\int_{M_3}A^{(1)}_\chi\wedge B^{(2)}_\rmV. 
\eea
Except for the last term, which is nothing but $S_3[\calA_{\rm sub}]$ given in (\ref{eq:discrete_anomaly}), $S_{\rm gauged}$ does not depend on the extension of fields to $M_3$ modulo $2\pi \im$. 
This means that the gauge invariance is satisfied by the anomaly inflow from the three-dimensional topological action $S_3[\calA_{\rm sub}]$, and this implies the 't~Hooft anomaly matching. 

Let us concretely check how the vacuum structure matches the 't~Hooft anomaly. By completing the square in terms of $\diff \phi$ in (\ref{eq:1flavor_bosonized}), we can easily find that the photon $\diff a$ gets the mass\footnote{The photon mass $m_\gamma^2$ in the general $(q, N)$ model arises from the fermion loop diagram at one-loop order. Compared to the Schwinger model,  the mass is enhanced by two factors. Charge at the vertices  is  replaced with   $e \rightarrow qe $, and there are 
$N$ fermions that can run in the loop, hence, $m_\gamma^2 = \frac{N q^2 e^2}{\pi }$.} 
\be
m_\gamma^2={q^2 e^2\over \pi}. 
\ee
To find how the anomaly is matched, we can pay attention only to the IR limit of the theory, and thus we can take the limit $e\to \infty$, i.e. the mass gap is infinite. 

We put our theory on a compact spacetime $M_2$, such as torus $T^2$. Since any one-point function with nontrivial charge under a symmetry does not develop the expectation value, we get 
\be
\langle \overline{\psi}_\rmR\psi_\rmL(x) \rangle_{M_2}\sim \langle \rme^{\im \phi(x)}\rangle_{M_2} =0
\ee
for $q>1$. This is the important difference when we compare it with $q=1$ case: When $q=1$, the one-instanton sector on $M_2$ gives the non-zero expectation value~\cite{Sachs:1991en}, 
\be
\langle \rme^{\im \phi(x)}\rangle_{M_2}\equiv v > 0, \quad (q=1). 
\ee
Next, we discuss the two-point correlation function, $\langle \rme^{\im \phi(x)} \rme^{-\im \phi(y)}\rangle_{M_2}$.  Equation of motion of $a$ says that $\diff \phi=0$, i.e. $\phi$ is constant, so that we find that 
\be
\langle \rme^{\im \phi(x)}\rme^{-\im \phi(y)}\rangle_{M_2}=v^2>0, 
\ee
whether $q>1$ or $q=1$. 

We now take the decompactification limit $M_2\to \mathbb{R}^2$, and also separate two points $|x-y|\to \infty$. The above discussion shows that 
\be
\lim_{|x-y|\to \infty}\lim_{M_2\to \mathbb{R}^2}\left(\langle  \rme^{\im \phi(x)}\rme^{-\im \phi(y)}\rangle_{M_2} - |\langle \rme^{\im\phi}\rangle_{M_2}|^2 \right)
=
\left\{\begin{array}{ccc}
v^2 &\quad (q>1),\\
0  &\quad (q=1). 
\end{array}
\right.
\ee
Therefore, the cluster decomposition holds for $q=1$ as shown explicitly in \cite{Sachs:1991en}, but this is not true for $q>1$. 
This is because, for $q>1$, $\mathbb{Z}_q$ chiral symmetry is spontaneously broken and the vacuum obtained by $M_2\to \mathbb{R}^2$ becomes the mixed state of those $q$ vacua. 
Each pure-state vacuum is labeled by $k=0,1,\ldots, q-1$, with 
\be
\langle \rme^{\im \phi(x)}\rangle_k=v\exp\left( {2\pi k-\theta\over q}\im \right), 
\ee
and 
\be
\lim_{M_2\to\mathbb{R}^2}\langle \mathcal{O}(x_1,\ldots,x_i)\rangle_{M_2}={1\over q}\sum_k \langle \mathcal{O}(x_1,\ldots,x_i)\rangle_k
\ee
for any local correlators. 
The existence of $q$ vacua does match the $\mathbb{Z}_q^{[1]}\times (\mathbb{Z}_q)_\rmR$ 't~Hooft anomaly. 
One of the main purpose of this paper is to obtain this fractionalized $\theta$ dependence with spontaneous chiral symmetry breaking using semiclassical approach with circle compactifications, following Refs.~\cite{Smilga:1993sn, Shifman:1994ce}. 

The interesting consequence of anomaly (\ref{eq:discrete_anomaly}) is that the partition function of these $q$ vacua are related as 
\be
(\mathbb{Z}_q)_\rmR: Z_k[B^{(2)}_\rmV]\mapsto Z_{k+1}[B^{(2)}_\rmV]=Z_k[B^{(2)}_\rmV]\exp\left(\im \int B^{(2)}_\rmV \right). 
\ee
The first relation is the very definition of the label $k$ of discrete chiral symmetry breaking, and the second relation represents the mixed 't~Hooft anomaly. 
This relation says that the $q$ vacua are different as symmetry-protected topological (SPT) phases protected by $\mathbb{Z}_q^{[1]}$, and the domain wall between $k$-th and $k'$-th vacua supports the charge $(k'-k)$ mod $q$ excitation under the $U(1)$ gauge symmetry (see, also, Refs.~\cite{Anber:2015kea, Sulejmanpasic:2016uwq, Komargodski:2017smk, Nishimura:2019umw} for nontrivial domain walls). 

\subsubsection{$q=1$: $N$-flavor Schwinger model and WZW model}

Next, let us consider the ordinary multi-flavor massless Schwinger model. 
This part is known in literatures, so we just briefly summarize it and include it in the analysis of general cases. 

In this case, the symmetry contains only the $0$-form symmetry, and it is the continuous chiral symmetry, $G=[SU(N)_\rmL\times SU(N)_\rmR]/(\mathbb{Z}_N)_\rmV$.  
Essentially, the 't~Hooft anomaly~(\ref{eq:tHooft_anomaly}) is just the perturbative non-Abelian chiral anomaly, determined by the $3$-dimensional Chern-Simons action:
\be
S_3[A_\rmR,A_\rmL]={1\over 4\pi}\int_{M_3} \tr\left[A_\rmR\left(\diff A_\rmR+{2\im \over 3}A_\rmR^2\right)-A_\rmL\left(\diff A_\rmL+{2\im \over 3}A_\rmL^2\right)\right]. 
\ee
It is important to notice that the spontaneous symmetry breaking is prohibited by Coleman-Mermin-Wagner theorem since the symmetry group $G$ is continuous~\cite{Coleman:1973ci, mermin1966absence}. 
This anomaly can be matched by $SU(N)_1$ WZW model~\cite{Wess:1971yu, Witten:1983tw,Witten:1983ar}, 
\be
S_{\rm WZW}={1\over 8\pi}\int_{M_2}\tr\left[\diff U\wedge \star \diff U^\dagger \right]+{\im\over 12\pi}\int_{M_3}\tr[(U^\dagger \diff U)^3], 
\ee
where $U$ is the $SU(N)$-valued field on $M_2$, which is extended to $M_3$ for the Wess-Zumino term. 
This can be explicitly obtained by non-Abelian bosonization to the multi-flavor Schwinger model, and taking the limit $e\to \infty$, to make the photon mass $m_\gamma^2=N e^2/\pi$ infinite. 
Therefore, the vacuum is unique, and the conformal field theory matches the 't~Hooft anomaly. 

The four-fermion interaction (\ref{eq:4Fermi_int}) corresponds to the double-trace terms $|\tr(U)|^2$, $|\tr(T^a U)|^2$. 
Requiring $\tr(U^k)=0$ on $M_2$ for $k=1,\ldots, N-1$ as a consequence of such deformations, the $SU(N)_1$ WZW theory can be continuously connected to the flag-manifold sigma model with the target space $SU(N)/U(1)^{N-1}$ with the specific theta angles~\cite{Tanizaki:2018xto}.  
The symmetry group is reduced to $G_{\rm sub}=SU(N)_\rmV/(\mathbb{Z}_N)_\rmV\times (\mathbb{Z}_N)_\rmR$, and the theta terms of the flag sigma model reproduces the discrete anomaly $S_{3}[\calA_{\rm sub}]$~\cite{Tanizaki:2018xto}. 

\subsubsection{General case: $q>1$ and $N>1$}

The non-Abelian bosonization maps $N$-flavor Dirac fermion to $U(N)_1$ WZW model~\cite{Witten:1983ar, Polyakov:1983tt, Polyakov:1984et}. 
The correspondence of the fermion bilinear operator is 
\be
\psi_\rmL \overline{\psi}_\rmR\sim U,  
\ee
where $U$ is the $U(N)$-valued scalar field. 
This tells that the element of the $0$-form symmetry,
\be
[(V_\rmL,V_\rmR,\rme^{\im \alpha_\rmR})]\in {SU(N)_\rmL\times SU(N)_\rmR\times (\mathbb{Z}_{Nq})_\rmR \over (\mathbb{Z}_N)_\rmV\times (\mathbb{Z}_N)_\rmR},
\ee
acts on $U$ as 
\be
U\mapsto V_\rmL U V_\rmR^\dagger \rme^{-\im \alpha_\rmR}. 
\ee
The bosonized action of the theory is given by 
\bea
S&=&{1\over 2e^2}\int_{M_2}|\diff a|^2+{1\over 8\pi}\int_{M_2}\tr\left(|\diff U|^2\right)\nonumber\\
&&+{\im \over 12\pi}\int_{M_3}\tr\left((U^{\dagger}\diff U)^3\right)+{ q\over 2\pi}\int_{M_3}\diff a\wedge \tr\left(U^{\dagger}\diff U\right). 
\label{eq:general_bosonization}
\eea
The first two terms are kinetic terms, and the third term is the level-$1$ Wess-Zumino term. The last term can be expressed as 
\be
\int_{M_2} a\wedge {1\over 2\pi}\diff (\ln \det(U)^q), 
\ee
and, in the limit $e\to \infty$, the $U(1)$ gauge field $a$ plays a role of the Lagrange multiplier field, so that $\det (U)\in \mathbb{Z}_q$ and it reproduces the ABJ anomaly $U(1)_\rmR \to (\mathbb{Z}_{Nq})_\rmR$.

Let us check that the bosonized action has the same anomaly $S_3[\calA]$, given in (\ref{eq:tHooft_anomaly_LR}), including the discrete factors of $G$. The gauge-invariance of the kinetic terms is evident, and thus let us concentrate on the last two topological terms. 
The covariant derivative $D$ on $U$ and $U^\dagger$ with background gauge fields $\calA$ are given as 
\be
DU=\diff U+\im \widetilde{A}_\rmL U-\im U (\widetilde{A}_\rmR+A^{(1)}_\chi),\quad 
DU^\dagger =\diff U^\dagger + \im (\widetilde{A}_\rmR+A^{(1)}_\chi) U^\dagger -\im U^\dagger \widetilde{A}_\rmL. 
\ee
The naive replacement $U^\dagger \diff U\to U^\dagger D U$, etc., is insufficient for the Wess-Zumino term, and we must find an appropriate local counterterm. With some trial-and-error, we can find that the following works well:
\bea
\Gamma_{\rm WZ}&\equiv&{1\over 12\pi}\tr\left((U^\dagger D U)^3\right)
-\im {q\over 2\pi}(\diff a+B^{(2)}_\rmV)\wedge \tr\left( U^\dagger D U\right) \nonumber\\
&&\hspace{-1em}+{1 \over 4\pi}\tr\left[(U D U^\dagger)\wedge \im(\widetilde{F}_\rmL-q B^{(2)}_\rmV)-(U^\dagger D U) \wedge \im(\widetilde{F}_\rmR+\diff A^{(1)}_\chi-q B^{(2)}_\rmV)\right]. 
\eea
The gauge invariance both under the ordinary and $1$-form transformations is manifest in this expression. 
A straightforward computation shows that 
\bea\hspace{-1em}
\Gamma_{\rm WZ}&=&
{1\over 12\pi}\tr[(U^\dagger U)^3] - {\im \over 2\pi}\diff a\wedge\left(\tr[U^\dagger \diff U]-q(N A^{(1)}_\chi+B^{(1)}_\rmR)\right)\nonumber\\
&&\hspace{-1em}+\diff \left({\im \over 4\pi}\tr\left[U^\dagger \diff U (\widetilde{A}_\rmR+A^{(1)}_\chi)-U\diff U^\dagger \widetilde{A}_\rmL+\im U^\dagger \widetilde{A}_\rmL U(\widetilde{A}_\rmR+A^{(1)}_\chi)\right]\right)\nonumber\\
&&\hspace{-1em}-{1\over 4\pi}\tr\left[(\widetilde{A}_\rmR+A^{(1)}_\chi)\diff\left(\diff (\widetilde{A}_\rmR+A^{(1)}_\chi)+{2\im \over 3}\widetilde{A}_\rmR^2\right)-\widetilde{A}_\rmL\left(\diff \widetilde{A}_\rmL+{2\im \over 3} \widetilde{A}_\rmL^2\right)\right]. 
\eea
The last line is equal to the topological action, $S_3[\calA]$, given in (\ref{eq:tHooft_anomaly_LR}).  
Other terms on the right-hand-side is defined on $M_2$ modulo $2\pi$. 
This shows that the gauge invariance is established by anomaly inflow from the three-dimensional bulk action, $S_3[\calA]$, and thus it has the same 't~Hooft anomaly with the massless Schwinger model.

\subsection{Anomaly under $S^1$ compactifications}
\label{sec:anomaly_S1}

In the later sections, we will establish the semiclassical understandings of nonperturbative phenomena in charge-$q$ $N$-flavor Schwinger model. 
In order to validate the semiclassical treatment, we need to put the theory on the cylinder with small compactification radius~\cite{Smilga:1993sn, Shifman:1994ce}. 
We have seen that the two-dimensional model has the 't~Hooft anomaly and it constrains the vacuum structures and the massless excitations. 
We would like to retain those 't~Hooft anomalies as much as possible, and we shall see that the appropriate flavor-twisted boundary condition plays an important role~\cite{Tanizaki:2017qhf, Tanizaki:2017mtm, Dunne:2018hog}. 

\subsubsection{Thermal compactification}

Let us put our theory on $M_2=M_1\times S^1\ni (x^1,x^2)=:(\tau,x)$, and assume that the size of $S^1$ is much smaller than that of $M_1$. 
We then obtain the effective field theory on $M_1$, and would like to understand its properties. 
We first consider the ordinary boundary condition\footnote{Here, we take the anti-periodic boundary condition for the fundamental fermion fields. We, however, would like to point out that the overall $U(1)$ phase of the boundary condition does not affect the physics since the $U(1)$ symmetry is gauged, and thus periodic and anti-periodic boundary conditions play the same role. In other words, since the local gauge-invariant operators are all bosonic, the above difference of boundary conditions does not change physics. This point will be discussed more in detail in Sec.~\ref{sec:holonomy}. } along the compactified direction $x\sim x+L$: 
\bea
\psi^f(\tau,x+L)&=&-\psi^f(\tau,x),\\
a(\tau,x+L)&=&a(\tau,x)+\diff \lambda(\tau), 
\eea
where $\lambda$ is a $2\pi$-periodic scalar on $\tau\in M_1$. 

To discuss the 't~Hooft anomaly of the $1$-dimensional effective theory on $M_1$, we first need to identify the symmetries. 
This boundary condition does not affect the $0$-form symmetry $G^{[0]}$. What is important for gauge theories is that Polyakov-loop operators become local gauge-invariant operators under $S^1$-compactification; 
\be
P(\tau)=\exp\left(\im \int_0^L a_2(\tau,x)\diff x\right). 
\ee
The $1$-form symmetry, $\mathbb{Z}_q^{[1]}$, in two-dimensions provides the additional $0$ -form symmetry, $\mathbb{Z}_q^{[0]}$, in one-dimensions, and it is given by 
\be
P(\tau)\mapsto \rme^{2\pi\im/q}P(\tau), 
\ee
so that the symmetry $G$ in two-dimensions induces 
\be
G\rightsquigarrow G_{1\rm d}={\mathbb{Z}_q^{[0]}}\times {SU(N)_\rmL\times SU(N)_\rmR\times (\mathbb{Z}_{q N})_\rmR\over (\mathbb{Z}_N)_\rmV\times (\mathbb{Z}_N)_\rmR}. 
\ee

In this thermally compactified theory, the $SU(N)_{\rmL/\rmR}$ chiral anomalies in two-dimensions disappear in one-dimensions and do not imply the anomaly matching condition. 
To see this, let us remind that, for instance, the $ SU(N)_\rmR$ two-dimensional anomalies take the form 
\be
{1\over 4\pi}\int_{M_3} \tr\left(A_\rmR\left(\diff A_\rmR+{2\im\over 3} A_\rmR^2\right)\right). 
\ee
When we gauge $SU(N)_\rmR$ symmetry as a symmetry of one-dimensional theory, $A_\rmR$ depends only on $\tau$ and should not show any $x$ dependence. 
As a consequence, the gauge fields on $M_3$ do not have $x$ dependence, and we identically obtain $A_\rmR \diff A_\rmR+{2\im\over 3}A_\rmR^3\equiv 0$ since it becomes the $3$-form of $2$-dimensional functions. 

The anomaly of higher-form symmetries is exceptional in this viewpoint~\cite{Gaiotto:2017yup}. 
Let us gauge $\mathbb{Z}_q^{[0]}$, and we denote its gauge field as 
\be
A_{\rmc(q)}=A_{\rmc (q),1}(\tau)\diff \tau. 
\ee
Although this is the one-form gauge field on $M_1$, it acts on Polyakov loops and thus its two-dimensional origin is $B^{(2)}_\rmV$:
\be
B^{(2)}_\rmV=A_{\rmc(q)}\wedge {\diff x\over L}. 
\ee
Substituting this expression into (\ref{eq:tHooft_anomaly}), we find that the discrete anomaly survives:
\be
\Omega_3^0={q\over 2\pi} (NA^{(1)}_\chi)\wedge A_{\rmc(q)}\wedge{\diff x\over L}, 
\ee
and then the anomaly inflow is controlled by the two-dimensional $\mathbb{Z}_q\times \mathbb{Z}_q$ Dijkgraaf-Witten action, 
\be
\int\Omega_3^0={q\over 2\pi}\int (N A^{(1)}_\chi)\wedge A_{\rmc(q)}. 
\label{eq:discrete_anomaly_pbc}
\ee
This describes the mixed anomaly of 
\be
\mathbb{Z}^{[0]}_q \times \left({\mathbb{Z}_{Nq}\over \mathbb{Z}_N}\right)_\rmR \subset G_{1 \rm d}. 
\ee

We therefore conclude that the discrete anomaly survives under thermal compactification when $q>1$. 
However, we completely lose the information about continuous chiral anomaly under this compactification. 
Especially, when $q=1$, we do not have the discrete anomaly, and the vacuum structure becomes completely trivial. 

\subsubsection{Flavor-twisted compactification}

If possible, we would like to keep the nontrivial structure of two-dimensional field theories as much as possible under $S^1$ compactification. 
If we take the thermal boundary condition for fermion fields, however, the information about continuous chiral symmetry is completely lost. 
Many recent studies of asymptotically free theories~\cite{Dunne:2016nmc, Dunne:2012ae,Unsal:2007jx,Unsal:2008ch, Unsal:2007vu, Kovtun:2007py,  Shifman:2008ja, Shifman:2009tp, Cossu:2009sq, Cossu:2013ora, Argyres:2012ka, Argyres:2012vv, Dunne:2012zk, Poppitz:2012sw, Anber:2013doa, Basar:2013sza, Cherman:2014ofa, Misumi:2014raa, Misumi:2014jua, Misumi:2014bsa, Dunne:2015ywa,Misumi:2016fno, Cherman:2016hcd, Fujimori:2016ljw, Fujimori:2017oab, Fujimori:2017osz, Fujimori:2018kqp, Sulejmanpasic:2016llc, Yamazaki:2017ulc, Itou:2018wkm, Buividovich:2017jea, Aitken:2017ayq}  suggest that we can keep the nontrivial vacuum structure of the original theory by taking the appropriate symmetry-twisted boundary condition. 
This is, indeed, noticed much earlier than these recent works in the context of multi-flavor Schwinger model by Shifman and Smilga~\cite{Shifman:1994ce}, but its full generality is appreciated by aforementioned works, and it is referred to as adiabatic continuity~\cite{Dunne:2016nmc}. 
We here provide its interpretation in view of 't~Hooft anomaly following Ref.~\cite{Tanizaki:2017qhf}. 

We put the flavor-twisted boundary condition on the fundamental fermion fields, 
\be
\psi'^f(\tau, x+L) =\rme^{2\pi \im f/N} \psi'^{f}(\tau,x),
\label{eq:twisted_bc}
\ee
and we simply denote this as $\psi'(\tau,x+L)=\Omega_F \psi'(\tau,x)$ introducing the diagonal matrix $\Omega_F=\mathrm{diag}(1,\rme^{2\pi\im/N},\ldots, \rme^{2\pi\im (N-1)/N})$. 
This is related to the fermion field with the periodic boundary condition as 
\be
\psi'^{f}(\tau,x)=\rme^{2\pi \im f x/NL}\psi^f(\tau,x), 
\ee
and the fermion kinetic term is given as 
\be
\int_{M_1}\diff \tau\int_0^L \diff x\sum_{f=1}^N \overline{\psi}^f \left(\gamma^\mu\p_\mu+\gamma^1 \im q a_1+\gamma^2\im\left(q a_2+{2\pi f\over NL}\right)\right)\psi^f. 
\ee
In this way, we can interpret the flavor-twisted boundary condition as the background $SU(N)_\rmV$ holonomy. 
We will use both descriptions in the following of the paper. 

In this boundary condition, the symmetry acting on the Polyakov loop $P(\tau)$ is not just $\mathbb{Z}_q^{[0]}$, but it is further extended to $\mathbb{Z}_{Nq}^{[0]}$ with the following transformation:
\be
a_2\mapsto a_2+{2\pi\over Nq L}, 
\label{eq:Znq_0form_sym_1}
\ee
and it is intertwined with the shift transformation, 
\bea
\psi^f\mapsto \psi^{f+1}\quad (f=1,\ldots, N-1),\quad \psi^N\mapsto \rme^{-2\pi \im x/L} \psi^1. 
\label{eq:Znq_0form_sym_2}
\eea
The ordinary flavor symmetry $G^{[0]}$ is explicitly broken to its maximal Abelian subgroup because of this boundary condition, 
\be
{U(1)_{\rmL}^{N-1}\times U(1)_\rmR^{N-1}\times (\mathbb{Z}_{Nq})_\rmR\over (\mathbb{Z}_N)_\rmV\times (\mathbb{Z}_N)_\rmR}\subset G^{[0]}. 
\ee
Then, the full symmetry group is 
\be
\mathbb{Z}_{Nq}^{[0]} \ltimes {U(1)_{\rmL}^{N-1}\times U(1)_\rmR^{N-1}\times (\mathbb{Z}_{Nq})_\rmR\over (\mathbb{Z}_N)_\rmV\times (\mathbb{Z}_N)_\rmR}. 
\label{twisted-full}
\ee
It is important for later applications that this symmetry group contains $\mathbb{Z}_{Nq}^{[0]}\times (\mathbb{Z}_{Nq})_\rmR$ as a subgroup. 

As we have discussed in gauging $\mathbb{Z}_q^{[0]}$, the $\mathbb{Z}_{Nq}^{[0]}$ background gauge field, $A_{\rmc (Nq)}$, is embedded into the two-form gauge field as 
\be
B^{(2)}_\rmV= A_{\rmc (Nq)}\wedge {\diff x\over L}. 
\ee
Substituting this into the anomaly-inflow action (\ref{eq:tHooft_anomaly}) in two-dimensions, we find that the effective theory on $M_1$ with the twisted boundary condition has the anomaly, which is determined by $\mathbb{Z}_{Nq}\times \mathbb{Z}_{Nq}$ Dijkgraaf-Witten action, 
\be
{Nq\over 2\pi}\int A^{(1)}_\chi\wedge A_{\rmc (Nq)}. 
\label{eq:discrete_anomaly_tbc}
\ee
Unlike the thermal boundary condition, this anomaly persists even for $q=1$ multi-flavor Schwinger model.


\section{Holonomy effective potentials of massless Schwinger models}\label{sec:holonomy}

As a preparation of quantum mechanical treatment in Sec.~\ref{sec:ChSSB}, we compute the effective potential of the Polyakov loop, $\exp(\im L a)$, by integrating out fermions. 
  Below, we describe two physically different compactification of the charge-$q$ $N$ flavor  Schwinger model: Thermal and flavor-twisted boundary conditions  denoted with $SU(N)_\rmV$ flavor matrix, $\Omega_F$. 

\subsection{Thermal boundary condition}

Let us first discuss that the overall $U(1)$ phase of fermion boundary conditions is unphysical in Schwinger models. More generally, this is true for Spin$^c$ gauge theories. 
In text-books, it is sometimes asserted that the thermal boundary conditions are necessarily anti-periodic for fermionic fields. 
Here, let us take a more general boundary condition on $\mathbb{R}\times S^1$, 
\be
\psi(\tau,x+L)=\rme^{\im \alpha} \psi(\tau,x),
\ee
and $\alpha=\pi$ corresponds to the usual thermal boundary condition. 
We perform the ``improper'' gauge transformation, 
\be
a\mapsto a'= a+{\alpha\over q L},\; \psi \mapsto \psi'=\rme^{-\im \alpha x/L} \psi, 
\ee
and then the form of the Lagrangian is not changed, but the fields satisfy the periodic boundary condition. 
Local gauge-invariant operators, $\star \diff a=\p_\tau a(\tau)$, $\overline{\psi}\psi$, etc., do not change under this transformation, and the only change on gauge-invariant operators appears as the phase of Polyakov loop, 
\be
\rme^{\im L a}\mapsto \rme^{\im \alpha/q}\rme^{\im L a}. 
\ee
Therefore, the computation with periodic boundary condition gives sufficient information to obtain the result with general $\alpha$, and the overall phase $\alpha$ is unphysical in this sense. 
In Hilbert space interpretation, these boundary conditions correspond to the computation of $ \tr(\rme^{-LH} \rme^{\im (\alpha +\pi)F} )$, where $F$  is fermion number. However, there are   no gauge invariant fermionic states in the Hilbert space of the model, and hence,  
\be
\tr(\rme^{-LH} \rme^{\im (\alpha+\pi) F} )=\tr(\rme^{-L H}), 
\ee
because $F\equiv 0$ on physical Hilbert space. 

We now compute the holonomy effective potential, using  periodic boundary condition for fermionic fields, $\alpha=0$. 
 Since this is $2$d $U(1)$ gauge theory, the holonomy potential is induced solely by fermions:
\be
V(a)=-\ln \mathrm{Det}(\gamma^{\mu}D_{\mu})=-N \ln\mathrm{Det}(\gamma^1 \p_\tau+\gamma^2(\p_x+\im a)). 
\ee
In the statistical-mechanics language, the potential takes the form: 
 \begin{align} 
 V(a)   & = - \frac{ N  }{L}  \int \frac{\diff  p}{2 \pi}    \left( \log (1 + \rme^{-L |{ p}| + \im L q a + \im \pi } )  + \rm c.c. \right) \cr 
 & =  \frac{ 2 N}{ \pi  L^2}  \sum_{n=1}^{\infty}  \frac{1} {n^2}     \cos(Lq a n)    \qquad \cr
 &= \min_{k\in \mathbb{Z}} \frac{ 2 N}{ \pi  L^2}  \left(  \frac{1}{4}   \Bigl(L q a  +  \pi (2k+ 1) \Bigr)^2  - \frac{\pi^2}{12} \right)
\label{1-loop-thermal}
 \end{align} 
 In the first line, the over-all minus sign is related to Pauli-exclusion principle. The $\rme^{\im \pi}$ arises from the periodic boundary conditions on fermions. 
In Appendix~\ref{sec:holonomy_bosonization}, we derive the same result in the bosonized theory. 

The minimal value of the potential, \eqref{1-loop-thermal},  is   the thermal free energy density: 
 \begin{align} 
 {\cal F}_{\rm thermal} & = -  \frac{ 2 N}{  L^2} \frac{\pi}{12},    \qquad
\label{free-energy}
 \end{align} 
and this is  the Stefan-Boltzmann  law for black-body radiation for $N$ species of Dirac fermions.

We make two comments on \eqref{1-loop-thermal}. 
\begin{itemize}
\item In the  usual Schwinger model,  $(q, N)= (1,1)$ and its multiflavor thermal version $(1, N>1)$,   the holonomy potential has a unique minimum in its fundamental domain located  at $L a=\pi$. 
\item  In the  $(q > 1, N \geq 1 )$ with  p.b.c. for all flavors,  the holonomy potential has $q$ minima 
in the fundamental domain  given by  $La= \frac{2 \pi }{q } (p + {1 \over 2} ) $,  with $ p=0,1, \ldots, q -1$.
These minima are separated by $L\Delta a= \frac{2\pi}{q}$.  
\end{itemize} 
 
 Unlike a purely  bosonic system  in which the 
$q  $-fold perturbative degeneracy would  generically be  lifted due to non-perturbative instanton effects, in the present case  
the degeneracy will survive.  This is guaranteed by the persistent mixed anomaly and is realized through fermionic zero-mode structure of fractional instantons. This is the subject matter of Sec.~\ref{sec:chiral_condensate_RS1_pbc}.

\subsection{Flavor-twisted boundary condition}
A boundary condition twisted by the  genuine global symmetry has crucial  physical consequences. 
Now, consider the  fermions with flavor-twisted boundary conditions: 
 \begin{align}
  \psi(\tau,x + L) &=   \Omega_F  \psi(\tau,x) 
  \label{bc}
\end{align}
In the operator formalism, this will correspond to  a grading and quantum distillation over the Hilbert space~\cite{Dunne:2018hog}.   
The holonomy potential is given by 
 \begin{align} 
 V_{\Omega_F}(a) & =  \frac{1}{ \pi  L^2}  \sum_{n=1}^{\infty}  \frac{1}{n^2}    \left(  \rme^{\im L q a n}  \tr ( \Omega_F^n)  +   \rm c.c. \right) ,    \qquad
\label{1-loop-tbc}
 \end{align} 
and we now substitute $\Omega_F=\mathrm{diag}(1,\rme^{2\pi\im/N},\ldots, \rme^{2\pi\im (N-1)/N})$. 
All  terms in \eqref{1-loop-tbc}, but $n=N n'$, are zero due to  $\Omega_F$-twist matrix. 
As a result, the potential takes the form:  
 \begin{align} 
 V_{\Omega_F}(a)    
 & =  \frac{2}{ \pi  L^2 } \frac{1}{N} \sum_{n'=1}^{\infty}  \frac{1 }{n'^2}     \cos { (N L q  a n')}  \cr
   \qquad
   &=  \min_{k\in \mathbb{Z}} \frac{ 2 }{ \pi  L^2 N } 
   \left(  \frac{1}{4}  \Bigl(q  N  L a +  \pi (2k + 1)\Bigr)^2  - \frac{\pi^2}{12} \right)  
\label{1-loop-tbc-2}
 \end{align} 
We make various comments for the minima of the holonomy potential: 
\begin{itemize} 
\item In the  $(q=1, N > 1 )$ with $\Omega_F$ twist, the holonomy potential has $N$ minima 
in the fundamental domain $L a\in [0, 2 \pi)$,  given by
$ L a= \frac{2 \pi }{ N} (p + {1 \over 2} ) $,  \; $ p=0,1, \ldots,  N-1$. 
 These minima are separated by $L \Delta a= \frac{2\pi}{N}$.  
\item For the most general   $(q > 1, N \geq 1 )$ with $\Omega_F$ twist, the holonomy potential has $q N$ minima 
in the fundamental domain, and those  minima are given by
$ L a= \frac{2 \pi }{q N} (p + {1 \over 2} ) $,  \;  $ p=0,1, \ldots, q N-1$, which are separated by  $L \Delta a= \frac{2\pi}{q N}$. 
\end{itemize}

Unlike a purely  bosonic theory in which the 
$q N $-fold perturbative degeneracy would  generically be  lifted due to instanton effects, in the present case 
 the $q N $-fold degeneracy is   not lifted due to  fermion zero mode structure of the instantons, and also as dictated by anomalies.   The mechanism that will be described is similar to 
 ${\cal N}=2$ supersymmetric quantum mechanics, in which, the degeneracy of the classical vacua 
persists despite  the instanton effects~\cite{Witten:1982df}.  
This is also explained by using Picard--Lefschetz theory applied to multi-instantons~\cite{Behtash:2015kva}. 
 This will be crucial in realizing the mixed anomalies in  reduced quantum mechanics.

The free energy density with  $\Omega_F $ twist is   given by 
 \begin{align} 
 {\cal F}_{\Omega_F} & = -  \frac{2}{N}  \frac{\pi}{12}  \frac{1}{L^2}  = \frac{ 1}{N^2} {\cal F}_{\rm thermal}  \qquad
\label{free-twisted}
 \end{align}

\section{Chiral condensate and Polyakov loop in Schwinger model on $\mathbb{R}\times S^1$}\label{sec:ChSSB}

Chiral symmetry breaking and chiral condensates in Schwinger model on ${\mathbb R}\times S^{1}$ were investigated in~\cite{Sachs:1991en,Smilga:1993sn,Shifman:1994ce}. In particular, Shifman and Smilga~\cite{Shifman:1994ce} investigated the $2$-flavor cases with the flavor-twisted boundary condition, with emphasis on the contributions from fractional instantons. It is of great importance to review their results and extend them to the charge-$q$ $N$-flavor cases.

Firstly, let us review the chiral condensate in the simplest case, or the massless $N=1$ Schwinger model with $q=1$ on ${\mathbb R}^{2}$.
The chiral condensate for this case is exactly calculated as
\be
\langle \overline{\psi}\psi \rangle \,=\, -{{m_\gamma}\over{2\pi}}{\rm e}^{\gamma}\,,
\ee
where $m_{\gamma}\equiv e/\sqrt{\pi}$ and $\gamma\simeq 0.5771$ is the Euler-Mascheroni constant.
There are several techniques to derive the chiral condensate~\cite{Manton:1985jm, Sachs:1991en, Nielsen:1976hs, Hortacsu:1979fg, Rothe:1978hx, Krasnikov:1980mc, Jayewardena:1988td}, including bosonization, functional integral around instanton backgrounds, and the cluster decomposition of the four-point correlators.

We here obtain the chiral condensates for charge-$q$ $N$-flavor Schwinger models with thermal b.c. (or, periodic b.c., p.b.c.) and ${\mathbb Z}_{N}$ t.b.c. on ${\mathbb R}\times S^{1}$ by use of quantum mechanical techniques.
We first note that we can choose $a_{1}=0$ by use of gauge degrees of freedom (i.e. temporal gauge).
Moreover, we here focus on the case 
\be
e L \ll 1,\label{eq:semiclassical_region}
\ee
and can drop the higher Kaluza-Klein (KK) modes of $a_{2}$. 
So, we will work on position-independent $a_{2}$ denoted just as $a_{2} \equiv a$ below.

The associated Dirac equation for general cases with charge-$q$ and $N$-flavor  is
\be
\left[\im {\p \over{\p t}} + \sigma_{3} \left(\im {\p\over{\p x}} -q a \right) \right] \psi= 0\,.
\ee
Here we denote the energy of $k$-th state as $E^{(k)}$ with assuming $\psi\sim \rme^{-\im E^{(k)} t} \psi_{k}(x)$.
We here call the compactified direction as $x^{2}=x$, and introduce a Minkowski time $t$ to study the system quantum-mechanically. 
The equation is then rewritten as
\be
E^{(k)} \psi_{k}(x) = -\sigma_{3} \left(\im{\diff \over{\diff  x}} -q a \right) \psi_{k}(x)\,.
\ee
The solution of eigenfunctions $\psi_{k}(x)$ depends on boundary conditions, charges and flavors,
thus we below discuss distinct cases separately.

An important consequence in Secs.~\ref{sec:chiral_condensate_RS1_pbc} and~\ref{sec:chiral_condensate_RS1_tbc}  is that the chiral condensate on $\mathbb{R}\times S^1$ with (\ref{eq:semiclassical_region}) behaves as 
\be
\langle \overline{\psi}\psi\rangle\sim {\#\over L}\exp\left(-{\#\over eL}\right),
\ee
where $\#$'s are numerical constants that depend on $q$, $N$, and boundary conditions. 
In this section, we will find their explicit forms by quantum mechanical computations, since this is important to understand the degeneracy of ground states. 
In Sec.~\ref{sec:fractional_quantum_instanton}, we shall reinterpret this behavior using the path-integral approach with semiclassical approximations. 
It is notable that the exponent of chiral condensates behaves as $\sim 1/e$, instead of the usual field-theoretic instanton action $\sim 1/e^2$. We shall interpret this as a manifestation of  ``quantum'' instanton in Sec.~\ref{sec:fractional_quantum_instanton}.

In Sec.~\ref{sec:chiral_condensate_RS1_pbc}, we construct theta vacua for cluster-decomposition properties about correlators of chiral condensates for thermal boundary conditions. 
In Sec.~\ref{sec:chiral_condensate_RS1_tbc}, we do the same analysis for flavor-twisted boundary condition, and we will see that the vacuum structures are different as a consequence of different anomalies. 
In Sec.~\ref{sec:Polyakov_loop}, however, the Polyakov-loop correlators are studied, and we show that they do not satisfy the cluster decomposition with theta vacua if the discrete chiral symmetry is spontaneously broken. 
We shall see in Sec.~\ref{sec:discrete_anomaly_qm} that this matches the 't~Hooft anomaly discussed in Sec.~\ref{sec:anomaly_S1}.

\subsection{Chiral condensate in thermal boundary condition}\label{sec:chiral_condensate_RS1_pbc}

\subsubsection{$q=1$, $N=1$ with thermal b.c.}
\begin{figure}[t]
\centering
\includegraphics[width=100mm]{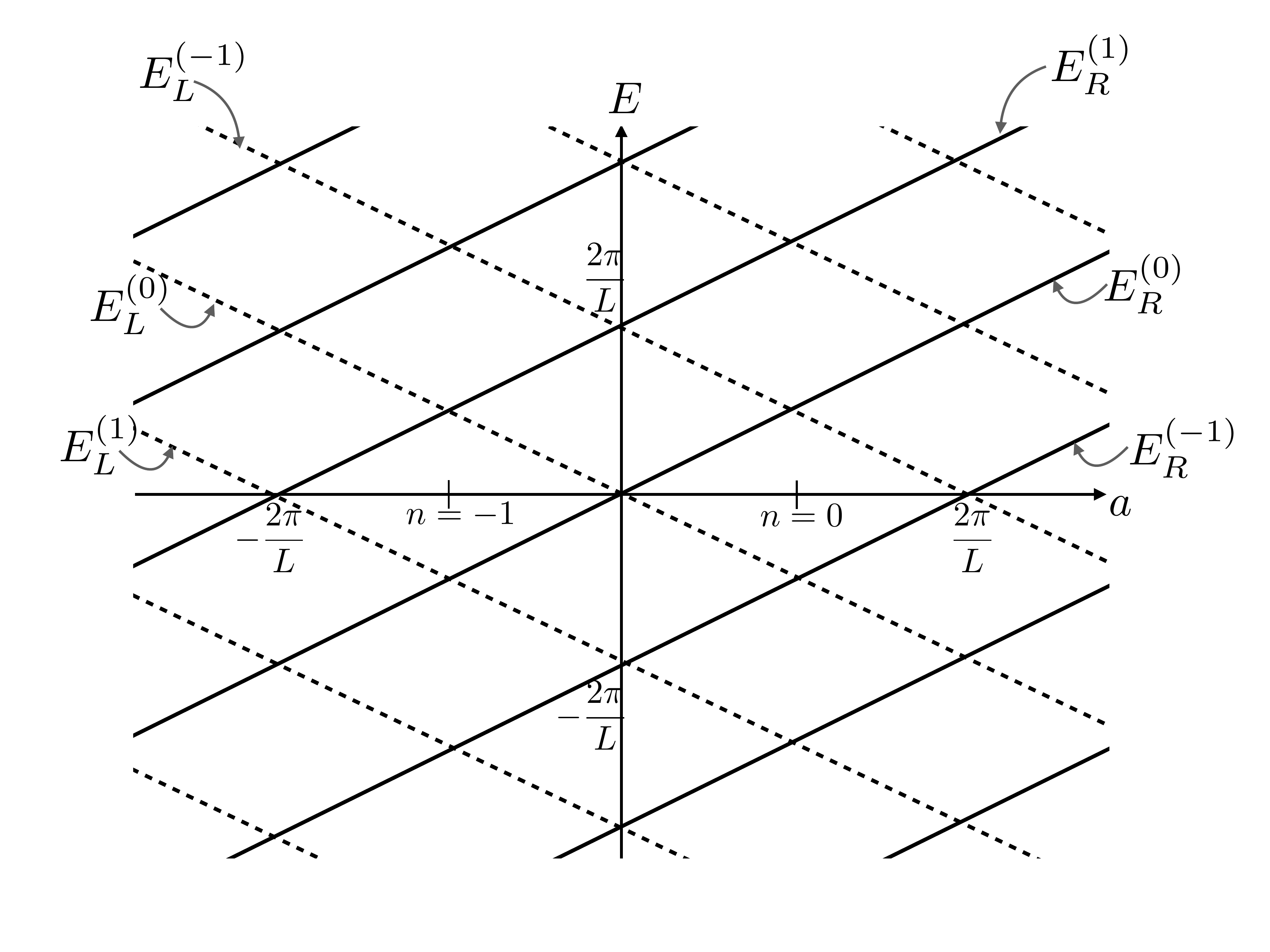}
\caption{One-particle energy levels as a function of $a$ for $q=1$, $N=1$ with periodic boundary condition.
The black solid line stands for $E^{(k)}_{\rm R}$ while the black broken line is $E^{(k)}_{\rm L}$. }
\label{fig:q1N1pbc}
\end{figure}

We review the results for this case by following the argument in the reference \cite{Shifman:1994ce}.
The eigenfunction satisfying the periodic boundary condition for this case is
\be
\psi_{k}(x) \,\propto\, {1\over \sqrt{L}}\exp\left( \im {2\pi k\over{L}} x  \right) \,,
\ee
and the one-particle energy of $k$-th level for the left-handed and right handed fermions is
\be
E^{(k)}_{\rm R} = {2\pi k\over{L}} + a\,,\quad\quad
E^{(k)}_{\rm L} = -{2\pi k\over{L}} - a\,.
\ee
When one of the $k$-th states for the left-handed fermion is filled, we denote the state as $|1_{\rm L}, k\rangle$.
These energies are depicted in Fig.~\ref{fig:q1N1pbc} as a function of $a$.
The black solid line stands for $E^{(k)}_{\rm R}$ while the black broken line is $E^{(k)}_{\rm L}$. 
It is notable that the periodicity of $a$ is 
\be
{2\pi\over{L}}\,,
\ee
which reflects the invariance under large gauge transformation,
\be
a \mapsto a + {2\pi\over{L}}\,,\quad\quad
\psi(x) \mapsto \rme^{-{2\pi \im\over{L}} x} \psi(x)\,.
\ee
The spectrum also indicates that the minimum of the induced potential $V_{\rm eff}(a)$ is
\be
a = {{\pi (2n+1)}\over{L}}\,,
\ee
with $n \in {\mathbb Z}$.
Since the effective potential $V_{\rm eff}(a)$ in the vicinity of the $n$-th minimum is derived from the regularized zero-point energy (Casimir energy) $E_{0}$, we take a sum over all the negative energy levels with the appropriate regularization with a small number $\epsilon$ as  
\begin{align}
E_{0} |_{ a \sim {{\pi (2n+1)}\over{L}} }
&= \sum_{k=-n}^{\infty} E_{\rm L}^{(k)} \exp(-\epsilon |E^{(k)}_{\rm L}|) +  \sum_{k=-\infty}^{-n-1} E_{\rm R}^{(k)} \exp(-\epsilon |E^{(k)}_{\rm R}|) 
\nonumber\\
&= {L\over{2\pi}} \left( a - {{\pi (2n+1)}\over{L}} \right)^{2}+ {\mathcal O}(\epsilon) + {\rm const.}\,,
\end{align}
which corresponds to $E_0=L V(a)$ in (\ref{1-loop-thermal}). 
Then one finds that the induced effective theory around the $n$-th potential minima is a simple harmonic oscillator,
\be
H_{\rm eff}  = -{\pi m_\gamma^2 \over{2L}}\left({\diff \over{\diff  a}}\right)^{2} + {L \over{2\pi }} \left( a - {{\pi (2n+1)}\over{L}} \right)^{2}\,,
\ee
with $m_{\gamma}^2 \equiv e^{2} /\pi $.
Therefore the eigenstate (wavefunction) of the $n$-th ground state is given by
\bea
\langle a|n\rangle &=& \left({L \over{ \pi^2 m_\gamma }}\right)^{1/4} \exp\left[-{1\over{2 \pi m_\gamma L}} \Bigl( L a - {{\pi (2n+1)}} \Bigr)^{2}\right]\nonumber\\
&&\times  \prod_{k=-n}^{+\infty}|1_{\rm L}, k\rangle    \prod_{k=-\infty}^{-n-1} |1_{\rm R}, k\rangle\,.
\label{eq:wavefunc_q1N1}
\eea
By shifting $a \to a +2\pi/L$ (large gauge transformation), one left-handed particle and one right-handed hole emerge as seen from Fig.~\ref{fig:q1N1pbc},
where we find out $\Delta Q =0$ and $\Delta Q_{5} =2$. 
It is nothing but manifestation of the $U(1)_{\rmA}$ axial (or ABJ) anomaly. 
The vacuum state invariant under the large gauge transformation is obtained as a linear combination of $|n \rangle$ with the vacuum angle $\theta$ as
\be
|\theta \rangle = \sum_{n} \rme^{\im n\theta} |n\rangle
\ee
The bilinear chiral condensate in this vacuum states is calculated as
\be
{\langle\theta | \overline{\psi}_{\rm L} \psi_{\rm R} | \theta \rangle \over{\langle\theta | \theta \rangle}}
= \rme^{\im \theta}\langle n-1| \overline{\psi}_{\rm L} \psi_{\rm R} | n \rangle 
= \rme^{\im \theta}{1\over{L}}\exp\left(-{\pi\over{Lm_{\gamma}}}\right)\,.
\ee
This is clearly seen from $\Delta Q_{5} =2$ under $a \to a +2\pi/L$.
It is also notable that the bilinear chiral condensate vanishes if $\Delta Q_{5} > 2$ under this shift.
We will see such cases for multi-flavor Schwinger models below.

It is instructive to calculate the four-point fermion correlator as
\begin{align}
\lim_{\tau \to \infty}{\langle\theta | \overline{\psi}_{\rm L} \psi_{\rm R} \rme^{-H\tau }\, \overline{\psi}_{\rm R} \psi_{\rm L} | \theta \rangle \over{\langle\theta | \theta \rangle}}
&= \langle n| \overline{\psi}_{\rm L} \psi_{\rm R} | n+1 \rangle   \langle n+1| \overline{\psi}_{\rm R} \psi_{\rm L} | n \rangle 
\nonumber\\
&= {1\over{L^{2}}}\exp\left(-{2\pi\over{Lm_{\gamma}}}\right)\,,
\end{align}
which clearly shows the cluster-decomposition property of the theta vacua~\cite{Sachs:1991en},
\be
\lim_{\tau\to \infty}\langle\theta | \overline{\psi}_{\rm L} \psi_{\rm R} (\tau)\, \overline{\psi}_{\rm R} \psi_{\rm L}(0) | \theta \rangle = |\langle\theta | \overline{\psi}_{\rm L} \psi_{\rm R} | \theta \rangle|^2.
\ee

\subsubsection{$q=1$, $N>1$ with thermal b.c.}
\begin{figure}[t]
\centering
\includegraphics[width=100mm]{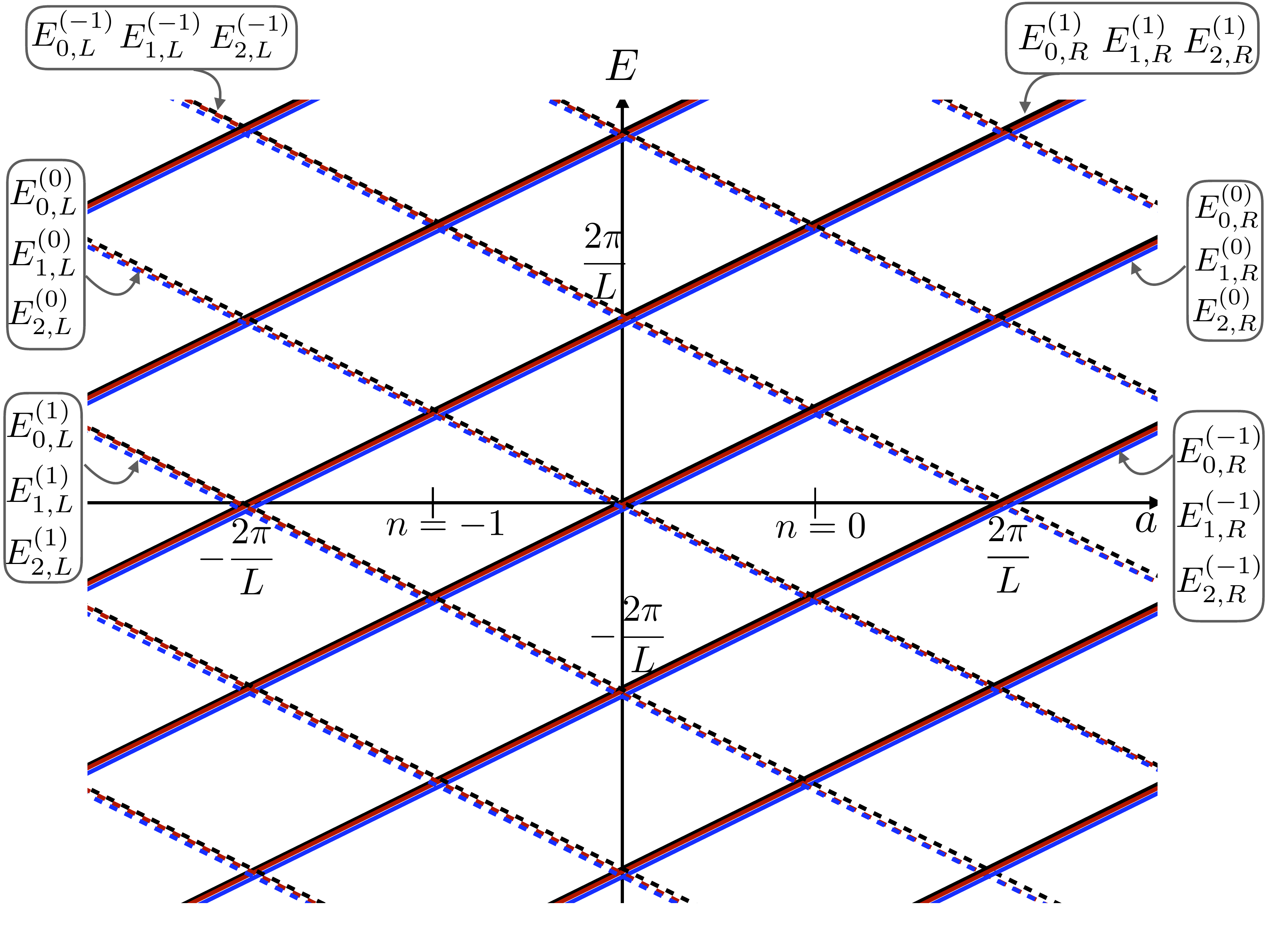}
\caption{One-particle energy levels as a function of $a$ for $q=1, N=3$ with periodic boundary condition.
Black, red and blue solid lines stands for $E^{(k)}_{f, {\rm R}}$ with $f=0,1,2$ while black, red and blue broken lines are $E^{(k)}_{f, {\rm L}}$ with $f=0,1,2$. 
}
\label{fig:q1N3pbc}
\end{figure}

The eigenfunction satisfying p.b.c. is
\be
\psi_{f, k} \propto {1\over \sqrt{L}}\exp\left( \im {2\pi k\over{L}} x  \right) \,,
\ee
where subscript $f$ is used to specify the flavor as $f=0,1,...,N-1$.
The one-particle energy of $k$-th state for the left-handed and right handed fermions for each flavor is
\be
E^{(k)}_{f,{\rm R}} = {2\pi k\over{L}} + a\,,\quad\quad
E^{(k)}_{f,{\rm L}} = -{2\pi k\over{L}} - a\,.
\ee
When one of the $k$-th states for the left-handed $f$-flavor is filled, we denote the state as $|1_{\rm L}^f, k\rangle$.
These energies for $N=3$ are shown in Fig.~\ref{fig:q1N3pbc}.
Black, red and blue solid lines stands for $E^{(k)}_{f, {\rm R}}$ with $f=0,1,2$ while black, red and blue broken lines are $E^{(k)}_{f, {\rm L}}$ with $f=0,1,2$. 
It is obvious that the energy levels for three flavors are degenerate for this case.
The periodicity of $a$ is again ${2\pi\over{L}}$, reflecting invariance under the large gauge transformation, $a \mapsto a + {2\pi\over{L}},\,\psi(x) \mapsto \rme^{-{2\pi \im\over{L}} x} \psi(x)$.
The minimum of the induced potential $V_{\rm eff}(a)$ is $a = {\pi (2n+1) \over{L}}$ with $n \in {\mathbb Z}$.
Then the induced effective Hamiltonian around the $n$-th potential minima is
\be
H_{\rm eff} = -{\pi m_\gamma^2\over{2NL}}\left({\diff \over{\diff a}}\right)^{2} + {NL \over{2\pi}} \left( a - {{\pi (2n+1)}\over{L}} \right)^{2}\,,
\ee
with $m_{\gamma}^2 \equiv N e^{2} /\pi $.
The eigenfunction of the $n$-th ground state is expressed as
\bea
\langle a|n\rangle &=& \left({N L} \over{ \pi^2 m_\gamma} \right)^{1/4} \exp\left[-{{N}\over{2\pi m_\gamma L}} \Bigl(L a - {{\pi (2n+1)}} \Bigr)^{2}\right] \nonumber\\
&&\times \prod_{f=0}^{N-1} \left(\prod_{k=-n}^{+\infty}|1_{\rm L}^{f}, k\rangle    \prod_{k=-\infty}^{-n-1} |1_{\rm R}^{f}, k\rangle \right)\,.
\label{eq:wavefunc_q1N_pbc}
\eea
By shifting $a \to a +2\pi/L$, $N$ left-handed particle and $N$ right-handed hole emerge,
where we have $\Delta Q =0$ and $\Delta Q_{5} =2N$. 
The vacuum states invariant under the large gauge transformation is obtained as a linear combination of $|n \rangle$ with the vacuum angle $\theta$ as $ |\theta \rangle = \sum_{n} \rme^{\im n\theta} |n\rangle$.
For this case one finds that the bilinear chiral condensate vanishes as
\be
\langle\theta | \overline{\psi}_{\rm L}^{f} \psi_{\rm R}^{f} | \theta \rangle  = 0\,,
\ee
since $\Delta Q_{5} =2N$ under $a \to a +2\pi/L$.
It means that the axial subgroup of $SU(N)_{\rm L}$ and $SU(N)_{\rm R}$ flavor symmetry is not broken, which is consistent with Coleman's theorem.

\subsubsection{$q>1$, $N=1$ with thermal b.c.}
\begin{figure}[t]
\centering
\includegraphics[width=120mm]{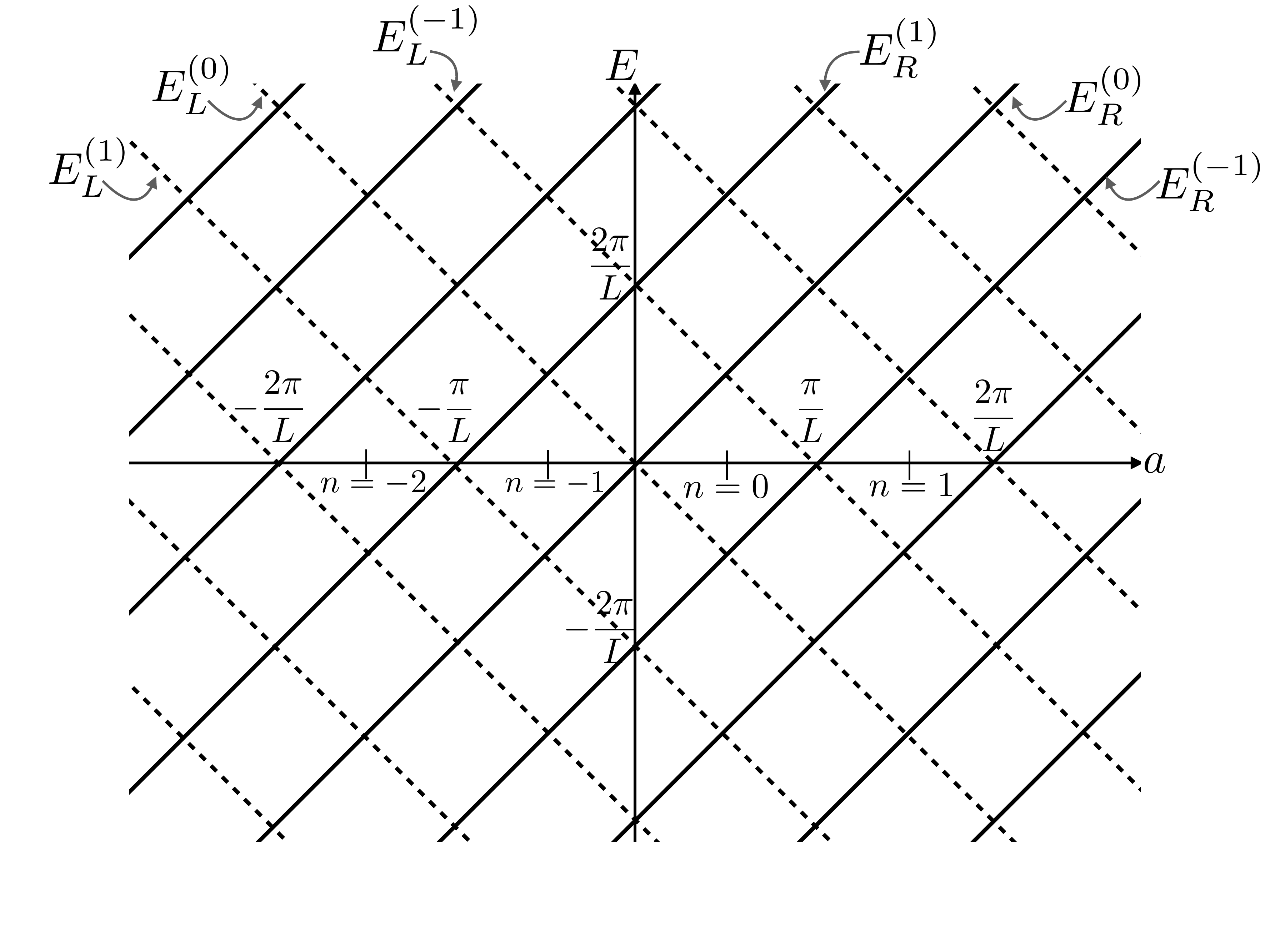}
\caption{One-particle energy levels as a function of $a$ for $q=2, N=1$ with periodic boundary condition.
A black solid line stands for $E^{(k)}_{\rm R}$ while a black broken line is $E^{(k)}_{\rm L}$. 
}
\label{fig:q2N1pbc}
\end{figure}

The eigenfunction satisfying the boundary condition is $ \exp\left( \im {2\pi k\over{L}} x  \right)$ and
the one-particle energy of $k$-th state for the left-handed and right-handed fermions is
\be
E^{(k)}_{\rm R} = {2\pi k\over{L}} + q a\,,\quad\quad
E^{(k)}_{\rm L} = -{2\pi k\over{L}} - q a\,.
\ee
These energies for $q=2$ are shown in Fig.~\ref{fig:q2N1pbc}.
A black solid line stands for $E^{(k)}_{\rm R}$ while a black broken line is $E^{(k)}_{\rm L}$. 
The periodicity of $a$ is 
\be
{2\pi\over{q L}},
\ee
which reflects the ${\mathbb Z}_{q}$ one-form symmetry \cite{Anber:2018jdf, Anber:2018xek, Armoni:2018bga}
\be
a \to a + {2\pi\over{q L}}\,, \quad\quad
\psi(x) \to \rme^{-{2\pi \im\over{L}} x} \psi(x)\,,
\label{eq:Zqsym}
\ee
which results in ${\mathbb Z}_{q}$ 0-form symmetry under compactification. 
The spectrum indicates that the minimum of the induced potential $V_{\rm eff}(a)$ is
\be
a = {{\pi (2n+1)}\over{q L}}\,.
\ee
and the induced effective hamiltonian around the $n$-th potential minimum is 
\be
H_{\rm eff} = -{\pi m_\gamma^2\over{2q^2 L}}{\diff^{2}\over{\diff a^{2}}} + {q^2 L \over{2 \pi}} \left( a - {{\pi (2n+1)}\over{q L}} \right)^{2}\,,
\ee
with $m_{\gamma}^2 \equiv q^2 e^{2} /\pi $.
The eigenfunction of the $n$-th ground state is expressed as
\bea
\langle a|n\rangle &=& \left(q^2 L \over{ \pi^2 m_\gamma} \right)^{1/4} \exp\left[-{q^2 \over{2 \pi m_\gamma L}} \left(L a - {{\pi (2n+1)}\over{q }} \right)^{2}\right]\nonumber\\
&&\times \prod_{k=-n}^{+\infty}|1_{\rm L}, k\rangle    \prod_{k=-\infty}^{-n-1} |1_{\rm R}, k\rangle \,.
\label{eq:wavefunc_qN1}
\eea
By shifting $a \to a +2\pi/(q L)$, one left-handed particle and one right-handed hole emerge,
where we have $\Delta Q =0$ and $\Delta Q_{5} =2$ as seen from Fig.~\ref{fig:q2N1pbc}. 

The physical vacuum states are constructed as linear combinations of $|n \rangle$ with the vacuum angle $\theta$ so that they are invariant under the large-gauge transformation, $a\mapsto a+2\pi/L$. We obtain $q$ different physical vacua as
\be
\widetilde {| \ell, \theta\rangle}  = 
\sum_{m \in\mathbb{Z}}\rme^{\im m\theta} |\ell+q m\rangle ,
\ee
for $\ell=0,1,\ldots, q-1$, and they are related under $\mathbb{Z}_q^{[1]}$ transformation as $\ell\mapsto \ell+1$.  
We take the linear combinations of these $q$ vacua so that they become eigenstates of $\mathbb{Z}_q^{[1]}$ symmetry:
\bea
|\theta,k\rangle&=&\sum_{\ell=0}^{q-1} \rme^{\im {\theta+2\pi k\over q}\ell} \; \widetilde {| \ell, \theta\rangle} 
\nonumber\\ 
&=&\sum_{\ell=0}^{q-1} \rme^{\im {\theta+2\pi k\over q}\ell}\left(\sum_{m \in\mathbb{Z}}\rme^{\im m\theta} |\ell+q m\rangle\right) \nonumber\\
&=&\sum_{n\in\mathbb{Z}}\rme^{\im {\theta+2\pi k\over q}n} |n\rangle, 
\label{eq:theta_vac_qN1}
\eea
with $k=0,q,\ldots, q-1$. 
It is notable that their dependence on vacuum angle $\theta$ is fractional, $\tilde{\theta}\equiv \theta/q$, and then $|\theta+2\pi,k\rangle=|\theta,k+1\rangle$.
The bilinear chiral condensate is thus calculated as
\be
{\langle\theta,k | \overline{\psi}_{\rm L} \psi_{\rm R} | \theta,k \rangle \over{\langle \theta,k | \theta,k \rangle}}
= \rme^{\im (\theta+2\pi k)/q}\langle n-1| \overline{\psi}_{\rm L} \psi_{\rm R} | n \rangle 
={1\over{L}} \rme^{\im (\theta+2\pi k)/q}\exp\left(-{\pi\over{L m_{\gamma}}}\right)\,.
\ee
This is clearly seen from $\Delta Q_{5} =2$ under $a \mapsto a +2\pi/(qL)$.
This condensate spontaneously breaks the discrete chiral symmetry $({\mathbb Z}_{q})_{\rm R}$ \cite{Anber:2018jdf, Anber:2018xek}. 
The emergence of fractionalized action $\pi/(L m_{\gamma})$ and fractionalized vacuum angle 
$\tilde{\theta}=\theta/q$ originates in the contribution of the fractional instantons to chiral condensate.
The four-point fermion correlator is also calculated as
\be
\lim_{\tau \to \infty } {\langle{\theta,k} | \overline{\psi}_{\rm L} \psi_{\rm R} e^{- H \tau}\,\overline{\psi}_{\rm R} \psi_{\rm L} | {\theta,k} \rangle \over{\langle {\theta,k} | {\theta,k} \rangle}}
= {1\over{L^{2}}}\exp\left(-{2\pi\over{Lm_{\gamma}}}\right)\,,
\ee
and thus the cluster decomposition is satisfied for these vacua. More generally, 
\be
\lim_{M_2\to \mathbb{R}\times S^1}\langle \mathcal{O}(x_1,\ldots, x_n)\rangle_{M_2}={1\over q}\sum_{k=0}^{q-1} {\langle \theta,k | \mathcal{O}(x_1,\ldots, x_n) | \theta, k\rangle \over \langle \theta,k|\theta,k\rangle}
\ee
for any two-dimensional local $n$-point functions $\mathcal{O}$, and the cluster decomposition holds for each sector $|\theta,k\rangle$.

\subsubsection{$q>1$, $N>1$ with thermal b.c.}
\begin{figure}[t]
\centering
\includegraphics[width=120mm]{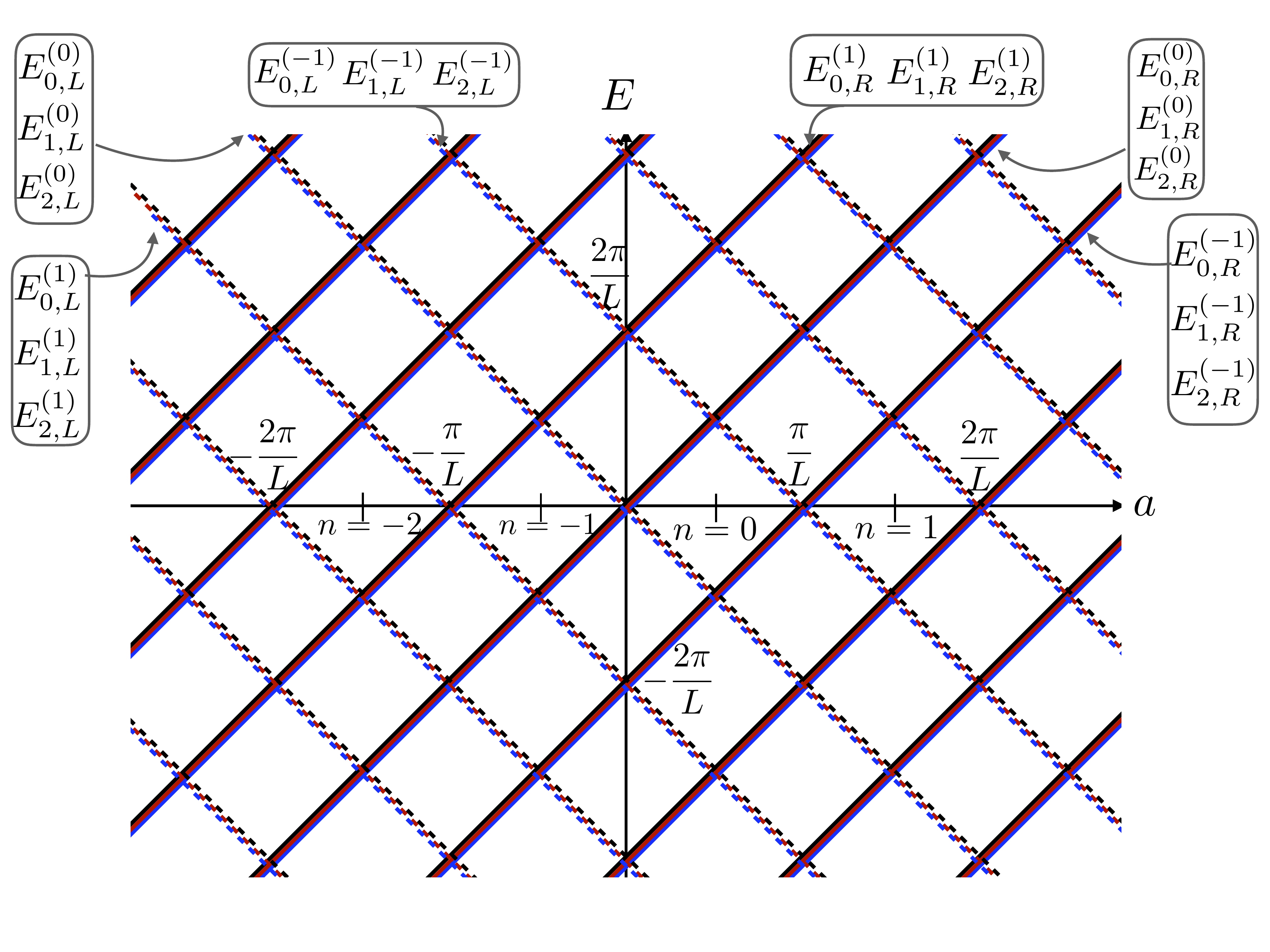}
\caption{One-particle energy levels as a function of $a$ for $q=2, N=3$ with periodic boundary condition.
Black, red and blue solid lines stands for $E^{(k)}_{f, {\rm R}}$ with $f=0,1,2$ while black, red and blue broken lines are $E^{(k)}_{f, {\rm L}}$ with $f=0,1,2$. 
}
\label{fig:q2N3pbc}
\end{figure}

The eigenfunction $\exp\left( \im {2\pi k\over{L}} x  \right)$ satisfying the p.b.c. leads to
the energy of $k$-th state for the right-handed and left-handed fermions for each flavor, 
$E^{(k)}_{f, {\rm R}} = {2\pi k\over{L}} + q a\,, E^{(k)}_{f,{\rm L}} = -{2\pi k\over{L}} - q a$. 
These one-particle  energies for $q=2, N=3$ are depicted in Fig.~\ref{fig:q2N3pbc}.
Black, red and blue solid lines stands for $E^{(k)}_{f, {\rm R}}$ with $f=0,1,2$ and black, red and blue broken lines are $E^{(k)}_{f, {\rm L}}$ with $f=0,1,2$,
where the energy levels for three flavors are degenerate for this case.
The periodicity ${2\pi\over{q L}}$ is a remnant of the ${\mathbb Z}_{q}$ one-form symmetry 
$a \mapsto a + {2\pi\over{q L}}\,, \psi(x) \mapsto \rme^{-{2\pi \im\over{L}} x} \psi(x)$. 
The spectrum indicates that the minimum of the induced potential $V_{\rm eff}(a)$ is
$a = {{\pi (2n+1)}\over{q L}}$ and the induced effective hamiltonian around the $n$-th potential minimum is given by 
\be
H_{\rm eff} = -{\pi m_\gamma^2\over{2q^2 N L}}{\diff^{2}\over{\diff a^{2}}} + {q^2 N L \over{2 \pi}} \left( a - {{\pi (2n+1)}\over{q L}} \right)^{2}\,,
\ee
where
\be
m_{\gamma}^2 \equiv N q^2 e^{2} /\pi\,.
\ee
The eigenfunction of the $n$-th ground state is expressed as
\bea
\langle a|n\rangle &=& \left({q^2 N L} \over{ \pi^2 m_\gamma} \right)^{1/4} \exp\left[-{{q^2 N}\over{2\pi m_\gamma L}} \Bigl(L a - {{\pi (2n+1)}\over q} \Bigr)^{2}\right] \nonumber\\
&&\times \prod_{f=0}^{N-1} \left(\prod_{k=-n}^{+\infty}|1_{\rm L}^{f}, k\rangle    \prod_{k=-\infty}^{-n-1} |1_{\rm R}^{f}, k\rangle \right)\,.
\label{eq:wavefunc_qN_pbc}
\eea
By shifting $a \to a +2\pi/(q L)$, $N$ left-handed particle and $N$ right-handed hole emerge,
where we have $\Delta Q =0$ and $\Delta Q_{5} =2N$ as seen from Fig.~\ref{fig:q2N3pbc}. 
The physical vacuum states are obtained as a linear combination of $|n \rangle$ with the vacuum angle $\theta$ as 
\be
|\theta, k \rangle = \sum_{n} \rme^{\im {\theta+2\pi k\over q}n} |n\rangle,
\label{eq:theta_vac_qN_pbc}
\ee
with $k=0,1,\ldots,q-1$.
The bilinear chiral condensate for this case vanishes due to $\Delta Q_{5} =2N$ under $a \to a +2\pi/(qL)$ as
\be
\langle\theta,k | \overline{\psi}_{\rm L}^{f} \psi_{\rm R}^{f} | \theta,k \rangle = 0\,.
\ee
The axial subgroup of $SU(N)_{\rm L}$ and $SU(N)_{\rm R}$ flavor symmetry is not broken, which is consistent with Coleman's theorem.
As shown in \cite{Anber:2018jdf, Anber:2018xek}, however, there emerges the determinant condensate ${\rm det}\overline{\psi}^{f}_{\rm L} \psi^{g}_{\rm R}$, which breaks 
the discrete chiral symmetry $({\mathbb Z}_{q})_{\rm R}$. 
This is exactly analogous to QCD with adjoint fermions on small $\mathbb{R}^3\times S^1$~\cite{Unsal:2007jx}. 
The reason of existence of the determinant condensate composed of $2N$ fermion operators again originates in $\Delta Q_{5} =2N$ under $a \to a +2\pi/(qL)$ in Fig.~\ref{fig:q2N3pbc}. 
We can compute its explicit form as
\begin{align}
{\langle {\theta,k}| {\rm det}\overline{\psi}^{f}_{\rm L} \psi^{g}_{\rm R} | {\theta,k} \rangle \over{ {\langle {\theta,k}| {\theta,k} \rangle}}}
&= \rme^{\im {\theta+2\pi k\over q}} N! \langle n-1 | \prod_{f} \overline{\psi}^{f}_{\rm L} \psi^{f}_{\rm R} | n \rangle
\nonumber\\
&= \rme^{\im {\theta+2\pi k\over q}} {N! \over{ L^{N}}} \exp\left(-{N\pi\over{Lm_{\gamma}}}\right)\,.
\label{eq:det_chiral_condensate_pbc}
\end{align}
This condensate are saturated by configuration with one fractional instanton.

\subsection{Chiral condensate in flavor-twisted boundary condition}
\label{sec:chiral_condensate_RS1_tbc}

\subsubsection{$q=1$, $N>1$ with ${\mathbb Z}_{N}$ twisted b.c.}\label{sec:NflavorSMQM}
\begin{figure}[t]
\centering
\includegraphics[width=120mm]{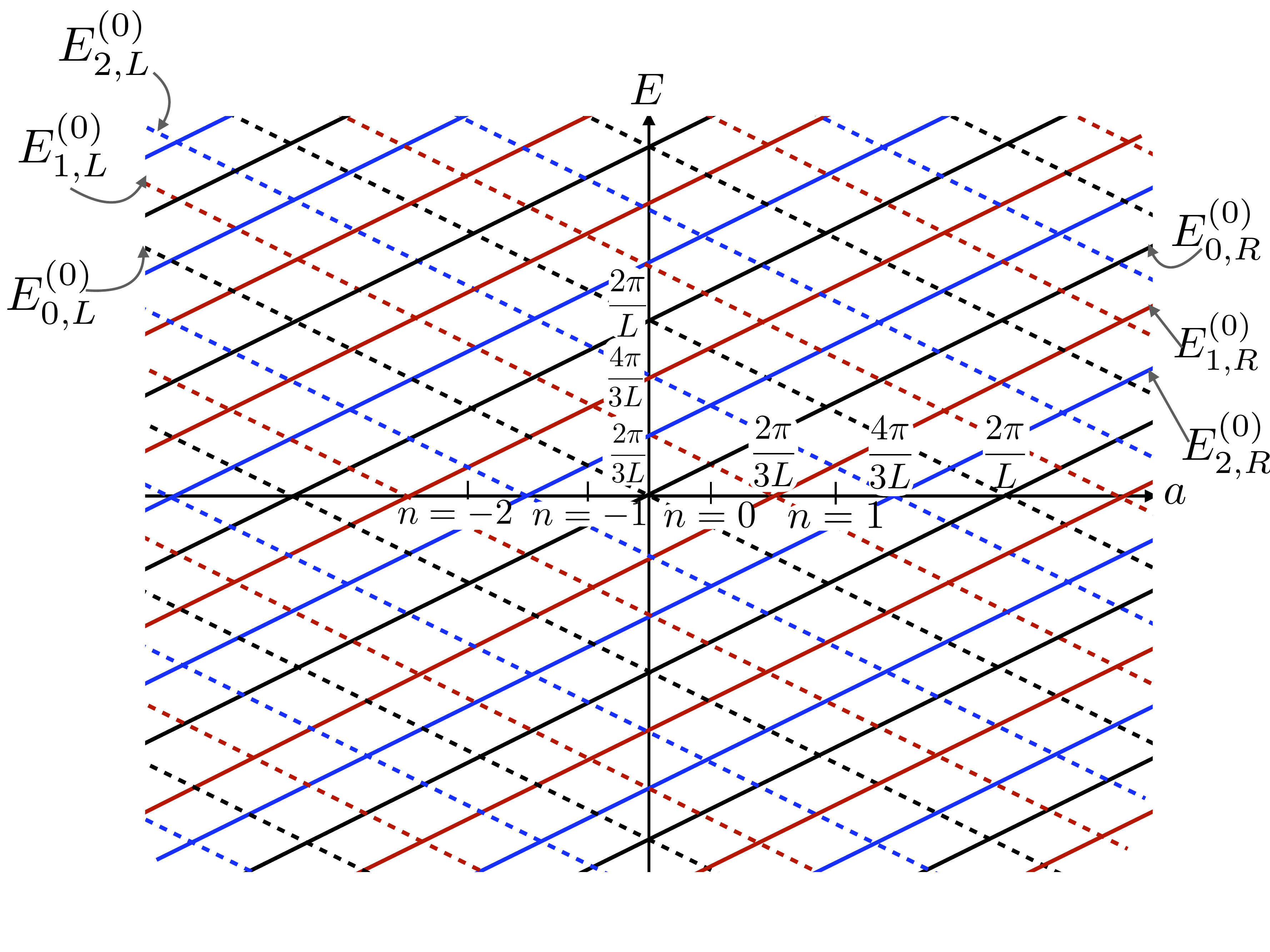}
\caption{One-particle energy levels as a function of $a$ for $q=1, N=3$ with ${\mathbb Z}_{N}$ twisted boundary condition.
Black, red and blue solid lines stands for $E^{(k)}_{f, {\rm R}}$ with $f=0,1,2$ while black, red and blue broken lines are $E^{(k)}_{f, {\rm L}}$ with $f=0,1,2$. 
}
\label{fig:q1N3tbc}
\end{figure}

The eigenfunction satisfying ${\mathbb Z}_{N}$ t.b.c. is
\be
\psi_{f, k} \propto {1\over \sqrt{L}} \exp\left( \im {2\pi (Nk-f)\over{N L}} x  \right) \,,
\ee
with $f=0,1,...,N-1$.
The one-particle energy of $k$-th state for the right-handed and left-handed fermions for each flavor is
\be
E^{(k)}_{f,{\rm R}} = {2\pi (Nk-f)\over{NL}} + a\,,\quad\quad
E^{(k)}_{f,{\rm L}} = -{2\pi (Nk-f)\over{NL}} - a\,.
\ee
When one of the $k$-th states for the left-handed $f$-flavor is filled, we denote the state as $|1_{\rm L}^f, k\rangle$.
These energies for $N=3$ are depicted in Fig.~\ref{fig:q1N3tbc}.
Black, red and blue solid lines stands for $E^{(k)}_{f, {\rm R}}$ with $f=0,1,2$ while black, red and blue broken lines are $E^{(k)}_{f, {\rm L}}$ with $f=0,1,2$. 
We note that the degeneracy of the energy levels for flavors is lifted in this case. 
The periodicity of $a$ is 
\be
{2\pi\over{NL}}\,,
\ee
reflecting the ${\mathbb Z}_{N}$ one-form symmetry, given in (\ref{eq:Znq_0form_sym_1}) and (\ref{eq:Znq_0form_sym_2}),  
\be
a \mapsto a + {2\pi\over{NL}},\quad\quad \psi_{f}(x) \mapsto \rme^{-{2\pi \im\over{NL}} x} \psi_{f+1}(x)\,,
\ee
which results in ${\mathbb Z}_{N}$ zero-form symmetry in the compactified theory.
The minimum of the induced potential $V_{\rm eff}(a)$ is 
\be
a = {\pi (2n+1) \over{NL}}\,,
\ee
with $n \in {\mathbb Z}$.
Then the induced effective Hamiltonian around the $n$-th potential minimum is
\be
H_{\rm eff} = -{\pi m_\gamma^{2}\over{2 N L}}\left({\diff \over{\diff a}}\right)^{2} + {N L \over{2\pi }} \left( a - {{\pi (2n+1)}\over{N L}} \right)^{2}\,,
\ee
with $m_{\gamma}^2 \equiv N e^{2} /\pi $.
Let us consider the eigenfunction of the $n$-th ground state. We here denote $n$ as $n=N\ell + j$ with $\ell \in {\mathbb Z}$ and $j=0,1,\ldots, N-1$, since the properties of the vacua depend on $n$ mod $N$. 
The eigenfunction is expressed as
\bea\hspace{-1em}
\langle a|n\rangle 
&=& \langle a|N\ell + j \rangle
\nonumber\\
&=& \left({N L} \over{ \pi^2 m_\gamma} \right)^{1/4} \exp\left[-{{N}\over{2\pi m_\gamma L}} \left( La - {{\pi (2(N\ell+j)+1)}\over{N}} \right)^{2}\right] \nonumber\\
&&\times \prod_{f=0}^{j} 
\left(\prod_{k=-\ell}^{\infty}|1_{\rm L}^{f}, k\rangle    
\prod_{k=-\infty}^{-\ell-1} |1_{\rm R}^{f}, k\rangle \right)
\prod_{f'=j+1}^{N-1} 
\left(\prod_{k=1-\ell}^{+\infty}|1_{\rm L}^{f'}, k\rangle    
\prod_{k=-\infty}^{-\ell} |1_{\rm R}^{f'}, k\rangle \right).
\eea
By shifting $a \to a +2\pi/(NL)$, one left-handed particle and one right-handed hole emerge,
where we have $\Delta Q =0$ and $\Delta Q_{5} =2$. 

The physical vacuum states are invariant under the large-gauge transformation, $a\mapsto a+2\pi/L$, and they are obtained as a linear combination of $|n \rangle$ with the vacuum angle $\theta$ as,
\be
\widetilde {| j, \theta\rangle}   = 
\sum_{\ell\in\mathbb{Z}}\rme^{\im \ell \theta}| N\ell+j \rangle. 
\ee
The eigenstates under $\mathbb{Z}_{N}^{[0]}$ shift-center symmetry are given by 
\bea
|\theta, k\rangle&=&  
\sum_{j=0}^{N-1} \rme^{\im {\theta+2\pi k\over N}\ell} \; \widetilde {| j, \theta\rangle} 
\nonumber\\ 
&=&
\sum_{n\in\mathbb{Z}}\rme^{\im n {\theta+2\pi k\over N}}|n\rangle, 
\eea
with $k=0,1,\ldots, N-1$, 
and their dependence on $\theta$ becomes fractional as $\tilde{\theta} = \theta/N$. 
The bilinear chiral condensate is thus calculated as
\bea
{\langle\theta, k | \overline{\psi}_{\rm L}^{f} \psi_{\rm R}^{f} | \theta,k \rangle \over{\langle{\theta, k} | {\theta,k} \rangle}}
&=& \rme^{\im {\theta+2\pi k\over N}}{\langle N\ell+f-1 | \overline{\psi}_{\rm L}^{f} \psi_{\rm R}^{f} | N\ell + f \rangle \over \sum_{f'} \langle N\ell+f| N\ell +f\rangle}\nonumber\\
&=&{1\over{N L}} \rme^{\im {\theta+2\pi k\over N}}\exp\left(-{\pi\over{N Lm_{\gamma}}}\right). 
\eea
This is due to $\Delta Q_{5} = 2$ under $a \to a +2\pi/(NL)$.
This result was first derived in \cite{Shifman:1994ce}.
This condensate breaks the discrete chiral symmetry $({\mathbb Z}_{N})_{\rm R}$. 
The emergence of  fractional vacuum angle 
$\tilde{\theta}=\theta/N$ originates in the contribution of the $1/N$ fractional instantons to chiral condensate.
We note that $q>1$, $N=1$ with p.b.c. and $q=1$, $N>1$ with ${\mathbb Z}_{N}$ t.b.c. share the properties including symmetries, chiral condensates and symmetry breaking patterns.
The four-point fermion correlator is
\be
\lim_{\tau \to \infty}{\langle{\theta,k} | \overline{\psi}_{\rm L}^{f} \psi_{\rm R}^{f} e^{- H \tau}\,  \overline{\psi}_{\rm R}^{f} \psi_{\rm L}^{f} | {\theta,k} \rangle \over{\langle {\theta,k} | {\theta,k} \rangle}}
= {1\over{N^{2}L^{2}}}\exp\left(-{2\pi\over{N Lm_{\gamma}}}\right)\,\,
\ee
which shows the cluster decomposition property\footnote{In the case of $(0+1)$d quantum mechanics, the breakdown of cluster decomposition means degenerate ground states as in the case of spontaneous symmetry breaking in higher dimensions, however, it does not lead the superselection rule unlike higher-dimensional case. This notice becomes important in discussion on Polyakov-loop correlators in Sec.~\ref{sec:Polyakov_loop}. Because of this special nature in $(0+1)$d, the degeneracy related to $(\mathbb{Z}_N)_\rmR$ discrete chiral symmetry does not mean the breakdown of $U(1)^{N-1}$ axial symmetry in the symmetry group (\ref{twisted-full}). This point will be discussed in more detail in Sec.~\ref{sec:twisted_wzw}. }.

\subsubsection{$q>1$, $N>1$ with ${\mathbb Z}_{N}$ twisted b.c.}\label{sec:chiral_condensate_qN_tbc}
\begin{figure}[t]
\centering
\includegraphics[width=120mm]{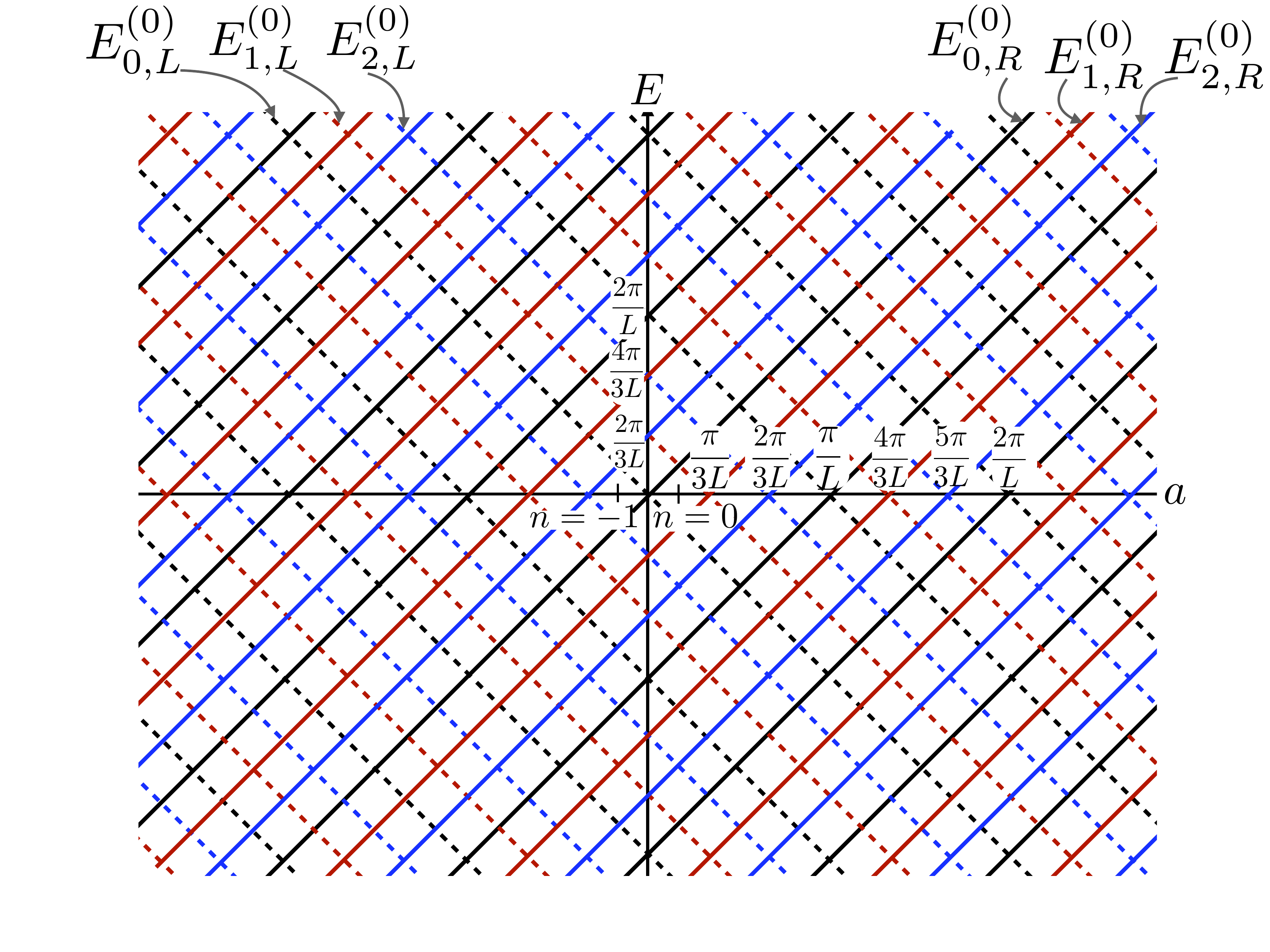}
\caption{One-particle energy levels as a function of $a$ for $q=2, N=3$ with ${\mathbb Z}_{N}$ twisted boundary condition.
Black, red and blue solid lines stands for $E^{(k)}_{f, {\rm R}}$ with $f=0,1,2$ while black, red and blue broken lines are $E^{(k)}_{f, {\rm L}}$ with $f=0,1,2$. 
}
\label{fig:q2N3tbc}
\end{figure}

The eigenfunction satisfying ${\mathbb Z}_{N}$ t.b.c. is $\exp\left( \im {2\pi k\over{N L}} x  \right)$ and the one-particle energy of $k$-th state for the right-handed and left-handed fermions for each flavor is
\be
E^{(k)}_{f,{\rm R}} = {2\pi (Nk-f)\over{NL}} + qa\,,\quad\quad
E^{(k)}_{f,{\rm L}} = -{2\pi (Nk-f)\over{NL}} - qa\,.
\ee
These energies for $q=2, N=3$ are depicted in Fig.~\ref{fig:q2N3tbc}.
Black, red and blue solid lines stands for $E^{(k)}_{f, {\rm R}}$ with $f=0,1,2$ while black, red and blue broken lines are $E^{(k)}_{f, {\rm L}}$ with $f=0,1,2$. 
We note that the degeneracy of the energy levels for flavors is again lifted in this case. 
The periodicity of $a$ is 
\be
{2\pi\over{qNL}}\,,
\ee
reflecting the ${\mathbb Z}_{qN}$ one-form symmetry 
\be
a \to a + {2\pi\over{qNL}},\quad\quad \psi_{f}(x) \to \rme^{-{2\pi \im\over{NL}} x} \psi_{f+1}(x)\,,
\ee
which results in ${\mathbb Z}_{qN}$ zero-form symmetry in the compactified theory.
The minimum of the induced potential $V_{\rm eff}(a)$ is 
\be
a = {\pi (2n+1) \over{qNL}}\,,
\ee
with $n \in {\mathbb Z}$.
Then the induced effective Hamiltonian around the $n$-th potential minimum is
\be
H_{\rm eff} = -{\pi m_\gamma^2\over{2q^2 N L}}\left({\diff \over{\diff a}}\right)^{2} + {q^2 N L \over{2\pi}} \left( a - {{\pi (2n+1)}\over{q N L}} \right)^{2}\,,
\ee
with $m_{\gamma}^2 \equiv N q^2 e^{2} /\pi$.
Let us consider the eigenfunction of the $n$-th ground state. We again denote $n$ as $n=N\ell + j$ with $\ell \in {\mathbb Z}$ and $j=0,1,2$. 
Then the eigenfunction is expressed as
\bea\hspace{-1em}
\langle a|n\rangle &=& \langle a|N\ell + j \rangle
\nonumber\\
&=& \left({q^2 N L} \over{ \pi^2 m_\gamma} \right)^{1/4} \exp\left[-{{q^2 N}\over{2\pi m_\gamma L}} \left( La - {{\pi (2n+1)}\over{qN}} \right)^{2}\right]\nonumber\\
&&\times \prod_{f=0}^{j} 
\left(\prod_{k=-\ell}^{\infty}|1_{\rm L}^{f}, k\rangle    
\prod_{k=-\infty}^{-\ell-1} |1_{\rm R}^{f}, k\rangle \right)
\prod_{f'=j+1}^{N-1} 
\left(\prod_{k=1-\ell}^{+\infty}|1_{\rm L}^{f'}, k\rangle    
\prod_{k=-\infty}^{-\ell} |1_{\rm R}^{f'}, k\rangle \right).
\label{eq:wavefunc_qN_tbc}
\eea
By shifting $a \to a +2\pi/(qNL)$, one left-handed particle and one right-handed hole emerge,
where we have $\Delta Q =0$ and $\Delta Q_{5} =2$. 
The physical vacuum state is obtained as a linear combination of $|n \rangle$ with the vacuum angle $\theta$ as 
\be
|\theta,k\rangle = \sum_{n} \rme^{\im n{\theta+2\pi k\over Nq}} |n\rangle,
\label{eq:theta_vac_qN_tbc}
\ee
with $k=0,1,\ldots,Nq-1$. 
This vacuum angle $\tilde{\theta}$ satisfies $\tilde{\theta} = \theta/(qN)$. The bilinear chiral condensate is calculated as
\bea
{\langle \theta,k | \overline{\psi}_{\rm L}^{f} \psi_{\rm R}^{f} | \theta,k \rangle
\over{\langle{\theta,k} | {\theta,k} \rangle}}
&=&\rme^{\im{\theta+2\pi k\over Nq}} {\langle N\ell +f-1 | \overline{\psi}_{\rm L}^{f} \psi_{\rm R}^{f} | N\ell+f \rangle \over \sum_{f'} \langle N\ell +f'| N\ell +f'\rangle}\nonumber\\
&=&{1\over{N L}} \rme^{\im{\theta+2\pi k\over Nq}}\exp\left(-{\pi\over{N Lm_{\gamma}}}\right). 
\label{eq:bilinear_condensate_qN_twisted}
\eea
This is due to $\Delta Q_{5} = 2$ under $a \to a +2\pi/(NL)$.
This condensate breaks the discrete chiral symmetry $({\mathbb Z}_{qN})_{\rm R}$. 
The emergence of fractionalized action $\pi/(N L m_{\gamma})$ and fractionalized vacuum angle 
$\tilde{\theta}=\theta/(qN)$ originates in the contribution of the $1/(qN)$ fractional instantons to chiral condensate. The four-point fermion correlator is
\begin{align}
\lim_{\tau \to \infty}{\langle{\theta,k} | \overline{\psi}_{\rm L}^{f} \psi_{\rm R}^{f} e^{- H \tau}\, \overline{\psi}_{\rm R}^{f} \psi_{\rm L}^{f} | {\theta,k} \rangle \over{\langle {\theta,k} | {\theta,k} \rangle}}
= {1\over{N^{2}L^{2}}}\exp\left(-{2\pi\over{N Lm_{\gamma}}}\right)\,\,
\end{align}
which shows the cluster-decomposition property.


\subsection{Chiral condensate for generic $L$}

The chiral condensate in Schwinger models on ${\mathbb R}\times S^{1}$ with arbitrary $e L$ was discussed in Refs.~\cite{Manton:1985jm, Sachs:1991en,Shifman:1994ce} via the bosonization, the four-point fermion correlators and the fracton path integral.
We below show the results extended to $q\geq 1$ and $N\geq 1$ with thermal and flavor-twisted boundary conditions and compare them to what we have obtained in the previous subsections.

Following the arguments in \cite{Sachs:1991en,Shifman:1994ce}, we find that the chiral condensate for generic $L$ for $q\geq 1$ and $N=1$ is expressed as
\begin{align}
|\langle \overline{\psi}_{\rm L} \psi_{\rm R} \rangle| 
&= {1\over{L}}\exp\left( -{\pi\over{Lm_{\gamma}}}\right)\exp\left[\gamma + {\pi\over{Lm_{\gamma}}} + \log{Lm_{\gamma}\over{4\pi}}-I(Lm_{\gamma})\right]
\nonumber\\
&={m_{\gamma}  \rme^{\gamma}\over{4\pi}} \exp[-I(Lm_{\gamma})] \,,
\end{align}
with
\be
I(x) = \int_{0}^{\infty} {dt\over{\sqrt{t^2 + x^{2}}}} \left(\coth{\sqrt{t^2 + x^2}\over{2}}-1\right)  =
\begin{cases}
\gamma + {\pi\over{x}} + \log{x\over{4\pi}} \quad\quad (x \sim 0)
\\
\quad\quad\quad0\quad\quad\quad\quad\quad (x \to \infty)
\end{cases}
\,,
\ee
with $m_{\gamma}^{2}=q^2 e^2/\pi$ \footnote{In \cite{Sachs:1991en}, the same function $I(x)$ is expressed as a distinct form $I(x)= \int_{0}^{\infty}{2\over{\rme^{x \cosh t} -1}} dt$.}.
This smoothly connects the chiral condensate $\rme^{-{\pi\over{Lm_{\gamma}}}} /L $ for small $L$ in the previous subsections to $m_{\gamma} \rme^{\gamma}/(4\pi)$ in $L\to \infty$ on ${\mathbb R}^{2}$.

For $q\geq 1$, $N>1$ with thermal boundary condition, we find that the chiral condensate vanishes even for $e L\not\ll 1$, 
\be
|\langle \overline{\psi}_{\rm L} \psi_{\rm R} \rangle| 
= 0\,,
\ee
which is consistent with the vanishing chiral condensate for small $L$ in the previous subsections.

\begin{figure}[t]
\centering\hspace{-4em}
\begin{minipage}{.45\textwidth}
\subfloat[$(q,N)=(1,1)$]{
\includegraphics[width=80mm]{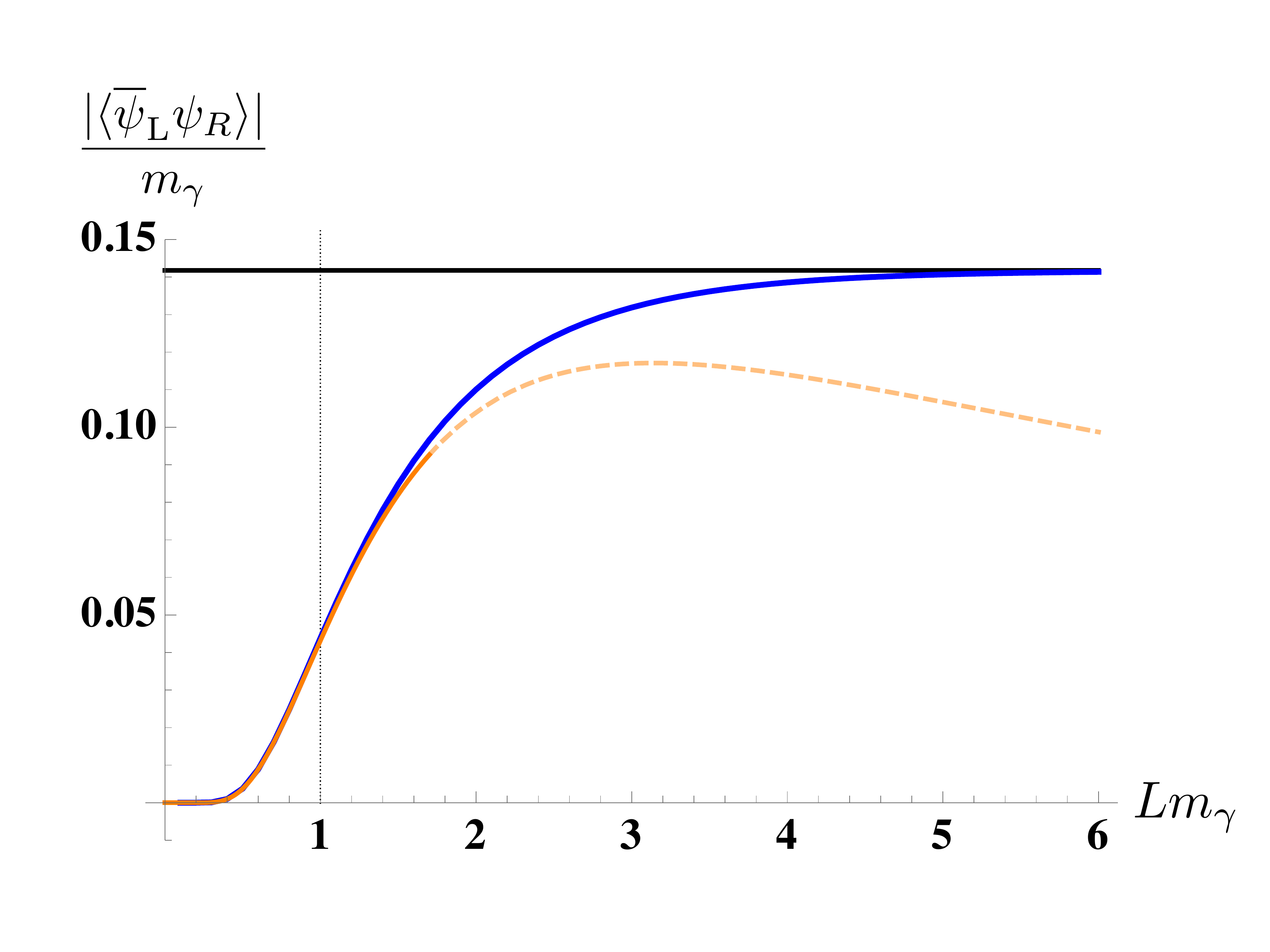}
\label{fig:q1N1_condensate}
}
\end{minipage}\hspace{2.5em}
\begin{minipage}{.45\textwidth}
\subfloat[$(q,N)=(1,3)$ with $\Omega_F$-twist]{
\includegraphics[width=80mm]{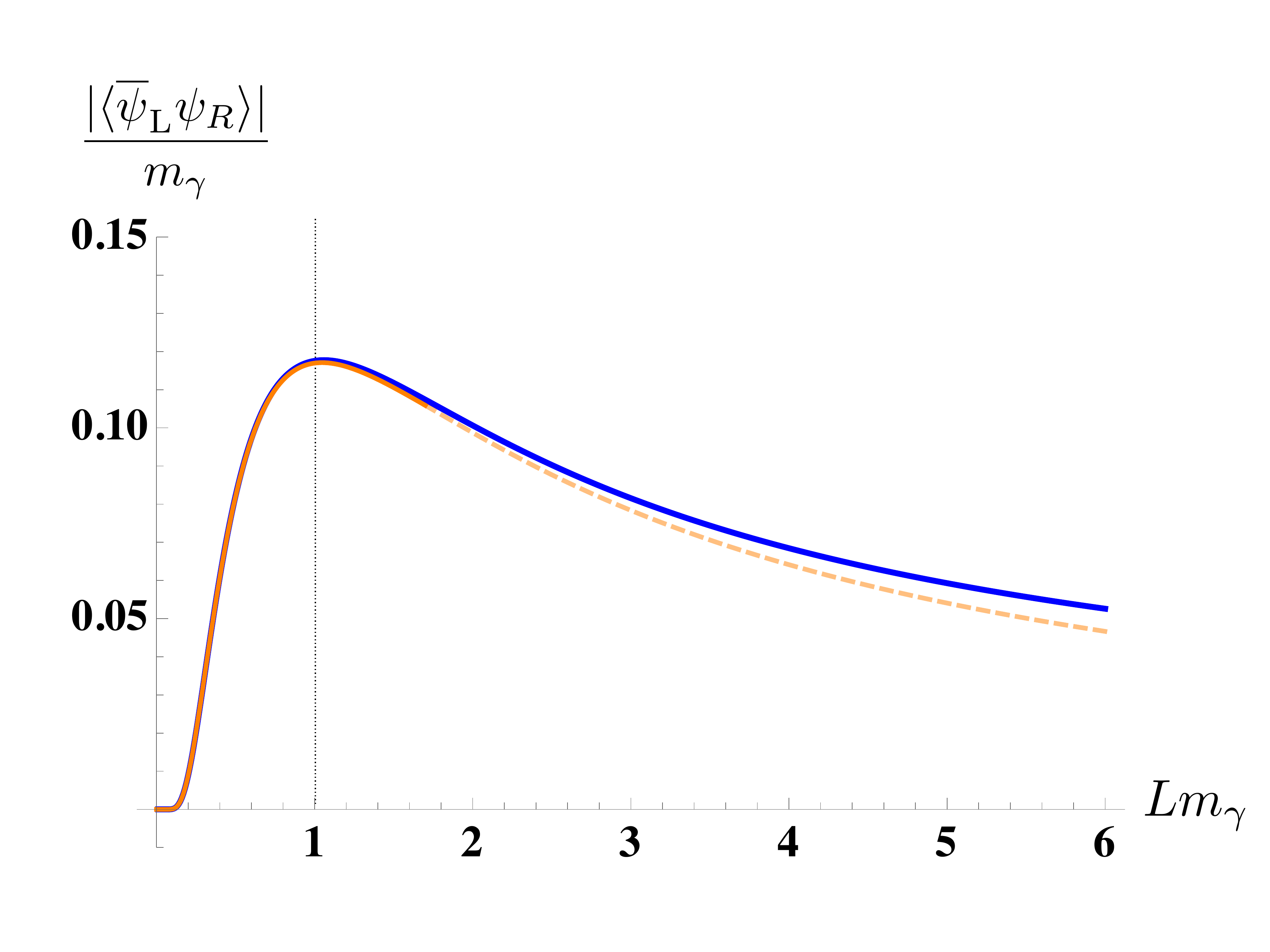}
\label{fig:q1N3_condensate}
}
\end{minipage}
\caption{
Behaviors of chiral condensate $|\langle{\psi}_\rmL\psi_\rmR\rangle|/m_\gamma$ as a function of $m_\gamma L$ for $(q,N)=(1,1)$ in the left figure (a) and for $(q,N)=(1,3)$ with the flavor-twisted boundary condition in the right figure (b).  
Blue solid curves are the exact results. The orange solid curves are the semiclassical result with $m_\gamma L<2$ and the yellow dashed curves are its extrapolation to $m_\gamma L>2$. 
}
\label{figure:chiral_condensate_exact_semiclassical}
\end{figure}

For $q\geq 1$, $N>1$ with the flavor-twisted boundary condition,
we find the chiral condensate for generic $L$,
\begin{align}
|\langle \overline{\psi}_{\rm L} \psi_{\rm R} \rangle| 
&= {1\over{NL}}\exp\left( -{\pi\over{NLm_{\gamma}}}\right)\exp\left[{1\over{N}}\left(\gamma + {\pi\over{Lm_{\gamma}}} + \log{Lm_{\gamma}\over{4\pi}}-I(Lm_{\gamma})\right)\right]
\nonumber\\
&= {1\over{NL}} \left( {Lm_{\gamma} \rme^{\gamma} \over{4\pi}} \right)^{1/N}  \exp\left[-{I(Lm_{\gamma})\over{N}}\right] \,,
\end{align}
with $m_{\gamma}^{2}=Nq^2 e^2 /\pi$.
This smoothly connects the chiral condensate $\rme^{-{\pi\over{NLm_{\gamma}}}} /(NL)$ for small $e L$ in the previous subsections to the scaling behavior $\langle \overline{\psi}_\rmL\psi_\rmR\rangle \sim L^{-(N-1)/N} \to 0$ in $L\to \infty$.
This is because the flavor-twisted boundary condition becomes irrelevant in a $L\to \infty$ limit,
where the chiral condensate vanishes for $N>1$ in Schwinger models. 
We note that the scaling exponent $(N-1)/N$ is nothing but the scaling dimension of the primary operator of $SU(N)_1$ WZW model.

In Fig.~\ref{figure:chiral_condensate_exact_semiclassical}, we compare the exact result for the chiral condensate (blue solid curves) with the result with fractional quantum-instanton for $(q,N)=(1,1)$ and $(q,N)=(1,3)$ with flavor-twisted boundary condition (orange solid and yellow dashed curves). 
We can see that when $m_\gamma L\ll 1$, those results behave in the exactly same way. 
Moreover, it is notable that, under the flavor-twisted boundary condition, the approximate expression of the chiral condensate $\rme^{-{\pi\over{NLm_{\gamma}}}} /(NL)$ which is in principle valid for small $e L$ behaves in a manner similar to the exact result even for large $eL$ as seen in Fig.~\ref{figure:chiral_condensate_exact_semiclassical}(b).


\subsection{Polyakov loop}\label{sec:Polyakov_loop}

In this section, we discuss the behavior of Polyakov loops on $\mathbb{R}\times S^1$. 
Since we neglect the higher KK modes for the gauge fields $a_2(\tau,x)=a(\tau)$, the Polyakov loop can be expressed as 
\be
P(\tau) = \exp\left(\im \int_0^L a_2(\tau,x)\diff x\right)= \exp(\im L a(\tau)). 
\ee
This is an order parameter of $\mathbb{Z}_q^{[0]}$ or $\mathbb{Z}_{Nq}^{[0]}$ (shift-)center symmetry depending on whether we take thermal or twisted boundary condition. 

\subsubsection{$q=1, N\ge1$ with thermal b.c.}

We first consider the case $q=1$ and $N\ge 1$ with periodic boundary condition. 
In this case, there is no nontrivial symmetry that acts on Polyakov loop, and thus its non-zero expectation value is naturally expected. 
This is still a good exercise to look at how we can compute the Polyakov loop, and its computation can be extended for other nontrivial cases. 

The $n$-th ground-state wave function is given in (\ref{eq:wavefunc_q1N1}) for $N=1$ and in (\ref{eq:wavefunc_q1N_pbc}) for $N>1$ with the periodic boundary condition. 
It is then easy to find that 
\bea
\langle n|P|n\rangle&=&\int \diff a\, \rme^{\im L a} \sqrt{{ NL\over \pi^2m_\gamma}}\rme^{-{ N\over \pi m_\gamma L} (La-\pi (2n+1))^2}\nonumber\\
&=&\int \diff a\, \rme^{\im L a+{\pi\im(2n+1)}} \sqrt{{NL\over \pi^2m_\gamma}}\rme^{-{N\over \pi m_\gamma L} (La)^2}\nonumber\\
&=&-\exp\left(-{\pi m_\gamma L\over 4 N}\right). 
\eea
and other matrix elements vanish, $\langle m|P|n\rangle=0$ with $m\not=n$, because of the mismatch of fermionic wave functions. 
Taking the $\theta$ vacuum, we obtain 
\be
{\langle \theta|P|\theta\rangle\over \langle \theta|\theta\rangle}=\langle n |P| n\rangle=
-\exp\left(-{\pi m_\gamma L\over 4N}\right). 
\ee
We can also check that the cluster decomposition holds as 
\be
\lim_{\tau\to \infty}{ \langle\theta|P(\tau)^\dagger P(0)|\theta\rangle\over \langle \theta|\theta\rangle}=\langle n|P^\dagger |n\rangle \langle n|P|n\rangle=\left|{\langle \theta| P|\theta\rangle\over \langle \theta|\theta\rangle }\right|^2, 
\ee
which again suggests the uniqueness of the ground state. 
This reproduces the result of \cite{Sachs:1991en}\footnote{In this section, we take the periodic boundary condition for Dirac fields instead of the anti-periodic boundary condition, and thus the Polyakov loop gets the negative sign while it is positive in Ref.~\cite{Sachs:1991en}. We again emphasize that this difference is not physical, since these boundary conditions are related by the shift of $a$ as $qa\to qa+\pi$.  } when $N=1$.

\subsubsection{$q>1, N\ge 1$ with thermal b.c.}

We next consider the case $q>1$ and $N\ge 1$ with periodic boundary condition. 
In this case, there exists $\mathbb{Z}_q^{[0]}$ symmetry that acts on Polyakov loop. 

The $n$-th ground-state wave function is given by (\ref{eq:wavefunc_qN1}) for $N=1$ and by (\ref{eq:wavefunc_qN_pbc}) for $N>1$ with periodic boundary condition. 
Therefore,
\bea
\langle n|P|n\rangle&=&\int \diff a\, \rme^{\im L a} \sqrt{{q^2 N L\over \pi^2m_\gamma}}\rme^{-{q^2 N\over \pi m_\gamma L} \left(La-{\pi (2n+1)\over q}\right)^2}\nonumber\\
&=&\int \diff a\, \rme^{\im L a+{\pi(2n+1)\over q}\im} \sqrt{{q^2 N L\over \pi^2m_\gamma}}\rme^{-{q^2 N\over \pi m_\gamma L} (La)^2}\nonumber\\
&=&\rme^{(2n+1)\pi\im/q}\exp\left(-{\pi m_\gamma L\over 4q^2 N}\right), 
\eea
and $\langle m|P|n\rangle =0$ for $m\not=n$. 
Unlike the chiral condensate, the Polyakov loop is not diagonalized by the theta vacua $|\theta,k\rangle$ given in (\ref{eq:theta_vac_qN1}).
Indeed, 
\bea
{\langle \theta, k'|P|\theta, k\rangle\over \langle \theta,k|\theta,k\rangle}&=&{1\over \langle \theta,k|\theta,k\rangle}\sum_n \rme^{-\im {\theta+2\pi k'\over q}}\rme^{\im{\theta+2\pi k\over q}} \langle n|P|n\rangle\nonumber\\
&=&{1\over \langle \theta,k|\theta,k\rangle}\rme^{\pi\im/q}\rme^{-\pi m_\gamma L/(4q^2 N)}\sum_n \rme^{{2\pi \im\over q}(1+k-k')n}\nonumber\\
&=&\delta_{k',k+1}\rme^{\pi\im/q}\rme^{-\pi m_\gamma L/(4q^2 N)}, 
\eea
and thus the diagonal elements vanish, $\langle \theta,k| P|\theta,k \rangle=0$. 
We emphasize that the Polyakov loop correlators converge to the non-zero constant as 
\be
\lim_{\tau\to \infty}{\langle \theta,k| P(\tau)^\dagger P(0) | \theta,k\rangle\over \langle \theta,k|\theta,k\rangle}=\exp\left(-{\pi m_\gamma L\over 2q^2 N}\right). 
\ee
Therefore, each theta vacuum $|\theta,k\rangle$ satisfies the cluster decomposition for two-dimensional local correlators, such as chiral condensates, but does not for Polyakov-loop correlators. 
We shall revisit this point in Sec.~\ref{sec:discrete_anomaly_qm}. 

The non-vanishing two-point correlator of Polyakov loops suggests that the string tension of test particles with charge-$1$ vanish, and the Wilson loop operator obeys perimeter law. 
This fact is not limited to the theory on $\mathbb{R}\times S^1$ with $eL\ll 1$. 
Indeed, using Abelian bosonizations, we can show that $\mathbb{Z}_q$ one-form symmetry is spontaneously broken even if the decompactification limit on $\mathbb{R}^2$ is taken. 
We emphasize that this phenomenon should not be understood by string breaking because all dynamical particles have charge $q$. The impossibility of string breaking by pair production is exactly why the theory has $\mathbb{Z}_q$ one-form symmetry: when it is possible, such loop operators are not order parameter for higher-form symmetries.  
Instead, this screening phenomenon should be understood by vacuum polarization.  
Although it was known for a long time that fractional charged particles show screening with massless dynamical fermions while confinement occurs without them~\cite{Schwinger:1962tp, Lowenstein:1971fc, Coleman:1975pw}, its correct interpretation as vacuum polarization was first clearly given in \cite{Iso:1988zi}, to our best knowledge. 
In Ref.~\cite{Hansson:1994ep}, this is interpreted as spontaneous $\mathbb{Z}_q$ symmetry breaking. 

This seems to be a rare example that realizes spontaneous breakdown of $\mathbb{Z}_n^{[d-1]}$ symmetry in $d$ spacetime dimensions, so let us explain this point in detail. Indeed, there is a folklore that the discrete $(d-1)$ form symmetry cannot be spontaneously broken~\cite{Gaiotto:2014kfa}, and our result disagrees with it. 
The origin of this folklore would come from another folklore that symmetry cannot be broken in quantum mechanics since the ground state is unique. 
However, uniqueness of the ground state can be shown only in limited situations, and there exist counterexamples: double-well quantum mechanics with infinite barrier, free particle on a circle with $\theta=\pi$, etc. 
One of the standard proof of the uniqueness is to apply Perron-Frobenius theorem to the imaginary-time Feymnan kernel in certain basis, and the sufficient condition for this to be true is that the classical potential is non-singular and that the path integral has no sign problem. 
In our situation, dynamical massless fermions disconnect topologically distinct sectors of the gauge fields, which breaks strong ergodicity, and this is indeed the origin of 't~Hooft anomaly to have multiple ground states in massless Schwinger model with discrete anomaly. 

We, still, would like to emphasize that, unlike the case $d\ge 2+1$, the spontaneous breakdown of one-form symmetry does not lead topological order in two dimensions. 
This is because it has a mixed anomaly with the $0$-form discrete chiral symmetry, which is also spontaneously broken, and the anomaly is saturated by having multiple vacua connected by discrete chiral transformation. For more details, see Sec.~\ref{sec:discrete_anomaly_qm}.

\subsubsection{$q\ge 1, N>1$ with $\mathbb{Z}_N$ twisted b.c.}

We next consider the case $q\ge 1$ and $N> 1$ with flavor twisted boundary condition. 
In this case, there exists $\mathbb{Z}_{qN}^{[0]}$ symmetry that acts on Polyakov loop. 

The $n$-th ground-state wave function is given by (\ref{eq:wavefunc_qN_tbc}). 
Therefore,
\bea
\langle n|P|n\rangle&=&\int \diff a\, \rme^{\im L a} \sqrt{{q^2 N L\over \pi^2m_\gamma}}\rme^{-{q^2 N\over \pi m_\gamma L} \left(La-{\pi (2n+1)\over qN}\right)^2}\nonumber\\
&=&\int \diff a\, \rme^{\im L a+{\pi(2n+1)\over qN}\im} \sqrt{{q^2 N L\over \pi^2m_\gamma}}\rme^{-{q^2 N\over \pi m_\gamma L} (La)^2}\nonumber\\
&=&\rme^{(2n+1)\pi\im/(qN)}\exp\left(-{\pi m_\gamma L\over 4q^2 N}\right), 
\eea
and $\langle m|P|n\rangle =0$ for $m\not=n$. 
Unlike the chiral condensate, the Polyakov loop is not diagonalized by the theta vacua $|\theta,k\rangle$ given in (\ref{eq:theta_vac_qN_tbc}).
Indeed, 
\bea
{\langle \theta, k'|P|\theta, k\rangle\over \langle \theta,k|\theta,k\rangle}&=&{1\over \langle \theta,k|\theta,k\rangle}\sum_n \rme^{-\im {\theta+2\pi k'\over qN}}\rme^{\im{\theta+2\pi k\over qN}} \langle n|P|n\rangle\nonumber\\
&=&{1\over \langle \theta,k|\theta,k\rangle}\rme^{\pi\im/(qN)}\rme^{-\pi m_\gamma L/(4q^2 N)}\sum_n \rme^{{2\pi \im\over qN}(1+k-k')n}\nonumber\\
&=&\delta_{k',k+1}\rme^{\pi\im/(qN)}\rme^{-\pi m_\gamma L/(4q^2 N)}, 
\eea
and thus the diagonal elements vanish, $\langle \theta,k| P|\theta,k \rangle=0$. 
We emphasize that the Polyakov loop correlators converge to the non-zero constant as 
\be
\lim_{\tau\to \infty}{\langle \theta,k| P(\tau)^\dagger P(0) | \theta,k\rangle\over \langle \theta,k|\theta,k\rangle}=\exp\left(-{\pi m_\gamma L\over 2q^2 N}\right). 
\ee
Again, each theta vacuum $|\theta,k\rangle$ satisfies the cluster decomposition for two-dimensional local correlators, such as chiral condensates, but does not for Polyakov-loop correlators. 

\subsection{Discrete anomaly matching}\label{sec:discrete_anomaly_qm}

Let us discuss how the discrete anomaly (\ref{eq:discrete_anomaly_pbc}) or (\ref{eq:discrete_anomaly_tbc}) is matched in the explicit construction of ground states. 

Let us denote $S$ and $C$ as the generators of $\mathbb{Z}_{Nq}^{[0]}$ and of $(\mathbb{Z}_{Nq})_\rmR$, respectively, and then the anomaly (\ref{eq:discrete_anomaly_tbc}) indicates that $\mathbb{Z}_{Nq}\times \mathbb{Z}_{Nq}$ symmetry is projectively realized:  $S^{Nq}=C^{Nq}=1$ and 
\be
SC = \rme^{-2\pi \im/Nq} CS. 
\label{eq:anomaly_algebra}
\ee
By definition, the $\theta$ vacuum $|\theta, k\rangle$ is the eigenstate of $\mathbb{Z}_{Nq}^{[0]}$, which satisfies
\be
S |\theta, k\rangle =\rme^{-2\pi \im k /Nq} |\theta, k\rangle. 
\ee
Since $k$ labels the phase of the chiral condensate, the discrete chiral transformation acts as 
\be
C|\theta, k\rangle =|\theta, k+1\rangle. 
\ee

Therefore, comparison between 
\be
SC|\theta, k\rangle = S|\theta, k+1\rangle = \rme^{-2\pi \im (k+1)/Nq} |\theta, k+1\rangle, 
\ee
and 
\be
CS|\theta, k\rangle = \rme^{-2\pi \im k/Nq} C|\theta, k\rangle = \rme^{-2\pi\im k/Nq} |\theta,k+1\rangle, 
\ee
shows that (\ref{eq:anomaly_algebra}) holds. 

The existence of projective phase forbids the simultaneous eigenstate under the center and discrete chiral symmetries. 
Since $|\theta,k\rangle$ is constructed as an eigenstate of the center symmetry, the Polyakov loop does not have the diagonal expectation value. Indeed,
\be
\langle \theta,k| P |\theta,k\rangle = \langle \theta,k| S^\dagger P S |\theta,k\rangle =\rme^{2\pi \im/Nq}\langle \theta,k| P |\theta,k\rangle,
\ee 
and this gives 
\be
\langle \theta,k|P|\theta,k\rangle =0. 
\ee
We can extend this discussion to see that the only possible nonzero amplitude is given by $\langle \theta,k+1|P|\theta,k\rangle$. 
In this sense, the Polyakov loop $P$ behaves in the same manner as $C$ up to an overall normalization. 
We can do the same argument about the chiral condensate $\overline{\psi}_\rmL \psi_\rmR$, and find that the only possible nonzero amplitude is given by $\langle \theta,k |\overline{\psi}_\rmL \psi_\rmR|\theta,k\rangle\sim \rme^{2\pi \im k/Nq}$. 
Again, up to an overall normalization, the chiral condensate behaves in the similar manner as $S$, and these relations are nothing but the consequence of mixed 't~Hooft anomaly.

\section{Quantum instanton on $\mathbb R \times S^1$ and chiral condensate}
\label{sec:fractional_quantum_instanton}

We usually call instantons as bosonic classical solutions with nontrivial topological charge. 
For well-definedness, we put the theory on the two-torus, $M_2=T^2=S^1_\beta\times S^1_L$, and let us discuss instantons of Schwinger model. 
For a given topological charge, 
\begin{align}
Q= \frac{1}{2\pi} \int \diff a  \in \mathbb Z, 
\end{align}
the $U(1)$ gauge field $a$ can be decomposed as~\cite{Sachs:1991en}
\be
a={2\pi Q\over \beta L}\tau\diff x+{2\pi\over \beta}h_1\diff \tau+{2\pi\over L}h_2\diff x+\star \diff \phi+\diff \lambda, 
\ee
where $h_{1,2}\in [0,1)$ denote constant holonomies, $\phi$ is the $\mathbb{R}$-valued scalar field without zero-mode, and $\lambda$ is the gauge parameter. 
Using this expression, the Maxwell action is bounded from below as 
\begin{align}
S = {1\over 2e^2}\int_{T^2} \diff^2 x \left((\Delta \phi)^2+\left(2\pi Q\over \beta L\right)^2\right) \ge \frac{2 \pi^2} {e^2 \beta L}Q^2.   
\label{action}
\end{align}
Since $e^2$ has mass dimension two, $\beta L$ in the denominator of  \eqref{action} is necessary to make the action dimensionless.  In the $\mathbb R^2$ limit, the instanton action actually vanishes.

In this section, we give a path-integral interpretation of the results in Sec.~\ref{sec:ChSSB}. 
There, we will see that ``quantum'' instanton plays an essential role to describe the spontaneous discrete chiral symmetry breaking in a semiclassical manner~\cite{Smilga:1993sn, Shifman:1994ce}. 
 
 \subsection{Fractional quantum instanton for thermal boundary condition} 

Let us take the periodic boundary condition for fermions. 
In (\ref{eq:det_chiral_condensate_pbc}), we obtain that $\det(\overline{\psi}_\rmL^f \psi_\rmR^{f'})$ condenses as the following matrix element is nonzero, 
\be
\langle n-1| \det\left(\overline{\psi}_\rmL^f \psi_\rmR^{f'}\right) |n\rangle \not = 0. 
\label{eq:det_cond}
\ee 
It breaks $\mathbb{Z}_q$ discrete chiral symmetry if $q>1$. 
Integrating out fermion fields, the holonomy potential is induced as \eqref{1-loop-thermal}, and we denote the effective potential for quantum mechanics of $a_2(\tau,x)=a(\tau)$ as 
\be
V_{\rm eff}(a)\equiv L(V(a)-\mathcal{F}_{\rm thermal})=\frac{q^2 N}{2 \pi L} \min_k \left(La + \frac{2 \pi (k+ \frac{1}{2})}{q} \right)^2. 
\ee 
The boundary condition for gauge fields $a(\tau)$ becomes 
\be
a(-\infty)={\pi(2n+1)\over qL},\quad 
a(+\infty)={\pi(2n-1)\over qL},
\ee 
by (\ref{eq:det_cond}), and this is also because the chiral condensate behaves as the generator of $\mathbb{Z}_q^{[0]}$ center symmetry at low energies due to mixed 't~Hooft anomaly as we have seen in Sec.~\ref{sec:discrete_anomaly_qm}.  
We can evaluate the lower bound of the Euclidean action by completion of the square (or, BPS trick) as follows:
\bea
S&=&\int_{-\infty}^{\infty}\diff \tau\left({L\over 2e^2}\left({\p a\over \p \tau}\right)^2+V_{\rm eff}(a)\right)\nonumber\\
&=&\int \diff \tau\left\{{\sqrt{L/2}\over e}{\p a\over \p \tau}\mp \sqrt{V_{\rm eff}(a)}\right\}^2\pm {\sqrt{2L}\over e}\int \diff \tau \sqrt{V_{\rm eff}(a)} {\p a\over \p \tau}\nonumber\\
&\ge & {\sqrt{2L}\over e}\left|\int \diff a \sqrt{V_{\rm eff}(a)}\right|. 
\eea 
The lower bound for the given boundary condition is saturated as  
  \begin{align}
  S_{\mathcal{F}}= \frac{\pi^{3/2} N^{1/2} }{  q e L} = \frac{\pi N } {m_\gamma  L},   
  \qquad {\rm where}  \;\;     m_\gamma = \frac{N^{1/2} q e}{\pi^{1/2} }. 
\end{align}
The topological charge of  this configurations is $Q=-1/q$. 
Unlike usual instanton, we balance the classical kinetic term and the quantum-induced potential to find this configuration with fractional topological charge, so let us call it fractional quantum instanton, or fracton\footnote{Note that this naming has no relation with fracton phases in condensed matter literatures. } following Ref.~\cite{Shifman:1994ce}. 
From the index theorem, we can deduce that each fracton supports $2N$ fermionic zero modes and this is consistent as we find it in the computation of the determinant condensate. 

\begin{figure}[t]\centering
\begin{minipage}{.4\textwidth}
\subfloat[$(q,N)=(1,1)$]{
\includegraphics[scale=0.2]{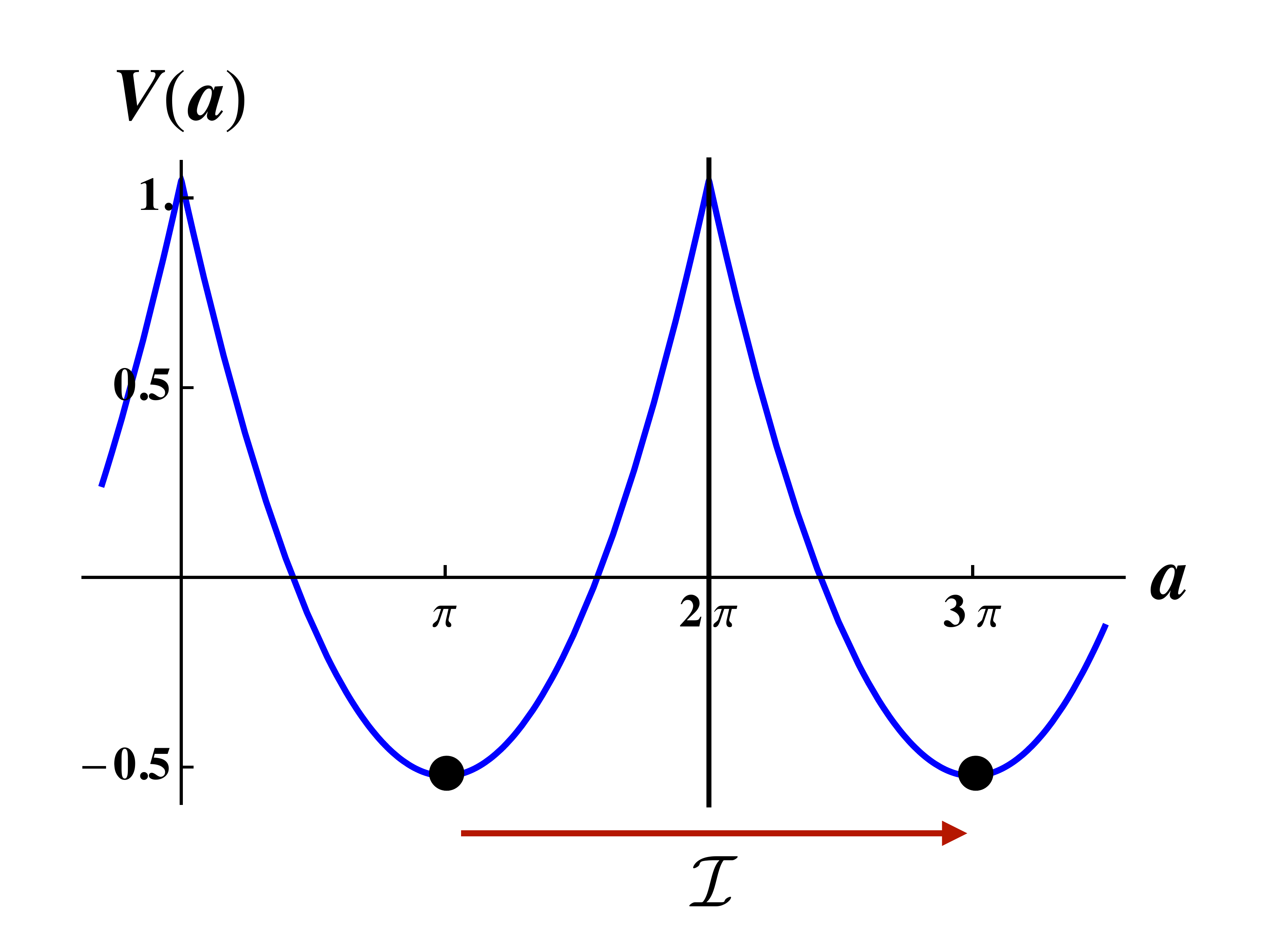}
\label{fig:q1N1_inst}
}
\end{minipage}\qquad\quad \qquad
\begin{minipage}{.4\textwidth}
\subfloat[$(q,N)=(2,1)$]{
\includegraphics[scale=0.2]{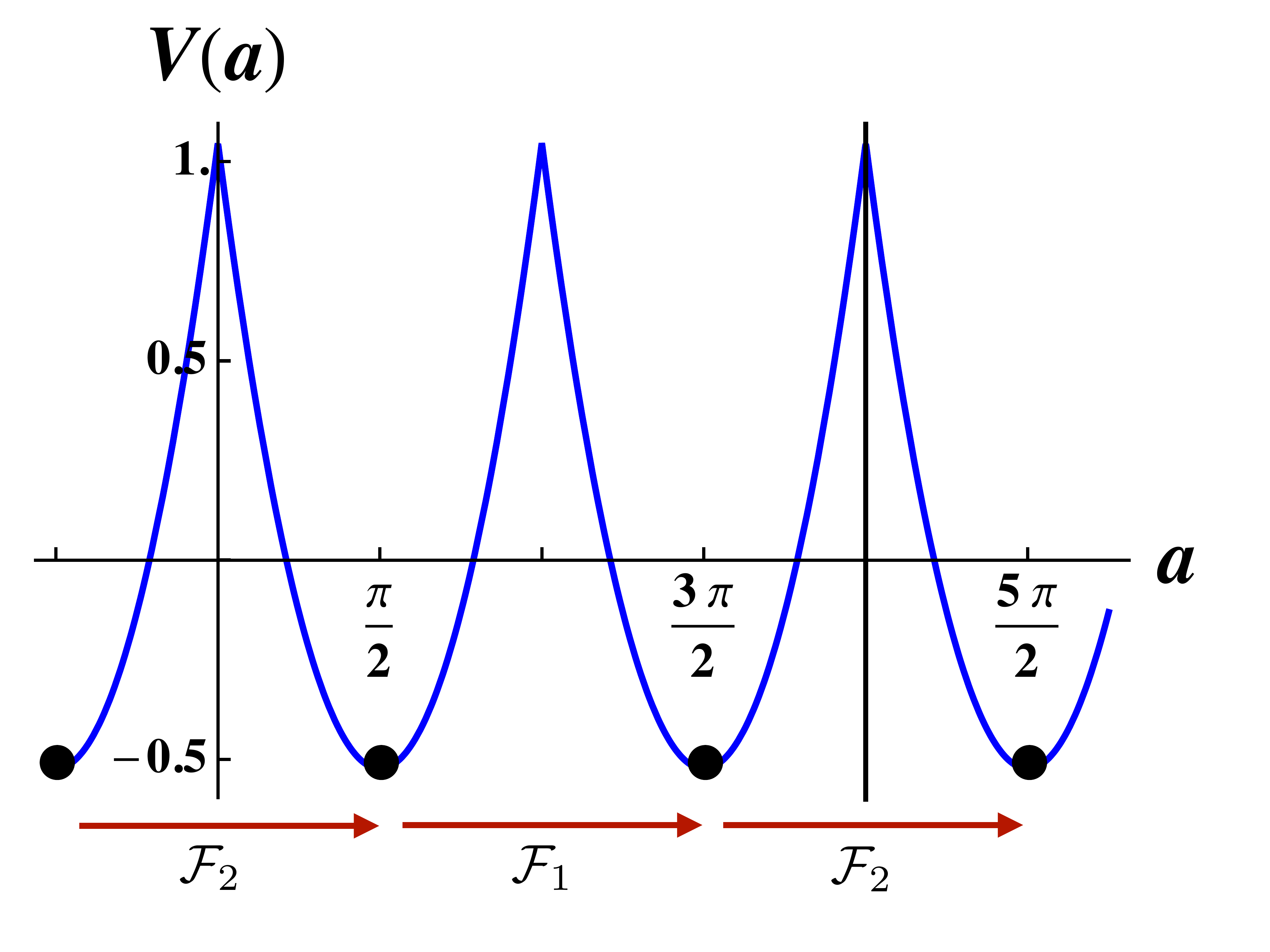}
\label{fig:q2N1_inst}
}\end{minipage}
\begin{minipage}{.4\textwidth}
\subfloat[$(q,N)=(1,3)$ with $\Omega_F$ t.b.c.]{
\includegraphics[scale=0.2]{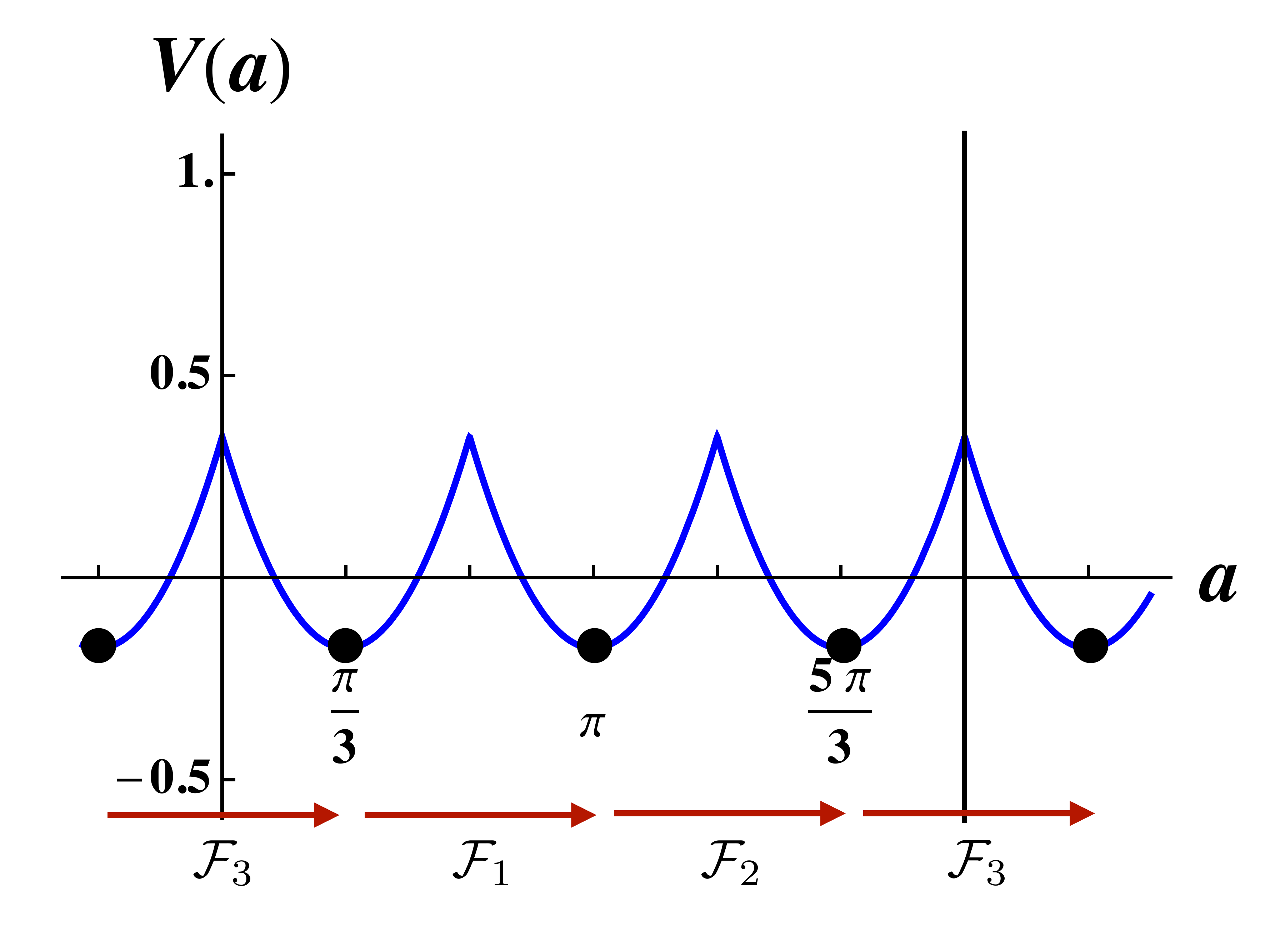}
\label{fig:q1N3_inst}
}\end{minipage}\qquad\quad \qquad
\begin{minipage}{.4\textwidth}
\subfloat[$(q,N)=(2,3)$ with $\Omega_F$ t.b.c.]{
\includegraphics[scale=0.2]{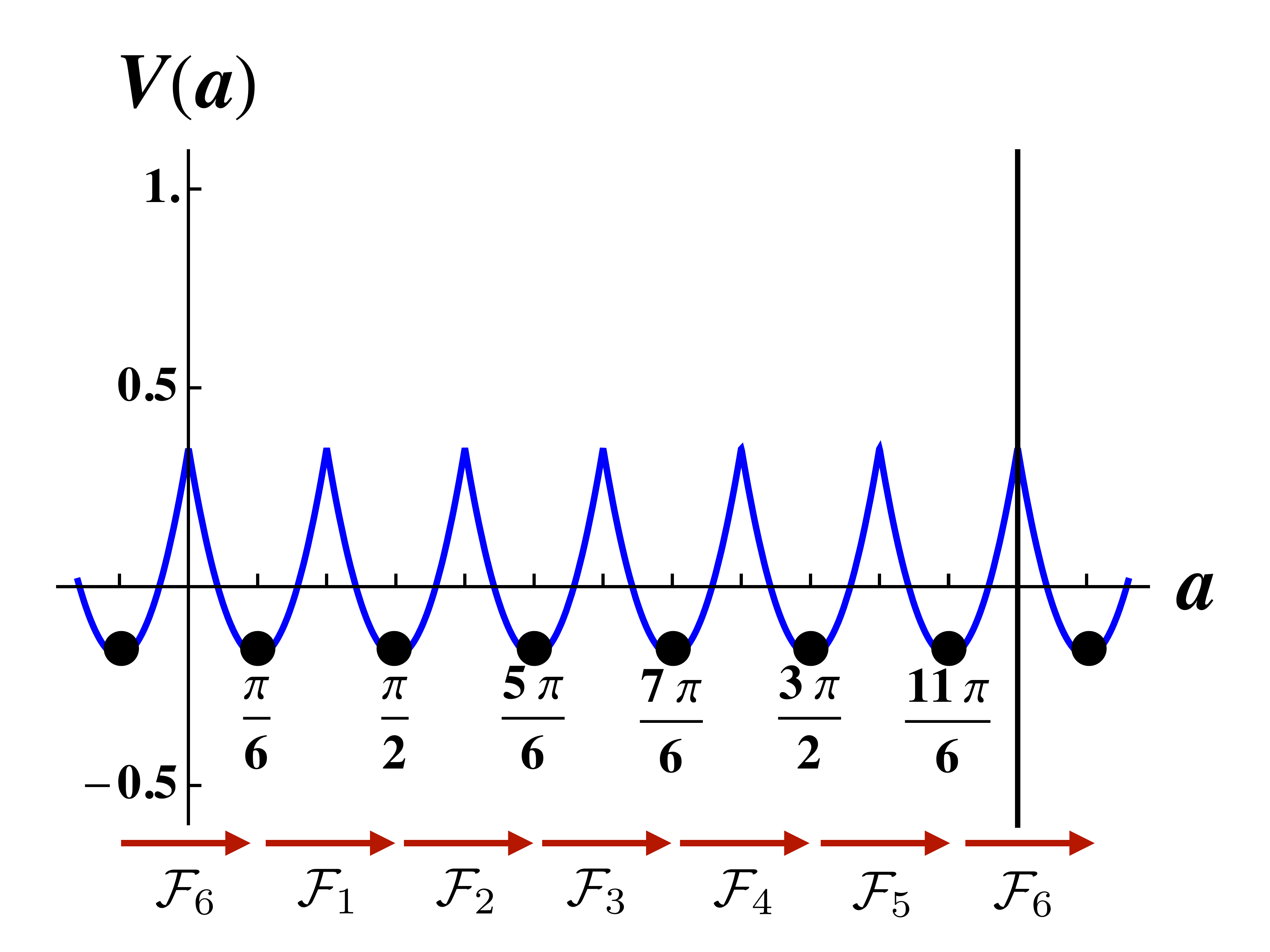}
\label{fig:q2N3_inst}
}\end{minipage} 
\caption{Quantum instantons in various setups: (a)~Quantum instanton for $(q,N)=(1,1)$, which contributes to $\langle \overline{\psi}_\rmL\psi_\rmR\rangle =L^{-1} \rme^{\im \theta}\rme^{-S_\mathcal{F}}$, as a consequence of ABJ anomaly. No symmetry breaking occurs. (b)~Fractional quantum instantons (= fractons) for $(q,N)=(2,1)$, which contributes to $\langle \overline{\psi}_\rmL\psi_\rmR\rangle=L^{-1}\rme^{\im \theta/q}\rme^{-S_{\mathcal{F}}}$. The fractional $\theta$ dependence reflects the spontaneous $\mathbb{Z}_q$ discrete chiral symmetry breaking. 
(c)(d)~Fractons for $(q,N)=(1,3)$ and $(q,N)=(2,3)$ with the flavor-twisted boundary condition $\Omega_F$, which contribute to $\langle \overline{\psi}^f_\rmL\psi^f_\rmR\rangle=(NL)^{-1}\rme^{\im \theta/Nq}\rme^{-S_{\mathcal{F}}}$. The fractional $\theta$ dependence reflects the spontaneous $\mathbb{Z}_{Nq}$ discrete chiral symmetry breaking. 
}
\label{fig:quantum_instanton}
\end{figure}

In Figs.~\ref{fig:q1N1_inst} and~\ref{fig:q2N1_inst}, we can see the quantum induced potentials and corresponding quantum instantons for $(q,N)=(1,1)$ and $(q,N)=(2,1)$, respectively. 
When $(q,N)=(1,1)$, we reproduce the result, $\langle \overline{\psi}_\rmL\psi_\rmR\rangle=L^{-1}\exp(-{\pi\over m_\gamma L}+\im \theta)$, in \cite{Sachs:1991en}, and this condensate is the consequence of ABJ anomaly. 
For charge-$2$ case in Fig.~\ref{fig:q2N1_inst}, fracton contributes to the fermion bilinear condensate, $\langle \overline{\psi}_\rmL\psi_\rmR\rangle=L^{-1}\exp(-{\pi\over m_\gamma L}+\im \theta/2)$, so that the $\theta$ dependence is fractionalized and $\mathbb{Z}_2$ discrete chiral symmetry is spontaneously broken. 

Since we balance the kinetic term of $O(1/e^2)$ and the quantum induced potential of $O(1)$, the fracton action $S_\mathcal{F}$ is of order $1/e$. This has the big difference with usual field-theoretic extended confugrations, which are typically of order $1/e^2$. 
It is notable that the fracton action $S_{\mathcal{F}}$ exactly gives the exponent of (\ref{eq:det_chiral_condensate_pbc}) and the fractional topological charge explains the $\theta$ dependence $\rme^{\im \theta/q}$. Therefore, we can directly observe the fracton contribution by measuring this condensate on $\mathbb{R}\times S^1$.

   \subsection{Fractional quantum instanton in flavor-twisted boundary condition} 
  With the insertion of a $\Omega_F$ twist,  we now have the quark-bilinear condensate (\ref{eq:bilinear_condensate_qN_twisted}) sourced by the matrix element
\be
\langle N\ell+f-1 | \overline{\psi}^f_\rmL \psi^f_\rmR | N \ell+f\rangle \not =0. 
\ee
Let us first integrate our fermions again when evaluating this amplitude on $\mathbb{R}\times S^1$, then the holonomy potential takes the form \eqref{1-loop-tbc-2}. We denote the effective potential for quantum mechanics of $a_2(\tau,x)=a(\tau)$ as 
\be
V_{\mathrm{eff},\Omega_F}(a)\equiv L\left(V_{\Omega_F}(a)-\mathcal{F}_{\Omega_F}\right)
=\frac{q^2 N}{2 \pi L} \min_k \left(La + \frac{2 \pi (k + \frac{1}{2}) }{qN}\right)^2, 
\ee 
and the boundary condition is 
\be
a(-\infty)={\pi (2(N\ell+f)+1)\over Nq},\quad 
a(+\infty)={\pi (2(N\ell+f)-1)\over Nq}. 
\ee
We now evaluate the lower bound of the Euclidean effective action with $V_{\mathrm{eff},\Omega_F}$ with the given boundary condition. 
The BPS trick gives 
\bea
S&=&\int_{-\infty}^{\infty}\diff \tau\left({L\over 2e^2}\left({\p a\over \p \tau}\right)^2+V_{\mathrm{eff},\Omega_F}(a)\right)\nonumber\\
&\ge & {\sqrt{2L}\over e}\left|\int \diff a \sqrt{V_{\mathrm{eff},\Omega_F}(a)}\right|. 
\eea 
Evaluating this lower bound for the given boundary condition, we obtain the fracton action,  
  \begin{align}
  S_{\mathcal{F}}=\frac{\pi^{3/2}}{ N^{3/2} q e L} = \frac{\pi} {N m_\gamma  L}, 
\end{align}
and this is nothing but the exponent of the chiral condensate (\ref{eq:bilinear_condensate_qN_twisted}). 
The topological charge of  this configurations is $Q=-1/Nq$, and this gives the fractional $\theta$ dependence, $\rme^{\im \theta/Nq}$. 
From the index theorem, we can deduce that each fracton supports $2$ fermionic zero modes and this is consistent as we find it in the computation of the fermion-bilinear condensate. 

The behavior of fractons can be seen in Figs.~\ref{fig:q1N3_inst} and~\ref{fig:q2N3_inst} for $(q,N)=(1,3)$ and $(q,N)=(2,3)$ with symmetry twisted boundary condition, respectively. 
When we take the twisted boundary condition, we can clearly see in the figure that the height of effective potential becomes shallow of $O(1/N)$, while the local fluctuation of $a$ feels the harmonic potential of $O(N)$. Let us compare the thermal and twisted boundary conditions in the table:
 \begin{center}
\begin{tabular}{ | l | l | l | l |  } 
 \hline
 & Barrier height  & Width $\Delta a$  & Fracton action \\ 
 \hline 
 Thermal  & $O(N)$ & $ 2\pi/q$  & $S_{\mathcal{F}} =  \frac{\pi N } {m_\gamma  L} $  \\  
 \hline
 $\Omega_F$ twisted  & $O(N^{-1})$& $ 2\pi/qN$   & $S_{\mathcal{F}} =  \frac{\pi  } { N m_\gamma  L} $   \\ 
 \hline
\end{tabular}
\end{center}
Differences of the potential barrier by $1/N^2$ and of the potential width by $1/N$ give the factor $\sqrt{1/N^2}\times 1/N=1/N^2$ in the fracton action in the twisted boundary condition compared with that of the thermal one.

This $N$ dependence of $S_\mathcal{F}$ affects the region of validity of our approximation for chiral condensate.  Note that the region of validity of the one-loop holonomy potential and the one of the semi-classical instanton analysis are parametrically different. 
The one-loop analysis  is reliable provided $L m_{\gamma}  \ll 1$. However, the semi-classical quantum-instanton analysis is reliable provided the quantum-instanton amplitude is small, $e^{-S_{\mathcal{F}}} \ll 1$.  
In the $N$-flavor   $\Omega_F$ twisted  case,  the smallness of this amplitude is valid provided ($ N L m_{\gamma}  \ll 1$), and then the region of validity $L \lesssim  \frac{1}{m_\gamma N}    \rightarrow 0 $ in the $N \rightarrow \infty$ limit. This is an  imprint of large-$N$ volume independence in the $N$ flavor twisted Schwinger model. In the thermal model, semi-classical approximation is always valid within the domain of validity of one-loop analysis.


\section{Effect of fermion mass and spontaneous $\mathsf{C}$ breaking at $\theta=\pi$}
\label{sec:massive_fermion}

In this section, we introduce the flavor-degenerate soft mass $m_\psi$ to Dirac fermions and discuss its physical effects by using symmetry, anomaly and global inconsistency, mass perturbation on $\mathbb{R}\times S^1$, and dilute fractional-quantum-instanton gas approximation also on $\mathbb{R}\times S^1$. 
Unlike massless case, the ground-state energy is affected by the $\theta$ angle, and the ground state is unique for generic $\theta$ angles. At $\theta=\pi$, we shall see that the charge-conjugation symmetry $\mathsf{C}$ is spontaneously broken when $qN \ge 3$ or $q\ge 2$. 
For $q=1$ and $N=2$, we shall see that $\mathsf{C}$ is spontaneous broken on $\mathbb{R}\times S^1$, but it is believed to go back to $SU(2)_1$ WZW model on $\mathbb{R}^2$ at $\theta=\pi$, which is related to the Haldane conjecture. 

\subsection{Anomaly and global inconsistency for massive Schwinger model}

Before the concrete analysis of massive Schwinger model, let us discuss the kinematical constraint by symmetry. 
We add the fermion mass term, 
\be
m_\psi \sum_f\int \diff^2 x (\overline{\psi}^f_\rmL\psi^f_\rmR+\overline{\psi}^f_\rmR\psi^f_\rmL), 
\ee
to the Lagrangian (\ref{eq:Lagrangian_Schwinger}), and we assume $m_\psi>0$ so that the $\theta$ angle has the definite meaning. 
This fermion mass term explicitly breaks the chiral symmetry $G^{[0]}$ in (\ref{eq:0form_symmetry}) to its vector-like subgroup completely, and we get 
\be
\mathbb{Z}_q^{[1]}\times {SU(N)_\rmV\over \mathbb{Z}_{N}}\subset G. 
\ee
Since this symmetry is vector like, this subgroup has no anomaly, and we do not have any interesting constraint on the vacuum structure so far.

Let us now consider the charge conjugation symmetry $\mathsf{C}$. This $\mathbb{Z}_2$ internal symmetry is generated by 
\be
\mathsf{C}: \psi \mapsto \im\gamma^2\overline{\psi}^T, \; \overline{\psi}\mapsto \psi^T \im\gamma^2,\; a\mapsto -a. 
\ee
In the chiral notation, $\mathsf{C}$ acts as 
\bea
&&\psi_\rmR\mapsto \overline{\psi}_\rmR,\; \psi_\rmL\mapsto -\overline{\psi}_\rmL, \nonumber\\
&&\overline{\psi}_\rmR\mapsto \psi_\rmR,\; \overline{\psi}_\rmL\mapsto -\psi_\rmL. 
\eea
Examples of $\mathsf{C}$-odd observables are $\diff a$, $\overline{\psi}_\rmR\psi_\rmL-\overline{\psi}_\rmL\psi_\rmR$. 
Especially, the $\theta$ term flip its sign under $\mathsf{C}$, and thus this is symmetry only when $\theta=0$ or $\theta=\pi$. 

The group structure including $\mathsf{C}$~\cite{Ohmori:2018qza} is given by the semidirect product,
\be
\left(\mathbb{Z}_q^{[1]}\times {SU(N)_\rmV\over \mathbb{Z}_{N}}\right)\rtimes (\mathbb{Z}_2)_{\mathsf{C}},
\ee
if $q\ge 3$ or $N\ge 3$, and otherwise it is given by the direct product,
\be
\left(\mathbb{Z}_q^{[1]}\times {SU(N)_\rmV\over \mathbb{Z}_{N}}\right)\times (\mathbb{Z}_2)_{\mathsf{C}}. 
\ee
Especially when $q=1$ and $N=2$, the group structure $SO(3)\times \mathbb{Z}_2=O(3)$ is the same with that of discrete chiral transformation. 

Let us discuss the mixed anomaly, and to find it we gauge $\mathbb{Z}_q^{[1]}\times PSU(N)_\rmV$ first, and we perform $\mathsf{C}$ after that~\cite{Komargodski:2017dmc}. 
The background gauge field consists of 
\begin{itemize}
\item $A_\rmV$: $SU(N)$ one-form gauge field, and 
\item $B_\rmV$: $\mathbb{Z}_{Nq}$ two-form gauge field. 
\end{itemize}
In the description of Sec.~\ref{sec:background_gauge_field}, we have to set $A_\rmL=A_\rmR=A_\rmV$, and set other gauge fields to be zero other than $B_\rmV$. 
All the terms in the gauged Lagrangian is manifestly gauge-invariant and $\mathsf{C}$-invariant, except for the $\theta$ term,
\be
{\im \theta\over 2\pi}\int (\diff a+B^{(2)}_\rmV). 
\ee
This shows that the gauged partition functions at $\theta=0$ and $\theta=\pi$ behave under $\mathsf{C}$ as 
\bea
Z_{M_2,\theta=0}[A_\rmV,B_\rmV]&\mapsto& Z_{M_2,\theta=0}[A_\rmV,B_\rmV],\\
Z_{M_2,\theta=\pi}[A_\rmV,B_\rmV]&\mapsto& Z_{M_2,\theta=\pi}[A_\rmV,B_\rmV]\exp\left(\im \int B_\rmV^{(2)}\right). 
\eea
Thus, the partition function at $\theta=\pi$ gets the $\mathbb{Z}_{Nq}$ phase under $\mathsf{C}$. We have to judge whether this is genuine anomaly or not by studying possible local counter terms. 

The only possible local counterterm is $\im k \int B_\rmV$ with $k=0,1,\ldots, Nq-1$. We then find
\be
Z_{M_2,\theta=\pi}[A_\rmV,B_\rmV]\rme^{\im k\int B^{(2)}_\rmV}\mapsto \left(Z_{M_2,\theta=\pi}[A_\rmV,B_\rmV]\rme^{\im k\int B^{(2)}_\rmV}\right)\exp\left(\im (1-2k)\int B_\rmV^{(2)}\right), 
\ee
and thus $\mathsf{C}$-invariance is established if and only if $2k-1=0$ mod $Nq$ for some $k$. 
If $Nq$ is even, no such counter term exists, and thus $\theta=\pi$ has a mixed 't~Hooft anomaly. 

When $Nq$ is odd, we can solve $2k=1$ mod $Nq$ as 
\be
k={Nq+1\over 2}, 
\ee
and thus there is no anomaly at $\theta=\pi$. However, since the coefficient of the local counterterm is quantized, we cannot take the simultaneous UV regularization so that the gauged partition functions at $\theta=0$ and $\theta=\pi$ are both $\mathsf{C}$ invariant. 
This is called global inconsistency condition~\cite{Gaiotto:2017yup, Tanizaki:2017bam, Kikuchi:2017pcp}. 
The matching condition for global inconsistency~\cite{Kikuchi:2017pcp} is 
\begin{itemize}
\item The vacua at $\theta=0$ and $\theta=\pi$ are different as symmetry-protected topological states, or
\item either of $\theta=0$ or $\theta=\pi$ has nontrivial vacuum as in the case of 't~Hooft anomaly. 
\end{itemize}

It is easy to see that the anomaly and global inconsistency is matched by SSB of $\mathsf{C}$ at $\theta=\pi$ in the massive Schwinger model with $qN\ge 2$ on $\mathbb{R}\times S^1$. 
Here, let us restrict our attention to the case with the $\Omega_F$-twisted boundary condition, so that the theory has $\mathbb{Z}_{Nq}^{[0]}$ symmetry.  
We have constructed the $\theta$ vacua, $|\theta,k\rangle$ with $k=0,1,\ldots, Nq-1$, in the massless limit in Sec.~\ref{sec:chiral_condensate_qN_tbc}, and the chiral condensates are given in (\ref{eq:bilinear_condensate_qN_twisted}). 
For sufficiently small mass, then, we find the ground-state energy $E_k(\theta)$ of $|\theta,k\rangle$ by mass perturbation as 
\bea
E_k(\theta)&=&-L m_\psi \sum_f{\langle \theta,k|(\overline{\psi}^f_\rmL\psi^f_\rmR+\overline{\psi}^f_\rmR\psi^f_\rmL)|k,\theta\rangle\over \langle \theta,k|\theta,k\rangle}\nonumber\\
&=&-2m_\psi \exp\left(-{\pi\over Nm_\gamma L}\right) \cos \left({\theta+2\pi k\over Nq}\right). 
\label{eq:multi_branch_energy}
\eea
We thus find that the ground state is unique for generic $\theta$, but $\theta=\pi$ is doubly degenerate if $Nq\ge 2$. 

Although our explicit computation of the ground-state energy is limited to the semiclassical regime $e L\ll 1$, the constraint via anomaly and global inconsistency is valid even when the decompactification limit $L\to \infty$ is taken. 
The ground-state energy on $\mathbb{R}^2$ with small fermion mass $m_\psi$ takes the complicated form as shown in Refs.~\cite{Coleman:1976uz, Hetrick:1995wq,Hetrick:1995yx}, but those results are indeed consistent with anomaly and global inconsistency. 

For $-\pi<\theta<\pi$, the ground state is uniquely determined to $k=0$, $|\theta,0\rangle$. As we have studied in Sec.~\ref{sec:Polyakov_loop}, the Polyakov loop vanishes for this vacuum,
\be
\langle \theta,0| P |\theta, 0\rangle=0. 
\ee
Unlike massless case, this is the unique vacuum, and thus $\mathbb{Z}_{Nq}^{[0]}$ symmetry is unbroken. 
This is consistent with the fact that the string tension does not vanish without massless fermions. 
If we study the two-point function of $n$-th Polyakov loop $P^n$, it roughly behaves in $\tau\to\infty$ as 
\bea
&&\langle\theta,0| P^{-n}(\tau) P^n(0)|\theta, 0\rangle\nonumber\\
&\sim& \exp\left[-\tau (E_n(\theta)-E_0(\theta))\right]\nonumber\\
&=& \exp\left[-2m_\psi \tau \,\rme^{-\pi/Nm_\gamma L}\left\{\cos\left({\theta+2\pi n\over Nq}\right)-\cos\left({\theta\over Nq}\right)\right\}\right], 
\eea
and it goes to zero unless $n=0$ mod $Nq$. 
The exponent, $E_n(\theta)-E_0(\theta)$, corresponds to the string tension of the charge-$n$ test particle. 
It would be worth to mention that $\theta=\pi$ is again exceptional in this regard, since the exponent vanishes at $\theta=\pi$ with $n=-1$ mod $Nq$. 
The corresponding statement is true for Wilson loops in the $2$d decompactification limit~\cite{Armoni:2018bga}.

\subsection{Dilute fractional-quantum-instanton gas  and multi-branched vacua}

Let us reproduce the above result as a semiclassical approximation of the path integral. 
We first discuss the holonomy potential. 
 Introducing mass term  modifies  $e^{-L |p| }  $ in \eqref{1-loop-thermal}  and \eqref{1-loop-tbc} into  $e^{-L \sqrt{p^2+m_\psi^2} } $. 
 As a result, the  thermal holonomy potential takes the form,
  \begin{align} 
 V_{\rm thermal}(a, m_\psi)    & =  \frac{ 2 N}{ \pi  L^2}  \sum_{n=1}^{\infty}  \frac{1}{n^2} 
   (m_\psi L n)K_1( m_\psi L n)   \cos(q a n) ,
\label{1-loop-thermal-massive}
 \end{align}  
and  similarly, in the  $\Omega_F$ background, the potential takes the form:  
  \begin{align} 
 V_{\rm \Omega_F}(a, m_\psi)    
 & =  \frac{2}{ \pi  L^2 } \frac{1}{N} \sum_{n'=1}^{\infty}  \frac{1}{n'^2}     (m_\psi L N n')K_1( m_\psi L N n')      \cos { (N q  a n')} . 
 \label{1-loop-tbc-2-massive}
 \end{align} 
 Here,  we denote $La$ simply by $a$, and  $ K_1( z)$  is the modified Bessel function of the first kind, whose asymptotic behaviors are
 \begin{align} 
 K_1(z) \sim \left\{  \begin{array}{ll} 
\frac{1}{z}+\frac{1}{4} z (2 \ln (z)+2 \gamma -1-2 \ln (2))+O\left(z^2\right),   &  \qquad z\rightarrow 0  \cr \cr
\sqrt{ \frac{\pi}{2z}} e^{-z} (1+ O(1/z)),  &  \qquad z\rightarrow \infty. 
 \end{array} \right.
 \end{align}

For concreteness, we consider the massive Schwinger model with $\Omega_F$ background in the following.
 The presence of mass smoothens the cusp that is present in the potential with massless fermions, but otherwise, the vacuum structures are extremely similar at this level: We have $Nq$ minima of the potential, which we denote as $a_j$ with $j=0,1,\ldots,Nq-1$. 
Writing $j=Nm+f$ with $m=0,\ldots, q-1$ and $f=0,\ldots, N-1$,  the perturbative ground state around $a_j$ corresponds to
\be
|a_j\rangle =\sum_{\ell\in \mathbb{Z}} \rme^{\im \theta \ell} |N (q\ell+m) +f\rangle,
\ee
in the notation of Sec.~\ref{sec:chiral_condensate_qN_tbc}. 
 
 The crucial difference between massless and massive cases is that, in the latter, the vacuum degeneracy will be lifted by quantum tunneling. In the massless case, this is prohibited since $|a_j\rangle$ with different $j$ have different charged under $\mathbb{Z}_{Nq}$ discrete chiral symmetry.  
In other words, the would-be tunneling process is fractional quantum instanton, which has $\Delta Q_5=2$, and it does not contribute to the partition function. 
Adding fermion mass, this symmetry is explicitly broken so that quantum tunneling process occurs. 

 The theory on small circle, $eL\ll 1$,  reduces to a simple quantum mechanical problem for the $a$ field.   This is the quantum mechanics of a particle moving  on a circle $S^1$ in the presence of a potential with $Nq$ minima 
 given in \eqref{1-loop-tbc-2-massive}  in its fundamental domain $a \in [0, 2\pi)$.  
There is no harm in simplifying the potential into 
\begin{align}
V(a) \sim   \cos(N q a) . 
\label{pot}
\end{align}
and there exists an Aharonov-Bohm flux passing through the center of the circle,    which captures  the  topological $\theta$ angle.

The partition function, $Z(\beta)=\tr[\exp(-\beta H)]$,  in the $\beta \rightarrow \infty$ limit  is dominated by the lowest-$Nq$ states. To all orders in perturbation theory, the harmonic states at the minima of the potential \eqref{pot}  are exactly degenerate. 
 To  simplify the discussion, we can forget about higher states in the spectrum  within Born-Oppenheimer approximations, 
 because instanton induced splittings  $\sim  \omega e^{-S_\mathcal{F}}$ is much smaller than perturbative gap in the spectrum $(\sim \omega)$. 
The path-integral representation of the partition function is 
\begin{align} 
 Z(\beta) & = \int_{a(\beta)= a(0)} \Diff a \exp(-S[a]),  
\end{align}
and this is a sum over paths obeying the periodic boundary condition, $a(\beta)= a(0)$. 
The maps $S^1_\beta \rightarrow S^1$ are classified by the winding numbers, $\pi_1(S^1)=\mathbb{Z}$, which is nothing but the topological charge $Q= \frac{1}{2\pi} \int \diff a \in {\mathbb Z}$ in $2$d language.

The essence  of the matter is following:  The partition function is a sum over integer topological charge configurations $Q \in \mathbb Z$,  yet, in the problem, there are classical solutions with fractional topological charges quantized in  $ 1/Nq$, which we call fractional quantum instanton with the action $S_\mathcal{F}$. 
A single fractional instanton does not contribute to the partition function because of the mismatch of boundary condition, and this introduces a constraint when considering the dilute gas approximation as a semiclassical approximation of $Z(\beta)$. 
 In the dilute fractional-instanton gas approximation of $Z(\beta)$, we must sum over all configurations with $n$ fracton, $ {\bar n }$ anti-fracton satisfying the constraint 
 $  n-\bar n  -W Nq=0$, where $W\in \Z $ is the winding number.   This guarantees that the gauge field is a connection of $U(1)$ bundle.

 In the dilute gas approximation, the partition function can therefore be written as 
\begin{align} 
 Z(\beta) 
&= N q \sum_{W \in \Z}  \sum_{n=0}^{\infty}   \sum_{\bar n =0}^{\infty}   \frac{1}{n!} \frac{1}{\bar n !}  \left( \beta K  \rme^{-S_{\mathcal{F}} + \im \theta/Nq}  \right)^{n}  \left( \beta K  \rme^{-S_{\mathcal{F}} - \im \theta/Nq}  \right)^{\bar n}    \delta_{n-\bar n  - W Nq} , 
\label{constrained}
\end{align}
where $K$ denotes the $1$-loop determinant multiplied by the volume of internal moduli, and it is given as $K=m_\psi$ for soft fermion mass. 
The constraint  $n-\bar n  - W qN =0$ guarantees that  the $\theta$ angle dependence  over any configuration contributing to path integral is  of the form $(\rme^{\im \theta/Nq})^{n -\bar n} = \rme^{\im W \theta}, \; W \in \mathbb Z $ as it must be.  
The overall factor $Nq$ is due to $Nq$ distinct classical minima in the fundamental domain. 

We can convert the constrained sum  \eqref{constrained} into a sum on the space of  representation of $\Z_{Nq}$, by using   ${\rm Hom}( \Z_{Nq}, U(1)) = \Z_{Nq}$,  and hence, 
\begin{align}  
\sum_{W \in \Z} \delta_{n-\bar n  - WNq}= \frac{1}{Nq}  \sum_{k=0}^{Nq-1}  
  \rme^{ \im 2 \pi   k (n-\bar n)/Nq} . 
\end{align}
We can now express the partition function as 
\begin{align} 
 Z(\beta)&= 
\sum_{k =0}^{Nq-1}   \sum_{n=0}^{\infty}   \sum_{\bar n =0}^{\infty}   \frac{1}{n!} \frac{1}{\bar n !}  \left( \beta K  \rme^{-S_\mathcal{F} + \im  \frac{ \theta+ 
2 \pi k} {Nq}}  \right)^{n }     \left( \beta K  \rme^{-S_\mathcal{F} -  \im \frac{\theta+ 2 \pi k} {Nq}}  \right)^{\bar n }    \cr 
&= \sum_{k =0}^{Nq-1}   \rme^{  2 \beta  K  \rme^{-S_\mathcal{F}} \cos  \frac{ \theta+ 2 \pi k} {Nq} }
\end{align}
This is the contribution of the $Nq$ lowest lying states to the partition function, which were degenerate to all orders in perturbation theory.
This reproduces the multi-branch structure (\ref{eq:multi_branch_energy}) of the ground state energies:
\begin{align} 
E_{\mathrm{G.S.}}(\theta)= \min_{k=0,1,\ldots,Nq-1} \left(-   2   m_\psi  \rme^{-S_\mathcal{F}} \cos  \frac{ \theta+ 2 \pi k} {Nq}\right).  
\end{align}
We can summarize the main result of this section as follows. 
\begin{itemize}
\item  The existence of fractional topological charge instantons along with the fact that 
 we need to perform a sum over  $U(1)$ gauge configurations with integer topological charges   leads to multi-branched observables  both in quantum mechanics  and corresponding QFT. 
 \item{If we refer to the action of winding number one configuration as $S_{W=1}$, the first non-perturbative saddle that contributes to the semi-classical expansion \eqref{constrained}  has action $2S_{\mathcal F} =    \frac{2 S_{W=1}}{Nq}$. In other words, 
 in the $W=0$ zero sector, minimal action is zero (perturbative vacuum), but non-perturbative saddles have action  $  \frac{ 2S_{W=1} n }{Nq}, n=1,2, \ldots$}
\end{itemize} 
 Note that, if we take $m_\psi \rightarrow 0$ limit, the vacuum family composed of $Nq$-branches  become degenerate.

 \section{Volume independence in $N\to\infty$ limit  and quantum distillations}
\label{sec:vol_indep}

A sub-class of  QFTs with large degrees of freedom 
       have properties which are independent of compactification radius of spacetime $\mathbb{R}^{d-1}\times S^1$. This property is called large-$N$ volume independence \cite{Kovtun:2007py}.     The extreme version of volume independence, where  space-time (lattice)  is reduced to a single   point,   is called Eguchi-Kawai reduction or large-$N$ reduction  \cite{Eguchi:1982nm, GonzalezArroyo:1982hz, Bhanot:1982sh}. 
       There are two 
       necessary and sufficient  conditions  for the validity of the volume independence. In a lattice formulation, these can be phrased as:  
\begin{itemize}
\item Translation symmetry of lattice  $L^d$  is not spontaneously broken. 
\item  Center symmetry is not spontaneously broken.
\end{itemize}
Provided these conditions are satisfied, expectation values and connected correlators of certain interesting  observables  
are independent of compactification radius. In the charge-$q$ Schwinger model, this includes flavor and center singlet observables. 

A  working realization of volume independence is a non-trivial phenomenon.  It demands absence of {\it any} phase transitions 
when the theory is compactified. For example, in 4d gauge theory, it demands the absence of deconfinement transition when the theory is compactified on $\R^3 \times S^1$.   
Deconfinement is a generic behavior with the thermal partition function. However, if one considers a graded partition function such as  $\tr [(-1)^F e^{-L H}]$, there is a special class of gauge theory in 4d that satisfies volume independences. These theories are gauge theories with adjoint fermions. 
 In these theories,  there must exist 
profound spectral  cancellations \cite{Basar:2013sza, Dunne:2018hog, Cherman:2018mya, Sulejmanpasic:2016llc}  to avoid phase transitions. 
This is a  version of quantum distillation, where only a subset of states contribute to the graded partition function.

Below, we comment on the realization of  volume independence in the charge-$q$ $N$-flavor Schwinger model in the large-$N$ limit.  
In particular, we will provide evidence that: 
\begin{itemize}
\item Thermal compactification of charge-$q \geq 1$ models  does not obey volume independence.  
\item $\Omega_F$ twisted compactification  respects volume independence.  
\end{itemize}

First, we provide a perturbative argument. In the thermal compactification, the KK modes  are quantized in the usual manner in units of 
$\frac{2 \pi}{L} $ as shown in Fig.~\ref{fig:q1N3pbc}.  In the twisted compactification, the usual KK-modes intertwine with the flavor-holonomy, and refined KK-modes are now quantized in units of 
$\frac{2 \pi}{LN} $, as shown in Fig.~\ref{fig:q1N3tbc}.
At fixed $L$, in the $N \rightarrow \infty$ limit, the modes become arbitrarily dense and form a continuum.  This is just like perturbative spectrum of the theory on $\R^2$ with one-flavor, and this is sometimes called flavor-momentum transmutation.

Next, let us now discuss the center-symmetry realization of the thermal boundary condition.  For simplicity, we discuss the case when 
we turn on  a soft mass term for the fermion $m_\psi$. In this case,   the minima of the holonomy potential remains unaltered, despite the fact that the potential is weakened. Furthermore,  the fermionic zero modes of the fractons are  lifted so that the transition amplitude between two consecutive minima is non-zero. 
 \begin{align}
 \langle n+1|e^{-\beta H} | n\rangle  \sim   e^{- S_\mathcal{F}}  \; m_\psi^{N}  =    e^{- \frac{\pi N } {m_\gamma  L}} \; m_\psi^{N}. 
  \label{fracton-amplitude}
 \end{align}
In charge-$q$ model with $N$ massive flavor, the one-loop potential 
 \eqref{1-loop-thermal-massive} implies that the $\mathbb Z_q$ center-symmetry is broken to all orders in perturbation theory. However, non-perturbatively, it is restored due to tunnelings between the $q$ perturbative minima.

In the limit $N \rightarrow \infty$,  however, we observe that the barrier in the potential  \eqref{1-loop-thermal-massive} becomes arbitrarily large while the location of the minima is unaltered. In this case,   
  \begin{align}
 \lim_{ N\rightarrow \infty} \left[  \langle n+1|e^{-\beta H} | n\rangle \sim    e^{- S_\mathcal{F}}  = 
 e^{- \frac{\pi N } {m_\gamma  L}} \right] =0 
  \label{fracton-amplitude2}
 \end{align}
 and the  tunneling amplitude is  zero.  Therefore, in the  $N \rightarrow \infty$ limit of charge-$q$ theory, the  $\mathbb Z_q$ center-symmetry is  spontaneously broken.   This implies that in the thermal case, large-$N$ volume independence fails.  
 Indeed, just inspecting the free energy density,  we observe the temperature dependence,  
 $ {\cal F}_{\rm thermal}  = -   N T^2 \frac{\pi}{6}   $  at leading order in $N$.

Lastly, we show that,  by using the twisted boundary condition with $\Omega_F$, we can 
 keep center-symmetry intact in the $N \rightarrow \infty$ thermodynamic limit, and volume independence is established.

As described in  Sec.~\ref{sec:chiral_condensate_qN_tbc}, 
 with the twist, there are $Nq$ type fractons, each of which has exactly two-fermionic zero modes.  
  Again, turning on a soft mass term for the fermion $m_\psi$,  
the chiral symmetry is explicitly broken.       The zero modes of fracton can be soaked up by the soft mass term and  the fracton amplitudes take the form ${\cal F}_j \sim   e^{- S_j }  \; m_\psi $.  

As described around \eqref{twisted-full},  the zero-form part $\Z_q^{[0]}$   of the 1-form center symmetry $\Z_q^{[1]}$ is   enhanced to  $\Z_{qN}^{[0]}$  in the compactified theory  due  to $\Omega_F$ background. 
  Below, we would like to discuss the realization of this symmetry.   Again, at finite-$N$, to all orders in perturbation theory, 
  the  $\Z_{qN}^{[0]}$ center-symmetry is broken, and non-perturbatively it is restored due to tunneling. The question is whether this restoration remains intact in the $N \rightarrow \infty$  limit. In this case, 
   \begin{align}
 \langle n+1|e^{-\beta H} | n\rangle  \sim      e^{- \frac{\pi } { N m_\gamma  L}} \; m_\psi,
  \label{fracton-amplitude}
 \end{align}
 which remains finite even for large-$N$, implying unbroken center at large-$N$. Despite the fact that this conclusion is correct, the reasoning is not. In fact, we cannot even  take $N \rightarrow \infty$ limit with the fracton action at fixed $L$,  because 
semi-classical  dilute instanton  gas approximation breaks down    when $S_{\mathcal{F}}=  \frac{\pi  } { N m_\gamma  L} \sim 1$.  

In the large-$N$ limit,  we can reliably state that 
$ V_{\Omega_F}(a)    \rightarrow 0$ and holonomy field $a$ direction becomes flat. The ground state wave function becomes the one of particle on a circle with no barriers in between. 
Since the tunneling becomes irrelevant in this regime,  the $\mathbb Z_{qN}$ center-symmetry is perturbatively  unbroken.  This guarantees that in the large-$N$ limit of the QED$_2$ with $\Omega_F$ twist, volume-independence holds. 

\begin{figure}[t]
\centering
\includegraphics[width=120mm]{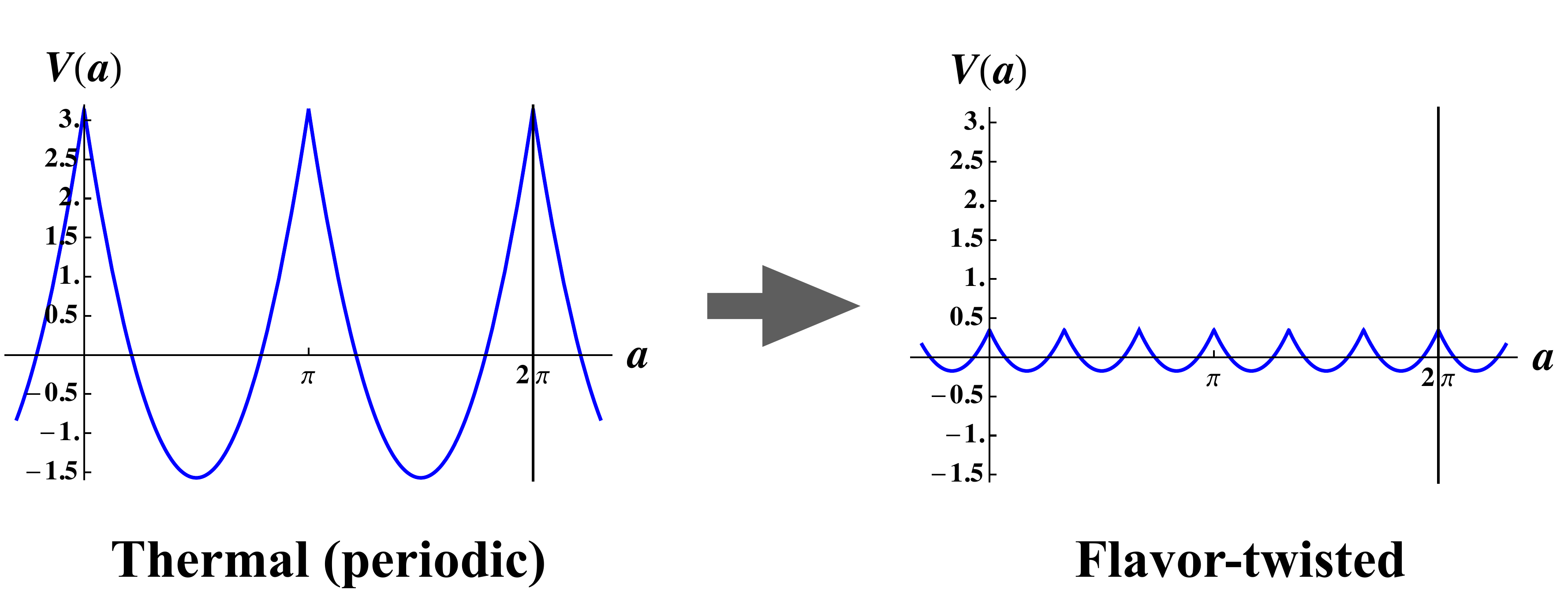}
\caption{The holonomy potential and free energy for thermal and twisted theory $q=2, N=3$. 
With the inclusion of the twist $\Omega_F$, potential  is suppressed by a factor $\frac{1}{N^2}$. The image of this process 
in   Hilbert space is  cancellation among states in the graded state sum, a process that we refer to as quantum distillation.  }
\label{fig:q2N3_sup}
\end{figure}

As a test, note the free energy in the $\Omega_F$ twisted background. In this case, free energy becomes  
$ {\cal F}_{\Omega_F}  = -   \frac{1}{N} T^2 \frac{\pi}{6}   $ at leading order in large-$N$ expansion.  Indeed, both $O(N^1)$ and $O(1)$  parts of the free energy is temperature independent. This can be clearly seen in Fig.~\ref{fig:q2N3_sup}.

\subsection{Interpretation as quantum distillation of Hilbert space}

So far, we have discussed the $O(N^{-2})$ suppression of the twisted free energy compared with the thermal free energy in microscopic viewpoint. 
In this section, we provide its reinterpretation in view of physical Hilbert space. 

In operator formalism, 
the thermal  and graded partition functions
correspond to 
\begin{align}
Z(\beta)&= \tr \Big[ e^{-\beta H}  \Big]  \cr 
Z_{\Omega_F} (\beta) &= \tr \Big[ e^{-\beta H} (-1)^F     \prod_{f=1}^{N} e^{i \frac{2\pi (f-1)}{N}  Q_f}  \Big],
\end{align} 
where $Q_f$ is the fermion number operator for species $f$, and $H$ is the Hamiltonian of QFT on $\R^1 $ or its compactification to  $S^1_\ell$.  Note that we consider  QFT on ${\bf T}^2= S^1_\beta \times S^1_\ell$ and assume boundary  condition is twisted in the $\beta$ circle. $S^1_\ell$ provides a regularization for $\R$.  
The Hilbert space data that enters to these partition functions are identical, and concerns the spectrum of the Hamiltonian $H= H_{\R}$.  However, in the graded sum, there are significant   cancellations among states. 

As an explicit example, consider   a state in the adjoint representation of the $SU(N)_V$, created by  
\begin{align} 
M_{f}^{f'}= \overline\psi_f \psi^{f'},  \qquad f, f'=1, \ldots, N 
\end{align}
In thermal case,   these state would contribute to free energy as $e^{ -\beta E_{\rm adj}} (N^2-1) $. 
However,  in the graded sum,  $\prod_{f=1}^{N} e^{i \frac{2\pi (f-1)}{N}  Q_f}$ assigns different phases to different states. 
For the adjoint representation, this assignment is 
\begin{align} 
M_{f}^{f'} \rightarrow   e^{ - 2 \pi i (f-f')/N} M_{f}^{f'} 
\label{adjoint}
\end{align} 
modifying the terms in the state sum into 
\begin{align} 
 (N^2-1) e^{-\beta E_{\rm adj} } \longrightarrow \left( \sum_{f, f'=1}^{N}    e^{ - 2 \pi i (f-f')/N} -1\right)   e^{-\beta E_{\rm adj}} =(-1) e^{-\beta E_{\rm adj}} 
\end{align} 
reducing the effect of $(N^2-1)$ degree of freedom to $(-1)$. 
This type of cancellation will occur in the  graded partition function for all states, 
except for the singlets, including the ground states.  

In order to understand the effect of these type of cancellation in thermodynamics,  we recall the relation between partition functions  and density of states following \cite{Cherman:2018mya}.  The inverse Laplace transform of the partition function is density of states. In the thermal case, the theory will exhibit a density of state expected from the Cardy formula for $2$d massless Dirac fermions,
\begin{align}
Z(\beta) \sim   \rme^{\frac{  N}{  \beta } \frac{\pi}{6} \ell }  \underbrace{\Longleftrightarrow}_{\rm Laplace \;  transform}  \rho(E) \sim \rme^{\sqrt { \frac{2 \pi}{3} N E \ell}},
\label{first-dos}
\end{align}
where the central charge is $N$. 
 In the graded case, we perform a graded sum over the Hilbert space of the theory. There are many cancellation and what is left out of these cancellations is the thermodynamic worth of the graded partition function. We find that 
\begin{align} 
Z_{\Omega_F} (\beta) \sim   \rme^{\frac{  1}{N  \beta } \frac{\pi}{6} \ell }  \underbrace{\Longleftrightarrow}_{\rm Laplace \;  transform}   \rho_{\rm twisted}(E) \sim \rme^{\sqrt { \frac{2 \pi}{3N}  E \ell}}
\label{second-dos}
\end{align}

The density of states  \eqref{first-dos} is the largest growth expected from a 2d-local QFT with $N$ degrees of freedom, corresponding to Stefan-Boltzmann growth.  In the twisted case,  the density of states will correspond to counting of states after cancellations triggered by the grading over Hilbert space.  
This growth  is clearly  extremely small compared to the thermal case. In particular, in the $N \rightarrow \infty$ limit, it is as if all states are eliminated after twisting and grading.  

Normally, the competition between the growth in the density of states against  the Boltzmann suppression   is responsible for the changes in the saddles of the partition function and phase transition. At large-$\beta$, only grounds states and low lying states contribute. But at low-$\beta$, many states may contribute on the same footing resulting possibly in phase transitions. 
 In the graded case, the growth of density of states is effectively diminished and only a hand full  of states contribute to state sum both at  large- and small-$\beta$.   This can and does prevent the possibility of phase transitions.  These type of spectral cancellations, reminiscent of supersymmetric QFTs, is at the very origin of the working realizations of  large-$N$ volume independence, both in QED$_2$ or 4d gauge theories such as QCD(adj).

%

\section{Twisted-compactification of $SU(N)_k$ Wess-Zumino-Witten model}\label{sec:twisted_wzw}

As we have explained in Sec.~\ref{sec:bosonization}, the low-energy behavior of charge-$1$ multi-flavor Schwinger model is described by the level-$1$ Wess-Zumino-Witten (WZW) model. 
More generally, the $SU(N)_k$ WZW model is defined as 
\be
S={1\over 2g}\int_{M_2}\tr[\diff U^\dagger \wedge \star \diff U]+ \im k \Gamma_{\rm WZ}[U], 
\ee
where the Wess-Zumino term is defined as 
\be
\Gamma_{\rm WZ}={1\over 12\pi}\int_{M_3} \tr[(U^\dagger \diff U)^3]. 
\ee
The RG flow has the conformal fixed point at $g=4\pi/k$. 

In this section, we show that this conformal behavior is responsible for the extra $\mathbb{Z}_N$ degeneracy of the ground states under the flavor-twisted boundary condition. 
We therefore consider the twisted boundary condition (\ref{eq:twisted_bc}), 
\be
u_{ij}(x,\tau+L)=\omega^{i-j} u_{ij}(x,\tau), 
\ee
with $U=[u_{ij}]$. The case $k=0$ with twisted boundary condition is studied in \cite{Cherman:2014ofa}.  The theory with this boundary condition has the $\mathbb{Z}_N$ symmetry, $u_{ij}\mapsto u_{i+1,j+1}$, and also the maximal Abelian subgroup of $G$. 

\subsection{Wess-Zumino term and topological $\theta$ terms}

Let us evaluate the Wess-Zumino term under the flavor-twisted boundary condition at small compactification radius. 

\subsubsection{$SU(2)$ case}

We first compute the $SU(2)$ case. In order to look at the classical vacua under the twisted boundary condition, it is convenient to take the Hopf coordinate,
\be
U=\begin{pmatrix}
\rme^{\im \phi}\sin\eta & \rme^{\im \xi}\cos\eta\\
-\rme^{-\im \xi}\cos \eta & \rme^{-\im \phi}\sin \eta
\end{pmatrix}. 
\ee
Here, $\phi$ and $\xi$ are $2\pi$ periodic variables, and $\eta\in [0,\pi/2]$. The twisted boundary condition says 
\be
\phi(x,\tau+L)=\phi(x,\tau),\;  \eta(x,\tau+L)=\eta(x,\tau), 
\ee
and 
\be
\xi(x,\tau+L)=\xi(x,\tau)+\pi. 
\ee
Thus, only $\xi$ should have nontrivial $\tau$ dependence  in the small compactification limit, i.e. it does not have KK zero mode. 

The kinetic term is given as 
\be
\tr[\p_\mu U^\dagger \p_\mu U]=2[\sin^2\eta (\p_\mu \phi)^2+\cos^2\eta (\p_\mu \xi)^2+(\p_\mu \eta)^2]. 
\ee
Because of the twisted boundary condition, the classical configuration of $\xi$ satisfies
\be
(\p_2\xi)^2 = \left({\pi\over L}\right)^2, 
\ee
and thus $\eta=\pi/2$ is energetically favored so that $\cos\eta=0$. 
We therefore approximate the kinetic term as 
\be
\tr[\p_\mu U^\dagger \p_{\mu} U]=2 (\p_\mu \phi)^2,  
\ee
by neglecting quantum fluctuations around $\xi=\pi \tau/L$ and $\eta=\pi/2$, i.e. 
\be
U\Bigr|_{M_2}=\begin{pmatrix}
\rme^{\im \phi}&0\\
0&\rme^{-\im \phi}
\end{pmatrix}. 
\ee 
Strictly speaking, this naive manipulation is subtle quantum mechanically~\cite{Fujimori:2016ljw}, and this is carefully considered in Ref.~\cite{Evslin:2018yfm} in the context of principal chiral model. It, however, turns out to be useful to understand the role of the WZ term under the twisted boundary condition. 
Especially, the structure of symmetry and 't~Hooft anomaly is maintained. 

Next, let us evaluate the WZ term setting $\phi=\phi(x)$, $\xi=\pi\tau/L$, and $\eta=\pi/2$ on $M_2=M_1\times S^1$. Using the Hopf coordinate,  the WZ term becomes 
\be
\Gamma_{\rm WZ}=-{1\over 2\pi}\int_{M_3}\sin (2\eta) \, \diff \phi \wedge \diff \xi \wedge \diff \eta. 
\ee
We take $M_3=M_2\times [0,1]$ and regard $M_2\times\{1\}$ as a point so that $\p M_3=M_2\times \{0\}\simeq M_2$. For this purpose, we need to set $U\bigr|_{M_2\times \{1\}}=\bm{1}$, while satisfying the boundary condition, and we pick up the following extension along $x^3\in [0,1]$,
\be
\begin{array}{cccccc}
x^3:&0& \longrightarrow& {1\over 2} &\longrightarrow & 1,\\
U:&\begin{pmatrix}
\rme^{\im \phi(x^1)}&0\\
0&\rme^{-\im \phi(x^1)}
\end{pmatrix} & \xrightarrow{\eta:{\pi\over 2}\to 0} & 
\begin{pmatrix}
0 & \rme^{\im \pi\tau\over L}\\
-\rme^{-\im \pi\tau\over L} & 0 
\end{pmatrix}& \xrightarrow{\eta:0\to {\pi\over2}} &
\bm{1}. 
\end{array} 
\ee
Using this parametrization, we get 
\be
\Gamma_{\rm WZ}= {\pi \over 2\pi}\int \diff \phi, 
\ee
and the theory becomes the particle on a circle with the theta angle $\theta=\pi k$:
\be
S={L\over g}\int |\diff \phi|^2+{\im \pi k\over 2\pi}\int \diff \phi. 
\ee

When $k$ is even, i.e. $k=0,2,\ldots$, the compactified WZW model has the unique ground state, and the energy gap is explained perturbatively with the formula $\sim g/2L$. 
When $k=0$, the model is nothing but the $SU(2)$ principal chiral model, and its perturbative nature of the energy gap is emphasized in Ref.~\cite{Evslin:2018yfm}. 

\subsubsection{$SU(N)$ case}

We extend the previous discussion to the case of general $N$. 
As in the $SU(2)$ case, we can argue that the classical moduli locates at the maximal torus $U(1)^{N-1}\subset SU(N)$ because of the flavor-twisted boundary condition, and we denote such a matrix as
\be
U=\begin{pmatrix}
\rme^{\im \phi_1}&0&0\\
0&\rme^{\im \phi_2}&0\\
0&0&\rme^{-\im(\phi_1+\phi_2)}
\end{pmatrix}
=\begin{pmatrix}
\rme^{\im \phi_1}&0&0\\
0&\rme^{-\im \phi_1}&0\\
0&0&1
\end{pmatrix}
\begin{pmatrix}
1&0&0\\
0&\rme^{\im (\phi_1+\phi_2)}&0\\
0&0&\rme^{-\im(\phi_1+\phi_2)}
\end{pmatrix}. 
\ee
That is, we regard the classical vacua $U$ as the successive multiplication of $2\times 2$ Cartan factors. 
Here, we explicitly work on $SU(3)$ case, but the generalization is straightforward. 

We consider the extension of $U$ to extra dimension successively. That is, we take $M_3=M_2\times ([0,1]\cup [1,2])$.  On $M_2\times [0,1]$, we parametrize $U$ as  
\be
U
=\begin{pmatrix}
\rme^{\im \phi_1}\sin\eta&\rme^{\im {2\pi \tau\over 3L}}\cos \eta&0\\
-\rme^{-\im{2\pi \tau\over 3L}}\cos \eta&\rme^{-\im \phi_1}\sin\eta&0\\
0&0&1
\end{pmatrix}
\begin{pmatrix}
1&0&0\\
0&\rme^{\im (\phi_1+\phi_2)}&0\\
0&0&\rme^{-\im(\phi_1+\phi_2)}
\end{pmatrix}, 
\ee
and the first $2\times 2$ factor is set identity at $x^3=1$ as we did in the $SU(2)$ case. On $M_2\times [1,2]$, we parametrize
\be
U
=\begin{pmatrix}
1&0&0\\
0&1&0\\
0&0&1
\end{pmatrix}
\begin{pmatrix}
1&0&0\\
0&\rme^{\im (\phi_1+\phi_2)}\sin \eta'&\rme^{\im {2\pi \tau\over 3L}}\cos \eta'\\
0&-\rme^{-\im {2\pi \tau\over 3L}}\cos \eta'&\rme^{-\im(\phi_1+\phi_2)} \sin\eta'
\end{pmatrix}, 
\ee
and perform the same trick for the second $2\times 2$ factor to set it identity at $x^3=2$. 
Since this is the inductive procedure, this computation also generalizes to $SU(N)$ cases in a straightforward manner. 

As a consequence, we find that 
\bea
\Gamma_{\rm WZ}&=&{2\pi\over N}{1\over 2\pi}\int [(\diff \phi_1)+(\diff \phi_1+\diff \phi_2)+\cdots+(\diff \phi_1+\cdots+\diff \phi_{N-1})\nonumber\\
&=&-{1\over 2\pi}\int \sum_{n=1}^{N-1} {2\pi n\over N}\diff \phi_n\qquad (\bmod 2\pi),  
\eea
and thus the problem becomes the $(N-1)$ particles on a circle with the theta angles $\theta_n=2\pi k n/N$. 
When the level of WZW model is $k=0$ modulo $N$, we again find that the ground state is unique and the energy gap at the leading order is explained perturbatively, $\Delta E\sim g/2L$, which generalizes the result of Ref.~\cite{Evslin:2018yfm}.

\subsection{Symmetry, Anomaly, and Energy spectrum}

The $\mathbb{Z}_N^{[0]}$ symmetry acts as 
\be
\mathbb{Z}_N^{[0]}:\phi_n\mapsto \phi_{n+1}\; (n=1,\ldots,N-2),\quad \phi_{N-1}\mapsto -(\phi_1+\cdots+\phi_{N-1}), 
\ee
which changes the $\theta$ angles as 
\bea
&&\theta_n\mapsto \theta_{n-1}-\theta_{N-1}=\theta_n-2\pi k\quad (n=2,\ldots,N-1),\\
&&\theta_1\mapsto -\theta_{N-1}=\theta_1-2\pi k. 
\eea
The difference of the action is 
\be
\Delta S= \im k \int (\diff \phi_1+\cdots+\diff \phi_{N-1})=0 \bmod 2\pi \im,
\ee
and thus this transformation is the symmetry. The theory also has the symmetry $U(1)^{N-1}\subset SU(N)_\rmR$, which acts on $\phi_n$ as 
\be
\phi_n\mapsto \phi_n+\alpha_n. 
\ee

Let us gauge the $U(1)^{N-1}$ symmetry by introducing $U(1)$ gauge fields $A_n$, and then the gauged action is 
\bea
S_{\rm gauged}&=&{L\over 2g}\int \left(\sum_{n=1}^{N-1}|\diff \phi_n+A_n|^2+\left|\sum_{n=1}^{N-1}(\diff \phi_n+A_n)\right|^2\right)\nonumber\\
&&-{\im k\over 2\pi} \sum_{n=1}^{N-1}{2\pi n\over N}\int (\diff \phi_n +A_n). 
\eea
We define the $\mathbb{Z}_N^{[0]}$ transformation on gauge fields as $A_n\mapsto A_{n+1}$ for $n=1,\ldots,N-2$, and $A_{N-1}\mapsto -(A_1+\cdots +A_{N-1})$.
The partition function $Z[A_n]$ changes under the $\mathbb{Z}_N^{[0]}$ transformation as 
\be
Z[A_n]\mapsto Z[A_n] \exp\left(-\im k\int (A_1+\cdots+ A_{N-1})\right),  
\ee
and this is the mixed 't~Hooft anomaly between $\mathbb{Z}_N^{[0]}$ and $U(1)^{N-1}$. 

We can find the energy $E_{\{m_n\}}$ of $U(1)^{N-1}$ charges $\{m_n\}\in \mathbb{Z}^{N-1}$ by~\cite{Tanizaki:2018xto, Kikuchi:2017pcp} 
\be
\exp(-\beta E_{\{m_n\}})=\int \Diff A_n Z[A_n] \exp\left(-\im \sum_{n}m_n \int A_n\right), 
\ee
where $\beta$ is the imaginary time, where $E_{\{m_n\}}$ are given by
\be
E_{\{m_n\}}={L\over 2g}\left[\sum_{n=1}^{N-1}\left(m_n-k{n\over N}\right)^2-{1\over N}\left(\sum_{n=1}^{N-1}\left(m_n-k{n\over N}\right)\right)^2\right]. 
\ee
Performing the $\mathbb{Z}_N^{[0]}$ transformation, the $U(1)^{N-1}$ charges are changed as 
\bea
m_{1}\mapsto -m_{N-1}+k,\; m_{n}\mapsto m_{n-1}-m_{N-1}+k \; (n=2,\ldots, N-2). 
\eea
This explains the $N$-fold degeneracy for $k\not=0$ mod $N$; for example, when $k=1$, the ground states are $N$-fold degenerate with 
\be
(m_1,\ldots,m_{N-1})=(0,\ldots, 0),\, (0,\ldots, 0, 1),\, (0,\ldots, 0, 1, 1), \ldots, (1,\ldots, 1). 
\ee
The coordinate expression of the state $|m_1,\ldots,m_{N-1}\rangle$ is given as 
\be
\langle \phi_1,\ldots, \phi_{N-1}|m_1,\ldots,m_{N-1}\rangle=\rme^{\im m_1\phi_1+\cdots+\im m_{N-1}\phi_{N-1}}, 
\ee
and thus 
\be
\langle m'_1,\ldots, m'_{N-1}|\rme^{\im \widehat{\phi}_n}|m_1,\ldots, m_{N-1}\rangle=\delta_{m'_1,m_1}\cdots \delta_{m'_n,m_n+1}\cdots \delta_{m'_{N-1},m_{N-1}}. 
\ee
Let us compute the vacuum expectation value of the fermion-bilinear scalar operator, 
\be
\overline{\psi}_\rmR \psi_\rmL\sim \rme^{\im \phi_1}+\cdots +\rme^{\im \phi_{N-1}}+\rme^{-\im(\phi_1+\cdots+\phi_{N-1})}. 
\ee
By taking the following superposition of the $N$ ground states, 
\be
|n\rangle= |0,\ldots,0\rangle+\omega^{n} |0,\ldots, 0,1\rangle +\cdots+\omega^{n(N-1)}|1,\ldots,1\rangle,
\ee
with $\omega=\rme^{2\pi\im/N}$, we find that  
\be
\langle n|\overline{\psi}_\rmR\psi_\rmL|n\rangle \sim \rme^{2\pi\im n/N}. 
\ee
We can compare this result with that of Sec.~\ref{sec:ChSSB}, especially Sec.~\ref{sec:NflavorSMQM}. 
Especially, we have clearly shown that the extra $\mathbb{Z}_N$ degeneracy of the ground states under the flavor-twisted boundary condition comes out of the $2$d conformal behavior.


\section{Conclusion and Outlook}\label{sec:conclusion}

What we have newly shown or obtained in the present work on the charge-$q$ $N$-flavor Schwinger model ($2$d QED) are summarized as follows;
\begin{itemize}
\item
full 't Hooft anomaly consistent with the results obtained analytically,
\item
$1/N^2$-suppressed holonomy potential for flavor-twisted cases under compactification,
\item
$Nq$ vacua and chiral condensate with fractional $\theta$ dependence for flavor-twisted cases,
\item
the expression of chiral condensate valid for all the range of the circumference
\item
the direct consequence of (fractional) quantum instantons on the physical quantities,
\item
the $Nq$-branch structure and the pattern of symmetry breaking for massive cases,
\item
new insights into the volume independence in the model,
\item
understanding on the WZW model as a dual theory of the Schwinger model. 
\end{itemize}
These results themselves are of great significance for understanding $2$d quantum field theory.
We below discuss their implications on other theories including $4$d QCD and string theory.

{\it 4d QCD}: It is worth while to investigate the vacuum structure of $4$d $N$-flavor QCD with ${\mathbb Z}_{N}$-twisted boundary condition on ${\mathbb R}^{3} \times S^{1}$ (${\mathbb Z}_{N}$-QCD) \cite{Kouno:2012zz, Sakai:2012ika, Kouno:2013zr, Kouno:2013mma, Kouno:2015sja, Iritani:2015ara, Cherman:2016hcd, Tanizaki:2017mtm} in comparison to our results for the $2$d $N$-flavor Schwinger model with the flavor twist on ${\mathbb R} \times S^{1}$.
In particular the $\theta$ vacuum structures in the two theories could share common properties.
Since the ${\mathbb Z}_{N}$-QCD results in the usual $N$-flavor QCD in a decompactification limit,
this avenue also has potential impact on the study of non-supersymmetric gauge theory on ${\mathbb R}^{4}$. 

{\it Brane configurations}: O$1^{-}$-$\overline{{\rm D}1}$ brane configurations corresponding 
to the charge-$2$ $8$-flavor Schwinger model are investigated in \cite{Armoni:2018bga}, where
$8$-flavor fermions and scalars are associated with eight dimensions transverse to $\overline{{\rm D}1}$ brane.
We can raise a question whether one can construct the brane configuration corresponding to the flavor-twisted Schwinger model, which has been the main topic in this work.
One here needs to consider how to introduce ${\mathbb Z}_{8}$ holonomy of the $U(1)$ gauge field in the brane configuration.
On the other hand, the brane configurations for the twisted ${\mathbb C}P^{N-1}$ and Grassmann models on ${\mathbb R} \times S^{1}$ were discussed in \cite{Misumi:2014bsa},
where multiple D$4$ branes are introduced to realize the T-dualized configurations. 
It is still an open problem whether we can apply a similar technique to the present problem.

{\it Resurgent structure}: One of striking properties in the charge-$q$ $N$-flavor Schwinger model is that fractional quantum instantons have direct consequence on physical quantities. The recent progress in the resurgent structure of quantum mechanics and field theories indicates the significance of fractional instantons and their composite objects called bions \cite{Argyres:2012vv, Argyres:2012ka, Dunne:2012zk, Dunne:2012ae}, whose contributions cancel out the imaginary ambiguities arising from perturbation series of physical quantities. It is quite intriguing to derive the bion contributions and compare them to the perturbative calculation in order to investigate the resurgent structure in the charge-$q$ $N$-flavor Schwinger model.

\acknowledgments
The authors thank S.~Sugimoto, M.~Anber, A.~Cherman, and T.~Sulejmanpasic for useful discussion. 
Especially, T.~M. appreciates discussion with S.~Sugimoto, and thanks the organizers of ``KEK Theory workshop 2018'' for giving them the opportunity.
The work of T.~M. was in part supported by the Japan Society for the Promotion of Science (JSPS) Grant-in-Aid for Scientific Research (KAKENHI) Grant Numbers 16K17677, 18H01217 and 19K03817.
The work of T.~M. was also supported by the Ministry of Education, Culture, Sports, Science, and Technology(MEXT)-Supported Program for the Strategic Research Foundation at Private Universities ``Topological Science'' (Grant No. S1511006).
The work of Y.~T. was partly supported by JSPS Overseas Research Fellowships. 
M.~U. acknowledges support from U.S. Department of Energy, Office of Science, Office of Nuclear Physics under Award Number DE-FG02-03ER41260.
This work was performed in part at ``Higher Symmetries conference 2019'' at Aspen Center for Physics, which is supported by National Science Foundation grant PHY-1607611. The authors appreciate their hospitality during the program.

\appendix

\section{Two-dimensional Dirac spinor}\label{sec:convention_spinor}

Throughout this paper, we take the following convention for the $2$-dimensional Dirac spinor in Euclidean metric.

Let $\gamma^{\mu}$ be $2\times 2$ matrices, satisfying 
\be
\{\gamma^{\mu},\gamma^{\nu}\}=2\delta^{\mu\nu},\; (\gamma^{\mu})^\dagger=\gamma^\mu, 
\ee
for $\mu,\nu=1,2$. These relations can be satisfied by using Pauli matrices as 
\be
\gamma^1=\sigma_1=
\begin{pmatrix}
0&1\\
1&0
\end{pmatrix},\; 
\gamma^2=\sigma_2=
\begin{pmatrix}
0&-\im \\
\im &0
\end{pmatrix}. 
\ee
We define the chirality matrix $\gamma$ by 
\be
\gamma=-\im \gamma^1\gamma^2, 
\ee
and this satisfies $\{\gamma^\mu,\gamma\}=0$, $\gamma^\dagger=\gamma$, and $(\gamma)^2=1$. Using the Pauli matrix, it is given by
\be
\gamma=\sigma_3=\begin{pmatrix}
1&0 \\
0 &-1
\end{pmatrix}.
\ee
The right/left-handed projectors $P_{\rmR/\rmL}$ are given by
\be
P_{\rmR}={1+ \gamma\over 2}=
\begin{pmatrix}
1&0 \\
0 &0
\end{pmatrix}, \; 
P_{\rmL}={1- \gamma\over 2}=
\begin{pmatrix}
0&0 \\
0 &1
\end{pmatrix},
\ee
respectively. 

The Dirac Lagrangian is given by 
\be
\mathcal{L}=\overline{\psi} (\slashed{\p}+m)\psi, 
\ee
where $\psi$ is the two-component spinor field and $\overline{\psi}$ is its conjugate field, $\slashed{\p}=\gamma^\mu \p_{\mu}$, and $m$ is the fermion mass. 
We define the right-handed fermion and its conjugate as 
\be
\psi_\rmR=P_\rmR \psi,\; \overline{\psi}_\rmR=\overline{\psi}P_\rmL,
\ee
and the left-handed ones as 
\be
\psi_\rmL=P_\rmL \psi,\; \overline{\psi}_\rmL=\overline{\psi}P_\rmR.
\ee
The Dirac Lagrangian is then written in the following form:
\bea
\mathcal{L}&=&\overline{\psi}_\rmR \slashed{\p}\psi_\rmR+\overline{\psi}_\rmL \slashed{\p}\psi_\rmL+m(\overline{\psi}_\rmR\psi_\rmL+\overline{\psi}_\rmL \psi_\rmR)\nonumber\\
&=& \overline{\psi}_\rmR (\p_1+\im \p_2)\psi_\rmR+\overline{\psi}_\rmL (\p_1-\im \p_2)\psi_\rmL+m(\overline{\psi}_\rmR\psi_\rmL+\overline{\psi}_\rmL \psi_\rmR). 
\eea
Here, we slightly abuse the notation: In the first line, $\psi_\rmR$ denotes the two-component spinor but it only has the upper component, while $\psi_\rmR$ in the second line denotes its upper component. The similar notice holds for others.

The motivation of the above definition for conjugate fields comes from the fact that, under the temporal reflection $\Theta:(x^1,x^2)\mapsto (-x^1,x^2)$, the reflection positivity $\langle \Theta(\mathcal{O})\mathcal{O}\rangle\ge 0$ is satisfied with the anti-linear operation $\Theta$ generated by 
\be
\Theta:\psi(x)\mapsto (\overline{\psi}(\Theta\cdot x)\gamma^1)^T,\; \overline{\psi}(x)\mapsto (\gamma^1 \psi(\Theta\cdot x))^T. 
\ee
With the above definition, we find that $\Theta:\psi_\rmR\mapsto (\overline{\psi}_\rmR\gamma^1)^T$ and $\psi_\rmL\mapsto (\overline{\psi}_\rmL\gamma^1)^T$ since $\gamma$ anti-commutes with $\gamma^\mu$. 
We note, in this notation, that $\gamma \psi_\rmR=\psi_\rmR$ but $\overline{\psi}_\rmR \gamma=-\overline{\psi}_\rmR$.

\section{Holonomy potential via bosonization} \label{sec:holonomy_bosonization}
Below, we show that the holonomy potential (\ref{1-loop-thermal}) can also be obtained by using  Abelian bosonization.  
In the original description, holonomy potential arises from integrating out fermions in the background $U(1)$ gauge field $a_2(\tau,x)=a(\tau)$, and is a one-loop effect. In the bosonized description, it is a tree level effect.

First, let us summarize the setup. 
For simplicity, we consider $N=1$-flavor charge-$q$ Schwinger model, and we set $\theta=0$.  The bosonized action is given in \eqref{eq:1flavor_bosonized}, and integration-by-part gives;
\be
S=\int_{M_2}\left({1\over 2e^2}|\diff a|^2+{1\over 8\pi}|\diff \phi|^2- {\im q\over 2\pi} \diff\phi\wedge  a\right).
\ee
Completing the square in terms of $\diff \phi$, we can find that the  gauge field gets the mass $m_\gamma=qe/\sqrt{\pi}$. 

We take $M_2=T^2\ni (\tau,x)$, with the identification $\tau+\beta\sim \tau$, $x+L\sim x$, and $L\ll \beta$. 
We will take $\beta\to \infty$ in the end, so we work on the temporal gauge $a_1\equiv 0$ (Precisely speaking, we cannot take temporal gauge naively at finite $\beta$, but this does not affect our discussion on the effective potential). 
Assuming $eL\ll 1$, we would like to derive the effective potential for the holonomy $\exp(\im L a_2)$.  
For that purpose, we put $a_2$ to be a constant, $a_2\equiv h$, so that 
\be
a=h\diff x. 
\ee
We denote the compact scalar field $\phi$ as 
\be
\phi(\tau,x)={2\pi n\over \beta}\tau+\widetilde{\phi}(\tau),
\ee
where $\tilde{\phi}$ is the $\mathbb{R}$-valued scalar, and we neglect its non-zero KK modes. 
We shall see that it is important to keep the winding number $n$ along the temporal direction even how $\beta$ is large. 

Substituting our approximation on the fields into the bosonized action, we obtain 
\bea
S&=&L\int\diff \tau \left({1\over 8\pi}\left({2\pi\over \beta}n+\dot{\widetilde{\phi}}\right)^2-{\im q\over 2\pi}\left({2\pi\over \beta}n+\dot{\widetilde{\phi}}\right)h\right)\nonumber\\
&=&L\int \diff \tau {1\over 8\pi}\dot{\widetilde{\phi}}^2+{L\over \beta}{\pi\over 2}n^2-\im L n q h,
\eea
and we drop the total-derivative terms to obtain the second line. The effective potential $V(h)$ is obtained as 
\bea
\exp(-\beta L V(h))&=&\sum_{n\in \mathbb{Z}}\int \Diff \widetilde{\phi} \exp(-S)
=\sum_n \exp\left[-{L\pi\over 2\beta}n^2+\im n q Lh\right]\nonumber\\
&=&\sum_k \exp \left[-L\beta {q^2\over 2\pi}\left(h-{2\pi k\over qL}\right)^2\right], 
\eea
up to an overall normalization constant, or constant shift of $V$. To find the last expression, we use Poisson summation formula. As a consequence, by taking $\beta\to \infty$, we obtain the effective action,
\be
V(a)=\min_k {q^2\over 2\pi}\left(a-{2\pi k\over qL}\right)^2. 
\ee
This coincides with (\ref{1-loop-thermal}) for $N=1$ by shifting $a\to a+{\pi/qL}$, and this shift corresponds to adding the imaginary chemical potential to make the fermion boundary condition to be periodic. 

\bibliographystyle{utphys}
\bibliography{./QFT,./refs,./lefschetz}
\end{document}